\documentclass[10pt]{article}       
\usepackage{cite}
\usepackage[utf8]{inputenc}
\usepackage{graphicx}
\usepackage{amsmath,amssymb}
\usepackage{mathptmx}      
\usepackage[varg]{txfonts}   

\if@mathematic
   \def\runinend{} 
   \def\floatcounterend{\enspace}
   \def\sectcounterend{}
\else
   \def\runinend{.}
   \def\floatcounterend{.\ }
   \def\sectcounterend{.}
\fi

\def\keywords#1{\par\addvspace\medskipamount{\rightskip=0pt plus1cm
\def\and{\ifhmode\unskip\nobreak\fi\ $\cdot$
}\noindent\keywordname\enspace\ignorespaces#1\par}}
\def\PACSname{\textbf{PACS}\enspace}
\def\PACS#1{\par\addvspace\medskipamount{\rightskip=0pt plus1cm
\def\and{\ifhmode\unskip\nobreak\fi\ $\cdot$
}\noindent\PACSname\ignorespaces#1\par}}

\def\bear{\be\begin{array}}
\def\eear{\end{array}\ee}
\newcommand{\abst}{\hspace*{1ex} }

\newcommand{\gsim}{\mbox{\raisebox{-0.3ex}{\footnotesize $\:\stackrel{>}{\sim}\:$}} }
\newcommand{\mytexttilde}{{\raise.17ex\hbox{$\scriptstyle\mathtt{\sim}$}}}
\newcommand{\kB}{k_{\mathrm{B}}}

\newcommand{\ha}{\frac12 }
\newcommand{\cosW}{\cos^2 \Theta_W}
\newcommand{\siW}{\sin \Theta_W}
\newcommand{\coW}{\cos \Theta_W}

\newcommand{\SU}{\mathit{SU}}
\newcommand{\gapprox}{\raisebox{-.2ex}{$\stackrel{\textstyle>}{\raisebox{-.6ex}[0ex][0ex]{$\sim$}}$}}

\newcommand{\bra}[1]{\langle{#1}|}
\newcommand{\braket}[1]{\langle{#1}\rangle}
\newcommand{\ket}[1]{|{#1}\rangle}

\newcommand{\lto}{\raisebox{-.1ex}{$\stackrel{\textstyle<}{\raisebox{-.6ex}[0ex][0ex]{$\to$}}$}}
\newcommand{\degK}{^\circ\mathrm{K}~}
\newcommand{\mbo}[1]{$#1$}
\newcommand{\lpl}{\Lambda_{\rm Pl}}
\newcommand{\MPl}{M_{\rm Pl}}
\newcommand{\mpl}{M_{\rm Pl}}
\newcommand{\tpl}{t_{\rm Pl}}
\newcommand{\Tpl}{T_{\rm Pl}}
\newcommand{\tCMB}{t_{\rm CMB}}
\newcommand{\ly}{~\ell \mathrm{y}}
\newcommand{\power}[1]{\times 10^{#1}}
\newcommand{\crn}{\nn \\ }
\newcommand{\bea}{\begin{eqnarray}}
\newcommand{\eea}{\end{eqnarray}}
\newcommand{\Ba}{\begin{eqnarray}}
\newcommand{\Ea}{\end{eqnarray}}
\newcommand{\al}{\alpha}
\newcommand{\mz}{M_Z^2}
\newcommand{\mw}{M_W^2}
\newcommand{\nn}{\nonumber}
\newcommand{\ra}{\rightarrow}

\newcommand{\eps}{\varepsilon}

\newcommand{\gv}{\mbox{GeV}}

\newcommand{\epo}{\,.}
\newcommand{\semis}{\,;\;\;}
\newcommand{\comas}{\,,\;\;}
\newcommand{\be}{\begin{equation}}
\newcommand{\ee}{\end{equation}}
\newcommand{\beq}{\begin{equation*}}
\newcommand{\eeq}{\end{equation*}}

\newcommand{\MSb}{$\overline{\rm MS}$ }

\newcommand{\MSbm}{\overline{\rm MS}}

\newcommand{\cL}{{\cal L}}
\newcommand{\sign}{\mbox{sign}}

\newcommand{\D}{\mathrm{d}}
\newcommand{\E}{\mathrm{e}}
\newcommand{\I}{\mathrm{i}}

\newcommand{\cO}{{\cal O}}

\newcommand{\munu}{{\mu \nu}}
\newcommand{\npb}{Nucl.\ Phys.\ B }

\newcommand{\pl}{Phys.\ Lett.\ }
\newcommand{\prd}{Phys.\ Rev.\ D }
\newcommand{\prl}{Phys.\ Rev.\ Lett.\ }
\newcommand{\pr}{Phys.\ Rev.\ }

\newcommand{\gmunuc}{g^{\mu\nu}}

\newcommand{\pamu}{\partial_\mu}
\newcommand{\panu}{\partial_\nu}

\newcommand{\Rexplicit}{\frac{2 }{ \eps} - \gamma + \ln 4 \pi + \ln \mu _0^2}
\begin{document}
\thispagestyle{empty}
\setcounter{page}{0}
\newpage

\title{
\begin{flushright}
\small
DESY~18-211,~~HU-EP-18/38\\[3mm]
\end{flushright}
The Hierarchy Problem and the Cosmological Constant Problem Revisited\\[3mm]
 {\normalsize \bf Higgs inflation and a new view on the SM of particle physics}
}


\author{Fred Jegerlehner\\
Humboldt-Universit\"at zu Berlin, Institut f\"ur Physik, \\ Newtonstrasse 15, D-12489 Berlin, Germany\\
Deutsches Elektronen-Synchrotron (DESY), \\ Platanenallee 6, D-15738 Zeuthen, Germany\\[6mm]
e-mail: {fjeger@physik.hu-berlin.de}           
}

\vfill



%

\maketitle

\begin{abstract}
We argue that the Standard Model (SM) in the Higgs phase does not
suffer from a ``hierarchy problem'' and that similarly the
``cosmological constant problem'' resolves itself if we understand the
SM as a low energy effective theory (LEET) emerging from a cutoff-medium at
the Planck scale.  We actually take serious Veltman's ``The Infrared -
Ultraviolet Connection'' addressing the issue of quadratic divergences
and the related huge radiative correction predicted by the SM in the
relationship between the bare and the renormalized theory, usually
called ``the hierarchy problem'' and claimed that this has to be
cured. We discuss these issues under the condition of a stable Higgs
vacuum, which allows extending the SM up to the Planck cutoff. The
bare Higgs boson mass then changes sign below the Planck scale, such
that the SM in the early universe is in the symmetric phase. The
cutoff enhanced Higgs mass term, as well as the quartically enhanced
cosmological constant term, provide a large positive dark energy that
triggers the inflation of the early universe. Reheating follows via
the decays of the four unstable heavy Higgs particles, predominantly
into top-antitop pairs, which at this stage are
massless. Preheating is suppressed in SM inflation since in the
symmetric phase bosonic decay channels are absent at tree level. The
coefficients of the shift between bare and renormalized Higgs mass as
well as of the shift between bare and renormalized vacuum energy
density exhibit close-by zeros at about $7.7 \times 10^{14}$ GeV and $3.1 \times
10^{15}$ GeV, respectively. The zero of the Higgs mass counter term
triggers the electroweak phase transition, from the low energy Higgs
phase and to the symmetric phase above the transition point. Since
inflation tunes the total energy density to take the critical value of
a flat universe and all contributing components are positive, it is
obvious that the cosmological constant today is naturally a
substantial fraction of the total critical density. Thus taking
cutoff enhanced corrections seriously the Higgs system provides
besides the masses of particles in the Higgs phase also dark energy,
inflation and reheating in the early universe.  The main unsolved
problem in our context remains the origin of dark matter. Higgs
inflation is possible and likely even unavoidable provided new physics
does not disturb the known relevant SM properties substantially. The
scenario highly favors understanding the SM and its main properties as
a natural structure emerging at long distance. This in particular
concerns the SM symmetry patterns and their consequences.\\[2mm]

\end{abstract}

\vfill

\noindent\rule{8cm}{0.5pt}\\
Invited talk at the Workshop ``Naturalness, Hierarchy and Fine Tuning''\\
RWTH Aachen, 28 February 2018 to 2 March 2018, Aachen,Germany

\newpage

\tableofcontents

\section{Prelude: Higgs inflation in a nutshell}
\label{sec:prelude}
\begin{figure}
\centering
\includegraphics[height=4cm]{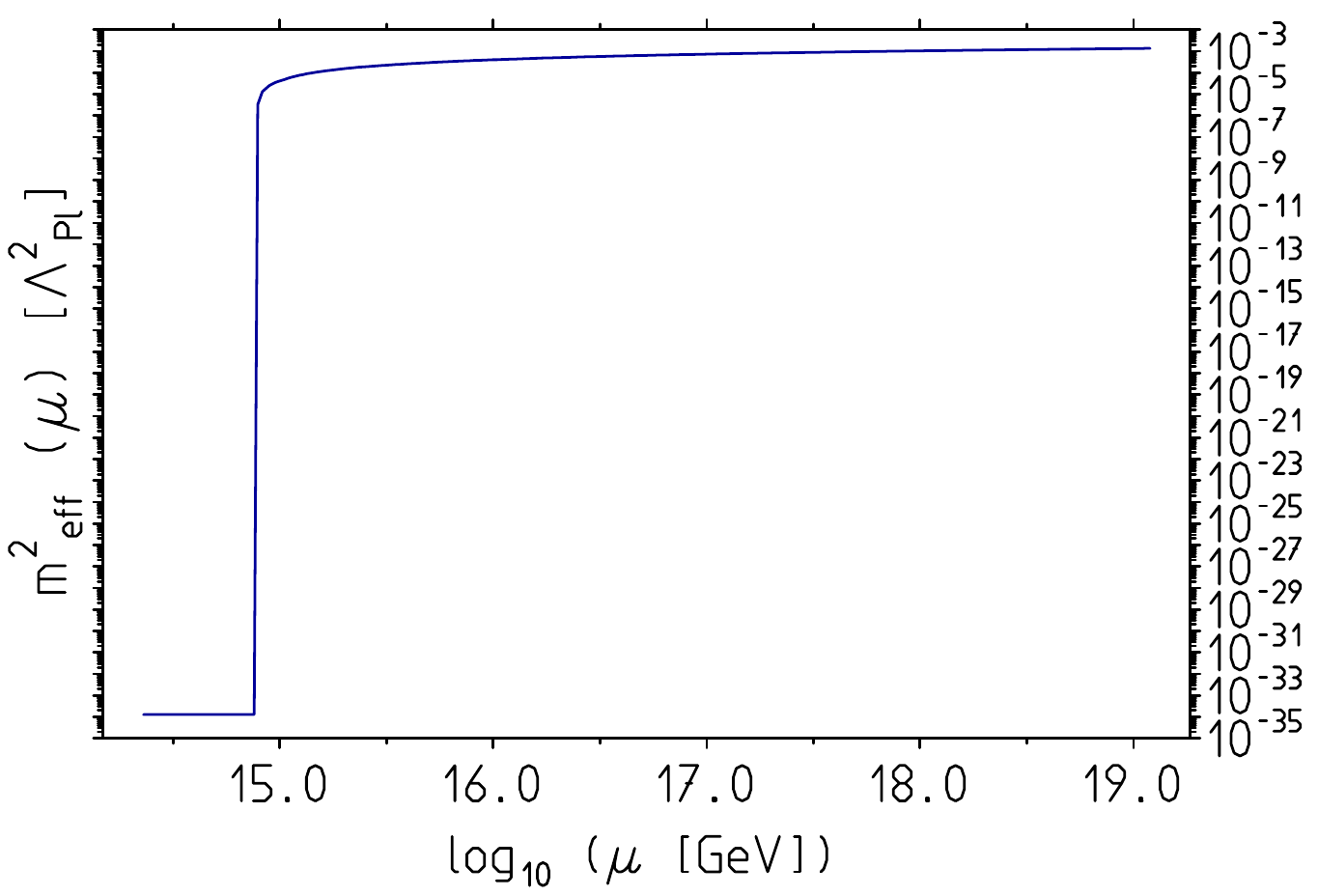}
\caption{The square of the effective Higgs-field mass as a
function of the log of the scale $\mu$, in units of $\lpl^2$. The
effective mass is given by the bare mass at short distances and by the
renormalized one at low energy.}
\label{fig:HPMup}
\end{figure}

In order to give a quick overview of what will be the essential
conclusion of the analysis I start with this prelude (see
~\cite{Jegerlehner:2013cta,Jegerlehner:2014mua}). The Standard Model
(SM) hierarchy problem~\cite{'tHooft:1979bh} is well known and
addressed very frequently to motivate Beyond the Standard Model (BSM)
scenarios in general and a supersymmetric extension of the SM in
particular. The renormalized Higgs boson mass is small, at the
ElectroWeak (EW) scale, the bare one is huge due to radiative
corrections growing quadratically with the ultraviolet (UV) cutoff,
which is assumed to be given by the Planck scale $\lpl
\sim 10^{19}~\gv$~\footnote{The Planck medium, which we may call
\textit{ether}, somehow gets shaped by gravity and quasi-particle
interactions emergent in the SM at low energies. It is characterized
by the well known fundamental Planck cutoff
\mbo{\lpl} or equivalently the Planck mass \mbo{\mpl},
which derive from the basic fundamental constants, the speed of light
\mbo{c} characterizing special relativity, the Planck constant
\mbo{\hbar} intrinsic to quantum physics and Newton's constant
\mbo{G_N} the dimensionful key parameter of gravity. \textbf{Unified} they provide an
intrinsic length \mbo{\ell_\mathrm{Pl}}, the Planck length, which also
translates into the Planck time \mbo{\tpl} and the Planck temperature
\mbo{\Tpl}:\\
Planck length: \mbo{\ell_\mathrm{Pl}=\sqrt{\frac{\hbar G_N}{c^3}}=1.616252(81)\power{-33}\mathrm{cm}\,,}\\
Planck time: \mbo{t_\mathrm{Pl}=\ell_\mathrm{Pl}/c=5.4\power{-44} \mathrm{sec}\,,}\\
Planck (energy) scale: \mbo{M_\mathrm{Pl}=\sqrt{\frac{c\hbar }{G_N}}=1.22\power{19}~\gv\,,}\\
Planck temperature:
\mbo{\frac{M_\mathrm{Pl}c^2}{\kB}=\sqrt{\frac{\hbar c^5}{G_N \kB^2}}=1.416786(71)\power{32\degK}\,.}\\
In our context, they define a shortest distance \mbo{\ell_\mathrm{Pl}}
and a beginning of time \mbo{\tpl\epo} i.e. $t>\tpl\epo$ The Planck era
energy scale equivalently is set by \mbo{E_\mathrm{Pl}=\lpl\equiv
\mpl} or temperature \mbo{\Tpl}, as for most time in the evolution
of the early universe, when elementary particle physics is at work and
before the epoch of formation of hadrons, particle processes are in
thermal equilibrium, with well-known exceptions during inflation and
the electroweak phase-transition.}. The cutoff dependence is illustrated in
Fig.~\ref{fig:HPMup} assuming the cutoff as a renormalization scale
and the SM Renormalization Group (RG) ruling the scale dependence of
the SM couplings (see below).
It is the RG improved version of Veltman's ``The Infrared -
Ultraviolet Connection''~\cite{Veltman:1980mj},
where the SM renormalization of the Higgs boson mass ($m_0$ the bare,
$m$ the renormalized mass)
\begin{equation}
m^2_0=m^2 +\delta m^2 \semis
\delta m^2=\frac{\lpl^2}{(16\pi^2)}\,C(\mu)\,,
\end{equation}
has been addressed (see
also~\cite{Decker:1979cw,Alsarhi:1991ji,Hamada:2012bp,Jones:2013aua}). The
coefficient function $C(\mu)$ depends on the dimensionless SM
couplings, which depend on the renormalization scale logarithmically
only. For an early discussion of the impact of running couplings to
the Higgs mass term and the problem of fine-tuning
see~\cite{Wetterich:1983bi}. The Higgs mass counterterm is huge when
we adopt the Planck scale as the cutoff to regulate UV
singularities. Is this a problem?  Is this unnatural? In the first
instance, it is a prediction of the SM! At low energy, we see what we
see (what is to be seen): the renormalizable, renormalized
SM~\cite{Glashow61,Weinberg67,QCD} as it describes close to all we
know up to LHC energies. But what does the SM look like if we go to
very high energies even to the Planck scale?  Not too far below the
Planck-scale, we start to see the bare theory i.e. the SM with its
bare short distance effective parameters, so in particular a very
heavy Higgs boson, which likely is moving at most very slowly.  The
potential energy
\bea
V(\phi)=\frac{m^2}{2}\phi^2+\frac{\lambda}{24}\phi^2
\label{potential}
\eea
then is large,
while the kinetic energy $\ha \dot{\phi}^2$ is small, as a
dedicated calculation shows. Here we have in mind the cosmological
solutions of Einstein's General Relativity Theory (GRT) for an
isotropic universe of constant spatial curvature, parametrized by the
Robertson-Walker metric\footnote{Einstein's field equation for the
metric tensor $g_\munu$, which incorporates the gravitational field,
is given by
\mbo{ G_\munu= \kappa T_\munu} where
\mbo{\kappa=\frac{8\pi G_N}{c^2}} is the effective interaction
constant,
\mbo{G_\munu=R_\munu-\ha R\, g_\munu} is the
Einstein curvature tensor (geometry) and \mbo{T_\munu} is the
energy-momentum tensor (matter and radiation). Cosmology is shaped by
Einstein gravity, which together with Weyl's
postulate, that radiation and matter (galaxies etc.) on the cosmological
scale behave like an ideal fluid, and the cosmological principle, assuming
isotropy of space (implying homogeneity), fixes the form of the metric
and of the energy-momentum tensor: 1) the metric (3-spaces of
constant curvature \mbo{k=\pm 1,0}) takes the form $\D s^2 = \left(c\D
t\right)^2-{a^2(t)}\,[{\D r^2}/({1-k r^2}) + r^2 \,
\D \Omega^2]$, where, in the comoving frame
$\D s =c\,\D t $ with \mbo{t} the \textit{cosmic time};
2) the energy-momentum tensor takes the form
$T^\munu=\left({\rho(t)}+{p(t)}\right)(t)\,u^\mu
u^\nu-{p(t)}\,g^{\mu\nu}\semis
u^\mu\doteq \frac{\D x^\mu}{\D s}$ where
\mbo{\rho(t)} is the density and \mbo{p(t)} the pressure of the fluid.
As a differential equation for the geometry factor $a(t)$  one obtains
Friedmann's equations (\ref{FReq}).
One needs \mbo{\rho(t)} and \mbo{p(t)} (which are related by an
equation of state characterizing the medium)
in order to get the radius of
the universe \mbo{a(t)} and its evolution in time. The Higgs potential contributes
$T_\munu = \Theta_\munu=V(\phi)\,g_\munu+ \mathrm{ \ derivative \
terms}$, where $\Theta_\munu$ is the symmetric energy-momentum tensor
of the SM (or extensions of it). Only a scalar potential can
contribute a term proportional to $g_\munu$, which mimics a
cosmological constant.}. The field $\phi$ is then a function of the
cosmic time $t$ only and $\dot{\phi}$ is the corresponding time
derivative. The Higgs boson contributes to the energy-momentum tensor
by providing the pressure $p=\ha\,\dot{\phi}^2-V(\phi)$ and the energy
density $\rho=\ha \, \dot{\phi}^2+V(\phi)\epo$ As we approach the
Planck scale (bare theory) the slow--roll condition
\mbo{\ha \, \dot{\phi}^2\ll V(\phi)} is satisfied during some window
in the time evolution and then $p\approx-V(\phi)\semis \rho \approx
+V(\phi)$ implies $p\approx-\rho$, which closely approximates the
equation of state $p_\Lambda=-\rho_\Lambda$ of Dark Energy (DE)
\mbo{\rho=\rho_\Lambda}, the equivalent of a Cosmological Constant
(CC) $\Lambda$. DE follows a very special equation of state, only
observed indirectly through Cosmic Microwave Background
(CMB)~\cite{Mather1990,Smoot1992,Bennett:2012zja,Ade:2013zuv,Adam:2015rua} pattern
and through Super Novae (SN) counts~\cite{Riess:1998cb,Perlmutter:1998np}. No lab system observation
so far has been reported to my knowledge, although statistical
mechanics system like the Ising model obey such a ground state
equation (see e.g.~\cite{Bass:2014lja}). Thus, as a consequence of the
hierarchy boost, the SM Higgs boson in the early universe delivers a
huge dark energy that is
inflating the universe and, which mimics strong anti-gravity at work.  The Friedmann equations together with energy
conservation read
\bea
(\frac{\dot{a}}{a})^2+k/a^2=\frac83\,\pi G_N \rho\semis \frac{\ddot{a}}{a}=-\frac43\, \pi
G_N \,(3p+\rho)\semis \dot{\rho}=-3\, \frac{\dot{a}}{a}\,(\rho+p)
\label{FReq}
\eea
and indeed if DE dominates we have $(3p+\rho)\approx -2\rho$ such that
we have an accelerated expansion $\ddot{a}/a >0$ and
\mbo{\frac{\D a}{a}=H(t)\,\D t} which implies an exponential growth
\mbo{a(t)=\exp Ht} of the radius $a(t)$ of the universe. \mbo{H(t)=\dot{a}(t)/a(t)} is
the Hubble constant where \mbo{H\propto
\sqrt{V(\phi)}} in a DE dominated era. Inflation stops quite quickly
as the field decays exponentially. The field equation
\bea
\ddot{\phi}+3H\dot{\phi}\simeq-V'(\phi)\,,
\label{fieldeq}
\eea
for a potential dominated by the mass term \mbo{V(\phi)\approx
\frac{m^2}{2}\,\phi^2} represents a harmonic oscillator with friction
and leads to Gaussian inflation as established by an analysis of the
CMB pattern by the Planck mission~\cite{Ade:2013ydc}. One of the
reasons why the inflation phenomenon must have happened in the early
universe is that the universe looks flat today, while a flat universe
in the absence of DE is exponentially unstable in its time
evolution. This because then formally the strong energy condition
$\rho+3p\geq 0$ holds, which implies $\ddot{a}/a\leq0$. So, different types
of solutions of the Friedmann equations at $\rho_\Lambda=0$ deviate
dramatically during the 13.8 billion years life of the universe after
the Big Bang. The ``flattenization'' by inflation is evident as the
curvature term $k/a^2(t)\sim k\,\exp(-2Ht) \to 0$ drops exponentially
independent of the curvature type. The latter, characterized by the
normalized curvature $ k=0,\pm 1$, distinguishes flat infinite,
spherical closed or hyperbolic open geometries. It is very important
to note that the CC given by the Higgs potential $V(\phi)$ in the
symmetric phase is \textbf{positive} in any case, very different from
the (much smaller) contribution from the Higgs-field vacuum
expectation value (VEV) in the broken phase, which is
negative~\cite{Dreitlein74}.  This already shows that dynamical
effects and in particular possible phase transitions make the CC
depend on time, although it does not depend explicitly on $a(t)$,
i.e., $\rho_\Lambda \propto a(t)^{0}$. As we know the matter density
scales like $\rho_m \propto a(t)^{-3}$ while the radiation density
decreases like $\rho_\gamma \propto a(t)^{-4}$ since radiation gets
red-shifted in addition to its spatial dilution\footnote{Matter here
includes dark matter (DM) and normal baryonic-matter (BM), the
non-relativistic stuff; radiation includes all relativistic degrees of
freedom: photons, neutrinos and at high energies other SM particles
besides the Higgs bosons, which get boosted to be heavy because of their missing
naturalness. Note that normal baryonic matter only emerges after the
QCD~\cite{QCD} hadronization phase-transition,i.e. after protons and neutrons have
been formed. In contrast, cold dark matter looks must have
existed much earlier not too long after Planck time.}.

Inflation tunes the total energy density to be that of a \textit{flat
space} (as if $k=0$), which according to (\ref{FReq}) for $k=0$ requires a specific ``critical'' energy density
\bea
{\rho_{0,{\rm crit}}=3H_0/(8\pi G_N)=\mu^4_{\rm crit}}\; \mathrm{ \ where \
}\; {\mu_{\rm
crit} \approx 0.00247~\mathrm{eV}}\epo
\label{rhocrit}
\eea
With $H_0$ the present Hubble constant
\mbo{\rho_{0,{\rm crit}}} is the present total energy density.
For an arbitrary mixture of dark energy, matter and radiation the
first of the equations (\ref{FReq}) reads
\bea
\rho=\rho_{0,{\rm crit}}\,\left[\Omega_\Lambda+\Omega_m\,\left(\frac{a_0}{a}\right)^3+\Omega_\gamma\,\left(\frac{a_0}{a}\right)^4\right]\,,
\eea
where $\Omega_i=\rho_{0,i}/\rho_{0,{\rm crit}}$ are the
present fractional densities and $a_0=a(t_0)$ the spatial metric scale
factor at present time $t_0$.
Including the curvature term $\Omega_k=-k/(a_0^2H_0^2)$ we have
\bea
\Omega_\Lambda+\Omega_m+\Omega_\gamma+\Omega_k=1
\label{omagaone}
\eea
as an exact equation, and when the curvature term is exponentially
suppressed we very accurately have
\bea
\Omega_\Lambda+\Omega_m+\Omega_\gamma \cong 1\,,
\eea
which is supported strongly by observation (CMB). Whatever constitutes
the universe, the curvature constant is $k=+1$, $k=0$ or $k=-1$
according to whether the present density $\rho_0$ is greater than,
equal to, or less than $\rho_{0,{\rm crit}}\epo$ A higher density
$\rho_0>\rho_{0,{\rm crit}}$ implies a re-contraction of the initially
(at the Big Bang) expanding universe, a lower density
$\rho_0<\rho_{0,{\rm crit}}$ would not be able to stop the
expansion forever.

We know that \mbo{\rho_{0\Lambda}=\mu^4_{0\Lambda}} today is about
\mbo{\mu_{0\Lambda}=0.00171~\mathrm{eV}} which in a flat universe must
be a fraction of the critical density, and actually has been
determined to amount to $69.2 \pm 1.2$ \%.  Since the non-DE
components drop with a power of the radius $a(t)$ as time goes on, the
asymptotic behavior is determined by $\rho_\Lambda$ solely. Friedmann
solutions in GRT with non-vanishing cosmological constant have been discussed
in~\cite{Felten:1986zz,Sahni:1999gb}.

We note that the large positive cosmological constant provided by the
SM Higgs sector, on the one hand, effectuates the inflation that is
needed to tune the total energy density into the critical density of
flat space. This tells us that on the other hand, the cosmological
constant has to be some fraction of the critical density, i.e. it is
self-tuned to be small. So also the cosmological constant problem may
turn out to get its natural explanation (see below).

Since inflation is strongly supported to have happened by observation, we
must assume the existence of an appropriate scalar field, and the
Higgs field is precisely such a field we need and within the SM it has
the properties that promote it to be the inflaton field. In contrast to
other inflation models, Higgs inflation is special because of almost all
properties are known or predictable! Below, I will argue that the SM in the
Higgs phase does not suffer from a ``hierarchy problem'' and that
similarly, the ``cosmological constant problem'' resolves itself if we
understand the SM as a low energy effective theory emerging from a
cutoff-medium at the Planck scale.

Adopting a bottom-up approach, I discuss these issues under the
condition of a stable Higgs vacuum, by predicting the behavior of the
SM when approaching the Planck era at high energies. SM Higgs
inflation as exposed in this prelude may look pretty simple but in
fact is rather subtle, because of the high sensitivity to the SM
parameters and high sensitivity to higher order SM effects. In any
case, my preconditions are: (i) a stable Higgs vacuum and a
sufficiently large Higgs field at \mbo{\MPl}, (ii) physics beyond the
SM should not spoil the main features of the SM. This means that SM
extensions like SuperSymmetry (SUSY), or Grand Unified Theories (GUT)
etc., pretending to solve the hierarchy problem and/or affecting the
SM RG-flow substantially, are to be excluded! Here we have to assume
that a kind of desert in the heavy particle spectrum is extending
effectively up to the Planck scale. This is not so far beyond the
``grand desert'' usually assumed to exist in the context of GUTs. This
does not exclude new physics that we know to exist, like dark matter,
Majorana neutrinos or axions, for example.

Slow-roll inflation in general has been investigated
in~\cite{Guth:1980zm,Starobinsky:1980te,Albrecht:1982wi,Linde:1981mu,Kolb:1990vq,Weinberg:2008zzc}
in the 80's mostly as a top-down approach. An alternative ``non-minimal
gravity'' Higgs-inflation approach has been advocated
in~\cite{Minkowski:1977aj,Zee:1978wi,Bezrukov:2007ep,Barbon:2009ya,Bezrukov:2010jz,Bezrukov:2014bra,Bezrukov:2014ipa}.
Yet a different ``eternal'' Higgs inflation ansatz has been
investigated within the context of superstring theory~\cite{Hamada:2015ria}. A
time-dependent cosmological constant has been obtained also in a model
which is based on a dilatation symmetry anomaly, where one
assumes the Newton ``constant'' to be a time-dependent dynamical degree of
freedom~\cite{quintessence}.

In this prelude, I have outlined what a correct interpretation of the
``hierarchy problem'' likely looks like, i.e. the predicted SM
hierarchy pattern is not a problem, rather it is the solution for
what we need to trigger inflation in the early universe. In the
following I will consider the hierarchy issue in a broader context and
discuss in some detail the intricacies of the cosmological constant
problem and Higgs boson inflation. I will try to convince the reader
that the Higgs boson inevitably delivers dark energy and the
consequent inflation is well supported by a self-consistent
perturbative SM
calculation~\cite{Jegerlehner:2013cta,Jegerlehner:2014mua}. The
approach is highly predictive and limited mainly by the uncertainties
of the knowledge of the SM parameters and the accuracy of the
perturbative calculations of the matching conditions between measured
and \MSb parameters and the \MSb renormalization group coefficients.

\section{The Higgs boson discovery -- the SM completion}
\label{sec:higgsfound}
With the discovery of the Higgs boson by ATLAS~\cite{ATLAS} and
CMS~\cite{CMS} in 2012 a last major but often questioned building
block of the electroweak SM has been
experimentally verified. The existence of an elementary scalar has
been found to be required to render the electroweak massive gauge
theory renormalizable in 1964 by Englert and
Brout~\cite{Englert:1964et} and by Higgs~\cite{Higgs:1964ia}. The key
mechanism turned out to be a Spontaneous Symmetry Breaking (SSB)
mechanism of the non-Abelian $\SU(2)_L$ gauge sector responsible for
the weak interactions. The corresponding Higgs mechanism generates
masses for all massive particles while not affecting the
renormalizable UV behavior of the massless unbroken theory. Now,
remarkably, the SM Higgs boson mass has been found in very special
mass range $ 125.18\pm0.16~\gv$, which seems to match the possibility
to extrapolate the SM up to the Planck scale. Knowing the Higgs mass
$M_H$ and using the mass coupling relation valid in the Higgs phase,
we also know the Higgs self-coupling $\lambda$ and hence the
renormalized Higgs potential $V=\frac{m^2}{2}\,H^2+\frac{\lambda}{24}
H^4\,,$ which is the object in our focus. Perturbativeness and vacuum
stability of the Higgs potential are the key issues in this context
(for early considerations
see~\cite{Cabibbo:1979ay,Hung:1979dn,Lindner:1985uk,Grzadkowski:1986zw,Lindner:1988ww,Sher:1988mj}).
\begin{figure}
\centering
\includegraphics[height=4cm]{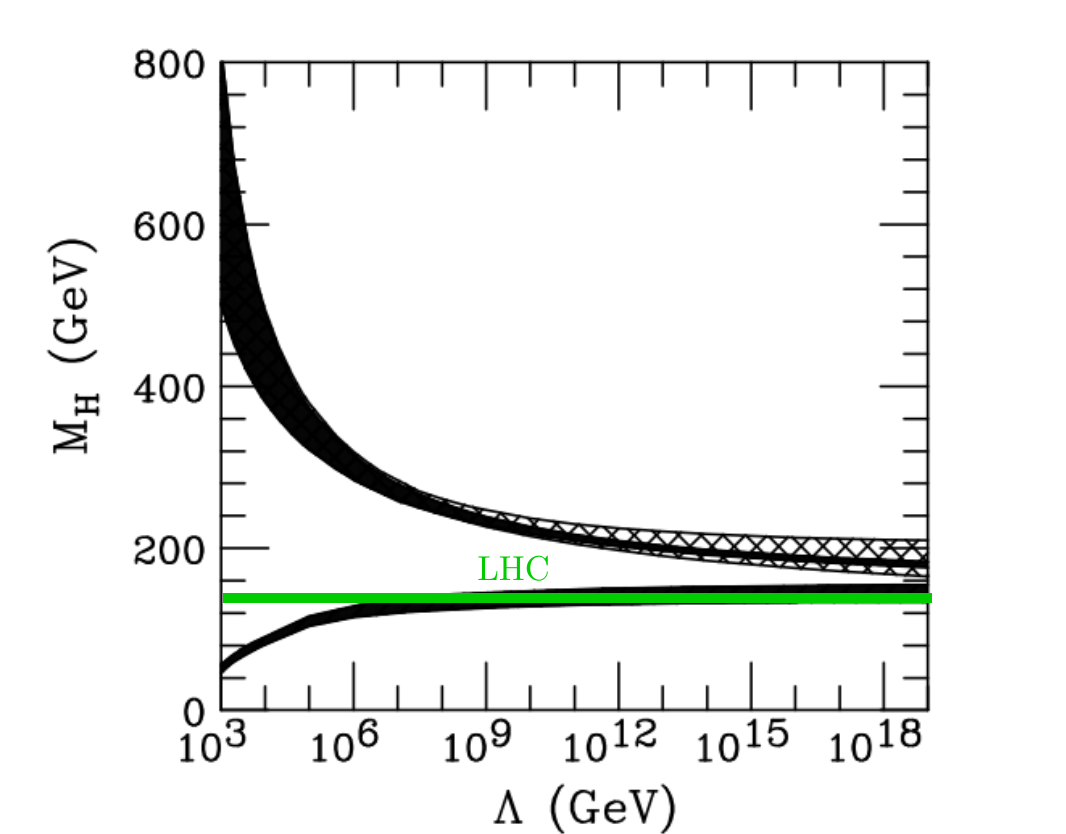}
\includegraphics[height=4cm]{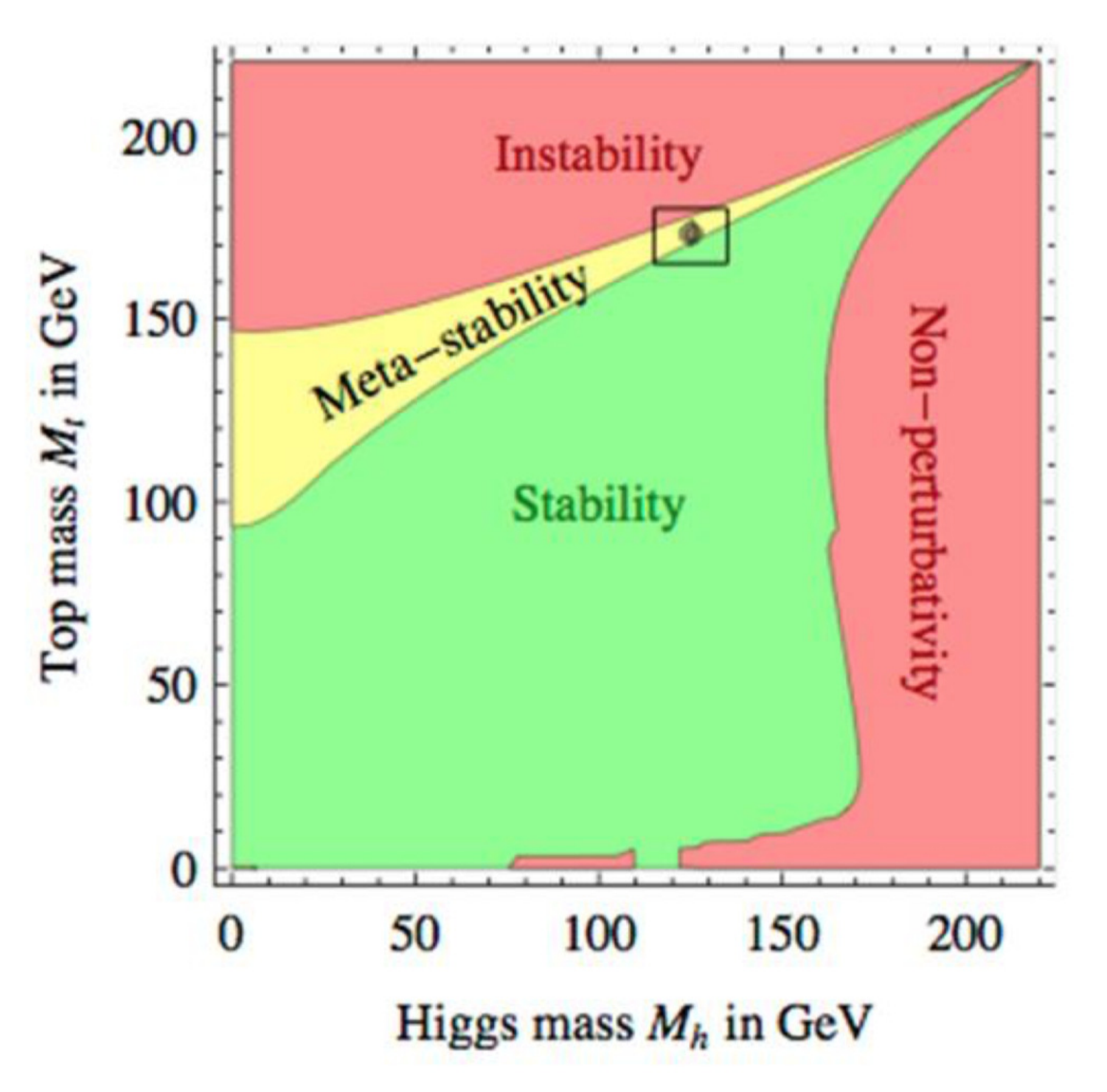}
\caption{Left: the SM Higgs system remains perturbative up to
\mbo{\lpl} if $M_H$ is light enough (upper bound= avoiding Landau
pole) and the Higgs potential remains stable (\mbo{\lambda >0}) if $M_H$
is not too light [parameters used:
\mbo{M_t=175[150-200]~\gv\semis
\alpha_s=0.118}]. [Reprinted with permission
from~\cite{Hambye:1996wb} http://dx.doi.org/10.1103/PhysRevD.55.7255.
Copyright (1996) by the American Physical Society]. Right: A plot of
the stability range in the $M_H-M_t$ plane. Courtesy of G. Degrassi et
al. http://dx.doi.org/10.1007/JHEP08(2012)098, (License: CC-BY-4.0).
Reproduced from~\cite{Degrassi:2012ry}}
\label{fig:HR}
\end{figure}
Fig.~\ref{fig:HR} adopted from an analysis by Hambye and Riesselmann
in 1996~\cite{Hambye:1996wb} illustrates the possible impact to have a
Higgs mass in a window extending up to the Planck scale. The
possibility that the SM may be extended right up to the Planck scale
has been analyzed also
in~\cite{Casas:1994qy,Espinosa:1995se,Schrempp:1996fb,IRS,Espinosa:2007qp,Ellis:2009tp,Hall}.
Later estimates have been improved after more precise SM parameters
like the QCD coupling $\alpha_s$ and top-quark mass $M_t$ became
available, see
e.g.~\cite{Shaposhnikov:2009pv,Holthausen:2011aa,Yukawa:3,Degrassi:2012ry,Moch12}. Given
the Higgs self-coupling, all relevant SM parameters are known.  While
the RG evolution equations in the symmetric phase of the SM have been
known for a long time to two loops, recently also the three-loop
coefficients in the \MSb scheme have been calculated~\cite{Mihaila:2012fm,Bednyakov:2012rb,Chetyrkin:2012rz}. The \MSb
input parameters, which are most suitable to parametrize the high
energy tail, have to be calculated via the matching conditions from
the experimentally measured ones
(see~\cite{Fleischer:1980ub,Sirlin:1985ux,Jegerlehner:2001fb,Jegerlehner:2012kn,Yukawa:3,Degrassi:2012ry,Moch12,Buttazzo:2013uya,Bednyakov:2015sca}
and references therein). The matching conditions are based on more or
less complete two-loop calculations that require the knowledge of the
two-loop renormalization counterterms. Besides the electromagnetic
vertex correction $\delta e$, all others are mass renormalization
counterterms $\delta M_b^2$ for bosons and $\delta M_f$ for fermions,
all given in terms of comparatively simple two-point functions, which
are completely known to two loops\footnote{A complete two-loop
calculation has been performed independently by Veretin and Kalmykov
in the context of~\cite{Jegerlehner:2001fb}. Two-loop integrals
exhibiting Higgs propagators have been expanded in $M_V^2/M_H^2$,
assuming the Higgs to be heavier than the $W$ and $Z$ bosons.  After
the Higgs mass has been found this expansion turned out to be
obsolete, and the relevant integrals had to be evaluated
numerically. This has later been performed partially in
~\cite{Jegerlehner:2012kn} and ``exact'' in~\cite{Kniehl:2015nwa} such
that a complete two-loop evaluation is available. For independent
calculations of the matching conditions also see~\cite{Martin:2018yow}
and references therein.}.

So far so good.  One important point is missing however: the physical
on-shell parameters are determined from experimental data by unfolding
the raw data from radiation and detector effects. They represent
pseudo-observables depending on theory input that relies on
approximations. One has to keep in mind that most LEP, Tevatron and
LHC results are based on incomplete two-loop calculations at
best. Exceptions are the extraction of $G_\mu$~\cite{Awramik:2003ee}
and of $\sin^2 \Theta^{\rm lept}_{\rm eff}$ for which complete
two-loop calculations exist~\cite{Awramik:2004ge}. Complete two-loop
calculations have not been available at LEP times. For Bhabha
scattering, which plays a key role for luminosity monitoring, the
two-loop QED corrections became available only lately (see
e.g.~\cite{Jadach:2018jjo} and references therein). Full two-loop
electroweak corrections to $2 \to 2$ processes either are still
missing or have not been available when parameters were extracted from
the data. There are persisting discrepancies at the 2 $\sigma$ level
in the determination of $\sin^2
\Theta^{\rm lept}_{\rm eff}$ between SLD and LEP as well as for
$\sin^2
\Theta^{\rm b\bar{b}}_{\rm eff}$ and so there remain questions about the
size of the estimated uncertainties of some of the input parameters.

Precision physics at LEP has been a great achievement also thanks to
continuous progress in QCD and electroweak higher order
calculations. Much progress has been achieved since. The SM is
established with unprecedented accuracy. Activities now
are focusing on LHC physics for obvious reasons. But more effort is
needed to keep alive gained expertise on electroweak precision
physics. Projects for future $e^+e^-$ colliders like the International
Linear Collider (ILC), which started with the unrealized TESLA
project~\cite{Accomando:1997wt}, and the Future Circular Collider
(FCC) project at CERN, are boosting efforts to reach much higher precision
in $Z$-peak, $W$-pair,  top-quark and Higgs-boson physics~\cite{FCCee11}.
The possibility that progress in this direction can establish the Higgs
vacuum to be stable and the Higgs boson to be the inflaton in fact may
be the strongest motivation to go on in realizing a next $e^+e^-$ machine
probing physics at the electroweak scale at much higher precision.

The most serious conceptual problem concerns the measurement of the
top-quark mass and the related determination of the top Yukawa
coupling. On the theory side, one is operating with the perturbative
concept of an on-shell top-quark mass, which is not an
observable. Usually, the on-shell mass is identified with the Monte
Carlo mass $M_t^{\rm MC}$. The latter is the kinematic mass parameter
used Monte Carlo event generators that are utilized to extract the
top-quark mass from top-quark production processes. The
non-perturbative color screening effects obscure the precise
determination of $M_t$ so far (see
e.g.~\cite{Moch12,Beneke:2016cbu}). In addition, most of the Monte
Carlo programs used at present do not take into account the
electroweak radiative corrections.

My quintessence: while the matching conditions
are known to two loops, the input parameters have not yet been
determined at the same level of accuracy, i.e. likely
reported errors are underestimated. These issues have to be reminded
before one can claim that metastability of the electroweak vacuum is a
proven fact.

\subsection{Matching conditions: \MSb parameters in terms of
physical parameters}
\label{ssec:match}
We want to solve the RG equations up to very high
scales, where mass effects are supposed to be negligible, in regions
where physical thresholds play no role. The mass independent \MSb
scheme is the most simple choice, and under these circumstances likely
a rather physical one, because it reflects the true UV structure of
the SM in terms of quasi-bare quantities.  In order to solve the \MSb
RG-equations we need the input \MSb values of the basic couplings
\mbo{g'\!,g,g_3,y_t} and \mbo{\lambda}. While
\mbo{g_3(M_Z^2)=\sqrt{4\,\pi \alpha_s(M_Z^2)}} is a standard \MSb
reference parameter, the key parameters \mbo{y_t} and \mbo{\lambda}
are known actually only through the related masses \mbo{M_t} and
\mbo{M_H}:
\bea
y_t^2=2\,\sqrt{2} G_\mu\, M_t^2\,(1+\delta_t(\al,\cdots))\semis
\lambda =3\,\sqrt{2}G_\mu\, M_H^2\,(1+\delta_H(\al,\cdots))\,,
\label{ytlabda}
\eea
where $\delta_i$ represent the corresponding radiative corrections.
The relations between the renormalized on-shell
masses and their \MSb versions are provided by the natural matching conditions
\mbo{m^2_{b\,\mathrm{ren}}=M^2_{b\,\mathrm{ren}}+ \left(\delta M_b^2- \delta m_b^2\right)\,,}
for bosons and \mbo{m_{f\,\mathrm{ren}}=M_{f\,\mathrm{ren}}+ \left(\delta M_f- \delta m_f\right)\,,}
for fermions\footnote{We denote
on-shell masses by capital, \MSb masses by lower case letters as
in~\cite{Jegerlehner:2013cta}}.
Formally we obtain them by writing the
relation between the renormalized and the bare masses (in the bare
Lagrangian) in the two schemes:
\bea
m^2_{b\,\mathrm{bare}} &\stackrel{\rm OS}{=}& M^2_{b\,\mathrm{ren}}+ \delta M_b^2\stackrel{\MSbm}{=}
 m^2_{b\,\mathrm{ren}}+ \delta m_b^2\,,\nn
\eea
for bosons and
\bea
m_{f\,\mathrm{bare}} &\stackrel{\rm OS}{=}& M_{f\,\mathrm{ren}}+ \delta M_f\stackrel{\MSbm}{=}
 m_{f\,\mathrm{ren}}+ \delta m_f\,,\nn
\eea
for fermions,
i.e. the mass shift is given by the \MSb finite part prescription
applied to the (UV singular) OS mass counterterm:\\
\bea
\left.m^2_{b}\right|_{\MSbm}=\left.M^2_{b}\right|_{\rm OS}+ \left.\delta
M_b^2\right|_{\Rexplicit \to \ln \mu^2}\,,
\eea
for bosons and
\bea
\left.m_{f}\right|_{\MSbm}=\left.M_{f}\right|_{\rm OS}+ \left.\delta
M_f\right|_{\Rexplicit \to \ln \mu^2}\,,
\eea
for fermions.

Here it is important to keep in mind: for the renormalized theory the
\MSb parametrization is not a parametrization in terms of observables
but serves as a convenient intermediate (auxiliary) parametrization. The
\MSb scheme is a purely perturbative concept (no non-perturbative
definition) and corresponding parameters are not measurable.

However, if we take the bare theory to be the physical one in the
sense of a low energy effective theory exhibiting a physical cutoff,
then the
\MSb parameters in the perturbative regime are representing the bare
parameters. Crucial for the extrapolation to the Planck scale, if
possible, it is to keep the relationships between bare, \MSb and
physical OS parameters gauge invariant and preserving the UV structure
(see e.g.~\cite{Fleischer:1980ub,Sirlin:1985ux}).

So in principle, on a conceptual level, we are confronted with a
well-defined problem of calculating the massive physical particle
self-energies exact to two loops, in addition to the e.m. vertex, in
order to find the appropriate input for the three-loop \MSb RG
running.

In the symmetric phase of the SM, except the \mbo{m^2}--term in the
Higgs potential, all masses vanish and the RG coefficient calculations
and the solution of the RG equations are straight forward. This insofar that
there are no special mass-coupling relations and the decoupling of
heavy states (the four Higgs scalars) is given as requested by the
Appelquist-Carazzone theorem (AC)~\cite{Appelquist:1974tg}, which in
the broken phase only holds in the QCD and QED sectors.

\begin{figure}
\centering
\includegraphics[scale=0.64]{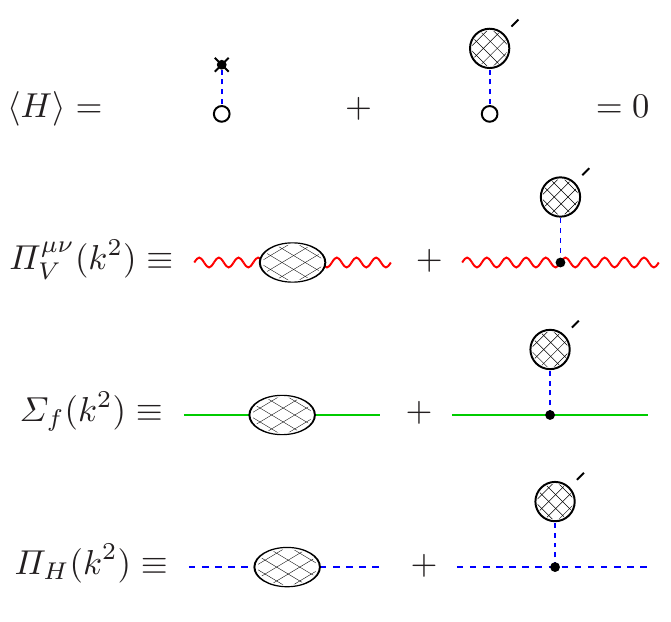}
\caption{Tadpoles show up in the broken phase as diagrams
contributing to the Higgs field VEV. They contribute to self-energies
as depicted. Tadpoles are gauge dependent and UV singular and have to
be taken into account as shown in order to preserve the gauge
symmetry~\cite{Fleischer:1980ub}.}
\label{Tadpoles}
\end{figure}

In contrast, in the Higgs phase of the SM, there are some tricky points
to be taken care of. \\
\abst \textbf{1)} The tadpole issue: if we
require the bare parameters, now represented by the \MSb parameters,
as physical we have to respect Ward-Takahashi and Slavnov-Taylor
identities. This requires to take into account tadpole contributions
according to Fig.~\ref{Tadpoles}, which mostly are omitted in
calculations (see
e.g.~\cite{Faisst:2004gn,Bednyakov:2015sca,Martin:2018yow}), because a
theorem~\cite{Taylor:1976ru,Kraus:1997bi} states that tadpoles cancel
in physical quantities provided they are expressed in terms of
physical quantities within the renormalized theory
(see~\cite{Jegerlehner:2012kn} for a recent discussion).\\
\abst \textbf{2)} The lack of decoupling issue: while in QED and QCD,
heavy particles decouple, within the SM heavy states do not decouple
when the mass-coupling relations come into play. Masses and couplings
are one-to-one interrelated, because all masses are generated by the
Higgs mechanism. For the given VEV $v$ a mass can only get large iff
the corresponding coupling gets large. The couplings are active at
scales below the related mass thresholds. Within the \MSb scheme,
which respects gauge invariance but has deficiencies like the lack of
decoupling of heavy states, decoupling has to be imposed by hand, and
one is working with effective field theory including only the active
flavors at the given scale, in place of full theory.  Most of the
matching analyses are inspired by techniques that are well established
in QCD. Typically, one is matching the $N_f$ with the $N_f+1$ flavor
effective QCD at the $N_f+1$ flavor threshold.  This is well justified
in the time-like regime where corresponding thresholds are
manifest. While physical observables are naturally sharing decoupling
properties, within the full SM the lack of automatic decoupling is a
serious shortcoming of the \MSb parametrization when applied below
the highest SM threshold at $2\,M_t$.

The most prominent non-decoupling effect is due to the large top
Yukawa coupling, which we know to be interrelated with the heavy
top-quark mass. It yields the leading correction of the EW
$\rho$-parameter, defined by the neutral to the charged current ratio of
the corresponding low energy effective Fermi couplings, which is given by
\bea
\rho=G_{\rm
NC}/G_{\rm CC}(0)=1+\frac{N_cG_\mu}{8\pi^2\sqrt{2}}
\left(m_t^2+m_b^2-\frac{2m_t^2m_b^2}{m_t^2-m_b^2}\ln\frac{m_t^2}{m_b^2}\right)
\approx 1+\frac{N_cy_t^2}{32\pi^2}\,,
\eea
where $N_c=3$ is the number of colors. It is quadratic in $y_t$ and
measures weak-isospin breaking by the Yukawa couplings of the heavy
fermions at zero momentum. So, the top Yukawa coupling is at work down
to zero momentum and not being active starting above the top mass threshold
only. This type of effect measured for the first time at LEP in 1995
far below the top-pair threshold, allowed to derive a bound on $M_t$
before the top-quark discovery at the Tevatron in 1996.

This shows that ``decoupling by hand'' cannot be applied for weak
contributions, i.e., we cannot parametrize and match together
effective theories by switching off fields of mass $M>\mu$ at
a given scale $\mu$.

Only a direct measurement of $y_t$ and $\lambda$ at a facility like
FCC-ee or ILC above the top-quark mass threshold can provide us the
precise input parameters we need.\\
\abst \textbf{3)} The Fermi constant issue:
the Higgs-field VEV $v$ determines the Fermi constant via
$$G_F=\frac{1}{\sqrt{2}v^2}\;\mathrm{ \ or \ }\;
\sqrt{2}G_\mu=v^{-2}=\frac{e^2}{4}\,\frac{M^2_Z}{M^2_W}\,\frac{1}{M_Z^2-M_W^2}\,.$$
For the on-shell counterterm we then have the relation ($c_W^2=M_W^2/M_Z^2\,,s_W^2=1-c_W^2$)
\bea
\frac{\delta G_F}{G_F}=2\,\frac{\delta v^{-1}}{v^{-1}}\;\mathrm{ \
and \ }\;
\frac{\delta v^{-1}}{v^{-1}}= \frac{\delta e}{e}
     -\frac{1}{2\,s_W^2}\,\big(s_W^2\,\frac{\delta
     M_W^2}{M_W^2}+c_W^2\,\frac{\delta M_Z^2}{M_Z^2}\big)\epo
\label{deltavinv}
\eea
Potentially, the Higgs-field VEV $v$ could be particularly affected by
non-decoupling effects. However, here we may take advantage of the
fact that tadpole contributions drop out in relations between physical
(on-shell) parameters and physical transition amplitudes. To be
compared are\\
\mbo{\bullet}~~ ~~low energy ~~~ -- $G_\mu=G_\mu(Q^2\approx 0)$ determined by the muon lifetime,\\
\mbo{\bullet}~~ $W$ mass scale --
$\hat{G}_\mu=G_\mu(Q^2\approx
M_W^2)=\frac{12\pi\Gamma_{W\ell\nu}}{\sqrt{2}M_W^3}$ given
by leptonic $W$ decay rate. Indeed $\hat{G}_\mu\approx G_\mu$ with
good accuracy as expected\footnote{A LO estimate with
$M_W=80.385\pm0.015~\gv$, $
\Gamma_W=2.085\pm0.042~\gv$ and  $ B(W\to\ell\nu_\ell)=10.80\pm0.09\%$ yields
$\hat{G}_\mu=1.15564(55)\power{-5}~\gv^{-2}$ to be compared with
$G_\mu=1.16637(1)\power{-5}~\gv^{-2}$, i.e. the on-shell Fermi
constant at scale $M_Z$ appears reduced by 0.92\% relative to $G_\mu\,.$}.

Finally we need the \MSb version of $G_F$. Again, evaluating
$G_{F\,{\rm bare}}=G_\mu+\delta G_F$ in both schemes we have $$G_F^{\MSbm}=G_\mu+ \left(\delta
\left.G_F\right|_{\rm OS}\right)_{\Rexplicit \to \ln \mu^2}\,.$$
We calculate it equivalently by using (\ref{deltavinv}) in the
respective schemes.
Then the \MSb top-quark Yukawa coupling is given by
\bea
y_t^{\overline{\rm MS}}(M_t^2)=
\sqrt{2}\,\frac{m_t(M_t^2)}{v^{\overline{\rm MS}}(M_t^2)}\semis v^{\overline{\rm MS}}(\mu^2)
=\left(\sqrt{2}\,G_{F}^{\overline{\rm MS}}\right)^{-1/2}(\mu^2)\,.
\eea
A good matching scale is $M_Z$ using $\alpha(M_Z),\alpha_s(M_Z)$ and
$\hat{G}_\mu$ as input. The non-perturbative contribution $\Delta
\alpha_{\rm had}^{(5)}(M_Z^2)$ to the shift in $\alpha(M_Z)$ is taken
from~\cite{Jegerlehner:2017zsb}. I have outlined the points that can
lead to slightly different input values for the \MSb parameters as
listed in Table~\ref{tab:MSbinp}.  Our evaluation is documented in
more detail in
Refs.~\cite{Jegerlehner:2012kn,Jegerlehner:2013cta,Jegerlehner:2014mua}. One
difference concerns the use of the Fermi constant, which in a number
of analyses is taken to be the low energy $G_\mu$, assumed not to be
running below the top-quark mass threshold, and often a difference
between its OS and \MSb version is not made (see
e.g. ~\cite{Degrassi:2012ry,Buttazzo:2013uya}). A second cause for a
difference is related to a different account of the tadpole
contributions (see
e.g. ~\cite{Bednyakov:2015sca,Martin:2018yow}). Given the \MSb input,
the RG-equations are then solved in the \MSb scheme to three loops by
also including four- and five-loop results in the QCD sector.
Concerning the RG-running, there is full agreement between the
different studies available in the literature for a given set of input
parameters.  The only parameter which does not agree within quoted
uncertainties is the top-quark Yukawa coupling $y_t$ (\ref{ytlabda}), which
apparently is the parameter which decides about the stability of the vacuum.

Another important point, which is only partially taken care of in
estimating the uncertainties of the physical input parameters, is the
following:\\
\abst \textbf{4)} The scheme dependence issue:
perturbative predictions are renormalization scheme dependent due to
truncation errors of the perturbative expansion. A typical example is
provided by the electromagnetic fine structure constant. At very low
energy the value in the Thomson limit $\alpha\simeq 1/137$ is adopted
as an input parameter, while at the $Z$ boson mass scale
$\alpha'=\alpha(M_Z)=\frac{\alpha}{1-\Delta \alpha(M_Z)}\simeq 1/127$
is a more appropriate input parameter, because it represents an
universal type large correction requiring RG resummation, and which
enters radiative correction calculations at many places. To one-loop
order $\alpha'=\alpha\,(1+a\,\alpha)$. If we calculate a
matrix-element including one-loop corrections $$M
^{(1)}=\alpha^n\,C\,(1+b\,\alpha)\,,$$ in terms of $\alpha'$ we obtain $${M'} ^{(1)}={\alpha'}
^n\,C\,(1+b'\,\alpha')\,,$$ and hence $${M'}
^{(1)} = M ^{(1)} +\delta M\,, $$ where, inserting $b'=b-n\,a\,,$
\be
\delta M
=\alpha^n\,C\,\left(\left(\frac12\,n(n-1)\,a^2+(n+1)\,ab'\right)\,\alpha^2
+ \cdots + a^{n+1} b' \alpha^{n+2}\right)\epo
\ee
Thus the result differs by $\delta M$. If we do not actually calculate
the higher orders $$\delta M= {M'} ^{(1)}-M ^{(1)}$$ must be
considered as an uncertainty due to unknown higher order
effects. Actually, results can differ non-negligibly when different
parameter sets are used as independent input parameters, like
$\alpha,M_W,M_Z,\cdots$, $G_\mu,M_W,M_Z,\cdots$ or the preferred LEP
parametrization in terms of the most precisely known parameters
$\alpha,G_\mu,M_Z,\cdots$, the scheme which is usually applied
(see e.g.~\cite{Jegerlehner:1989ye}).

I am mentioning the scheme dependence uncertainties here because, to my
knowledge, they are not fully taken into account in standard extractions of
on-shell parameters from the data. One more reason why strong statements
concerning a proof of metastability and ``just failing vacuum stability'',
it seems to me can hardly be justified in view of the strong
sensitivity of vacuum stability on precise input parameters.

I remind that the most important difference between my ``cutoff extended''
SM and the most often discussed metastability path to high energies
(discussed within the framework of the renormalized SM) lies in the
fact the I have to take care that my
\MSb parametrization is equivalent to the bare
parametrization. This is what makes the inclusion of tadpoles
mandatory. We have explicitly checked in~\cite{Jegerlehner:2001fb}
that only by including tadpoles the \MSb RG equations in the broken
phase agree with the ones in the symmetric phase\footnote{Obviously,
in electroweak theory, what is called \MSb scheme may refer to two
different versions, depending on whether tadpoles are dropped or
not~\cite{Faisst:2004gn}. This only concerns calculations in the
broken phase where mass effects play a role, as for the matching
conditions (see~\cite{Jegerlehner:2012kn}). Note that RG coefficients
are calculated in the massless symmetric phase where they are
unambiguous, since tadpoles are absent.}.

\subsection{The SM running parameters}
\label{ssec:run}
\begin{figure}
\centering
\includegraphics[height=5cm]{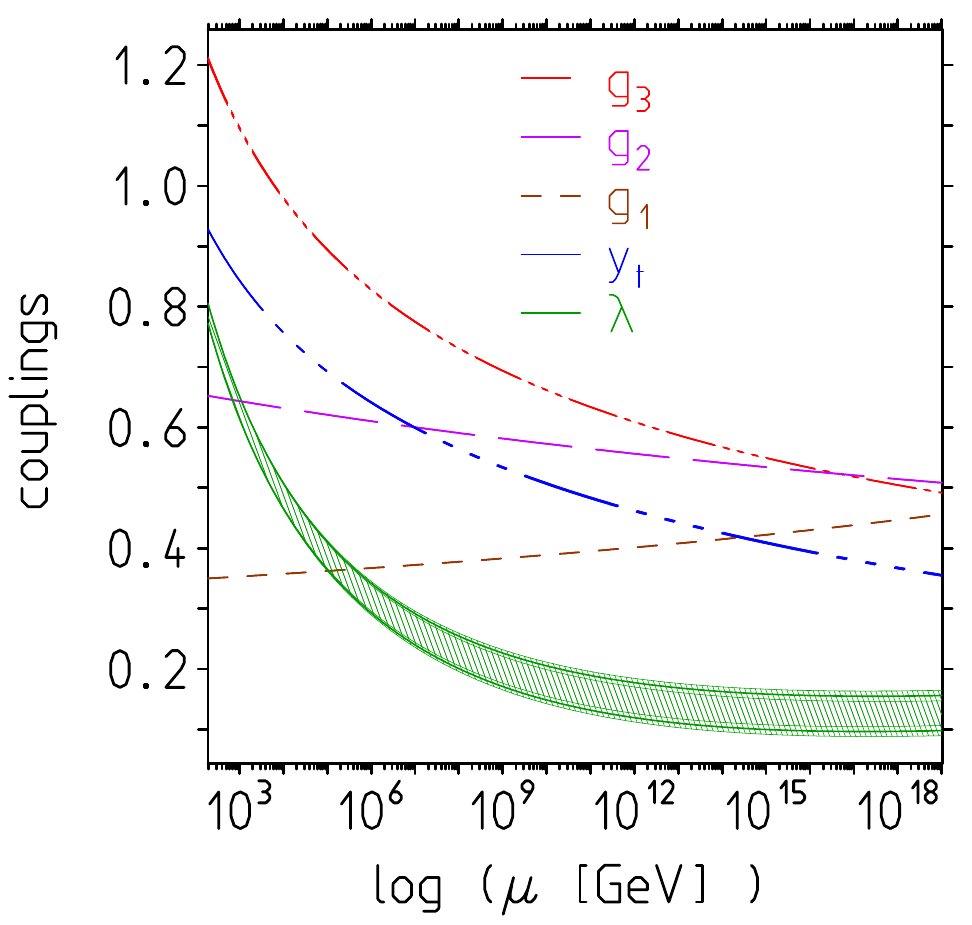}
\caption{The SM dimensionless couplings in the \MSb scheme as a
function of the log of the renormalization scale for $
M_H=124-126~\gv$ (shaded band). Input parameters as in Table~\ref{tab:MSbinp}}
\label{fig:SMrunning}
\end{figure}

In Fig.~\ref{fig:SMrunning} we plot the evolution of the SM couplings
as a function of the log of the energy scale.
As we learn from Fig.~\ref{fig:SMrunning} the amazing thing is that
the perturbation expansion turns out to work up to the Planck scale!
In our analysis, for the input parameters specified below, we have no
Landau pole or other singularities and the Higgs potential remains
stable. This likely opens a new gate to precision cosmology of the
early
universe~\cite{Yukawa:3,Degrassi:2012ry,Jegerlehner:2013cta}. The
remarkable interrelations between SM couplings may be summarized as
follows: the \mbo{U(1)_Y} coupling \mbo{g_1} is screening (IR free),
the \mbo{\SU(2)_L} coupling \mbo{g_2} and the \mbo{\SU(3)_c} coupling
\mbo{g_3} are anti-screening (UV free=Asymptotic Freedom (AF)) as
expected~\cite{AF}.  In contrast the top-quark Yukawa coupling
{\mbo{y_t}} and the Higgs self-coupling {\mbo{\lambda}}, which are
screening if standalone (IR free, like QED), within the SM change
their behavior from IR free to UV free, such that perturbation theory
works the better the higher the energy in these couplings as
well. What happens is that QCD effects dominate the behavior of the
top-quark Yukawa coupling RG provided $g_3>\frac{3}{4}\,y_t$ in the
gaugeless (\mbo{g_1,g_2=0}) approximation, which is
satisfied. Similarly, the top-quark Yukawa effect dominates the Higgs
coupling RG provided $\lambda<
\frac32\,(\sqrt{5}-1)\,y_t^2\,,$ which also holds in the gaugeless
(\mbo{g_1,g_2=0}) limit. These conditions are satisfied in the SM with
the given parameters and extend to higher orders as far as these are
known. We note that the Abelian hypercharge coupling $g_1$, although
increasing with energy, stays small up to $\lpl$ such that it does not
affect perturbativeness. Note that in spite of its increasing behavior
$g_1<g_3<g_2$ at Planck scale. Interestingly there we have an inverted
$g_3<g_2$ hierarchy of the non-Abelian gauge couplings. In the focus
is the Higgs self-coupling, because it may not stay positive
\mbo{\lambda > 0} up to
\mbo{\lpl}. In fact a 3 $\sigma$ significance for meta-stability is
claimed e.g. in~\cite{Yukawa:3,Degrassi:2012ry} (see right panel of
Fig.~\ref{fig:HR}). Calculating previously missing two-loop
contributions to the matching conditions the significance for missing
stability could be reduced to a 1 $\sigma$ gap
in~\cite{Bednyakov:2015sca}. The existence of a zero in $\lambda(\mu)$
crucially depends on the precise size of the top-quark Yukawa coupling
\mbo{y_t}, which actually seems to decide about the stability of our
world. Note that
\mbo{\lambda=0} would be an essential singularity! Uncertainties here
have to be reduced by more precise input parameters and better
established EW matching conditions. For our input parameters, the Higgs
coupling decreases up to the zero of the beta--function
\mbo{\beta_\lambda} at \mbo{\mu_\lambda \sim 3.5\power{17}~\gv},  where
$\lambda$ is small but still positive and above $\mu_\lambda$
increases with energy up to \mbo{\mu=\MPl}.

I think our discussion shows that ATLAS and CMS results may
have revolutionized particle physics in an unexpected way, namely showing
that the SM has higher self-consistency (conspiracy) than expected and
previous arguments for the existence of new physics may turn out not
to be compelling. Also, the absence so far of any established new physics signal at
the LHC may indicate that commonly accepted expectations may not be
satisfied. On the one hand, it seems to look completely implausible to
assume the SM to be essentially valid up to Planck energies, on the
other hand, the flood of speculations about physics beyond the SM
have been of no avail. Within the context of GUTs, a large gap in the
particle spectrum, the ``grand desert'' up to the GUT scale at about
$10^{16}~\gv$, still is a widely accepted hypothesis. So why the
``grand desert'' could not extend a little further namely to
$10^{19}~\gv$?  The central issue for the future is the very delicate
``acting together'' between SM couplings, which makes the precision determination
of SM parameters more important than ever. This mainly challenges
accelerator physics, the LHC experiments, and the future ILC and FCC-ee
projects (top-quark and Higgs-boson factories), which could improve
the precision values for \mbo{\lambda}, \mbo{y_t} and
\mbo{\alpha_s}. Still important are lower energy hadron facilities,
which should provide more precise hadronic cross sections in order to reduce
hadronic uncertainties in \mbo{\alpha(M_Z)} and
\mbo{\alpha_2(M_Z)}. This could open a new gate to precision cosmology
of the early universe, in case the Higgs boson inflation scenario outlined
in Sect.~\ref{sec:prelude} could be hardened.

\section{Thoughts on guiding principles and paradigms in particle physics}
\label{sec:paradigms}
In last decades ``solving the hierarchy problem'' has been a strong
motivation to find possible extensions of the SM. Guiding principles
often have played an important role in progress in science although
they afterward turned out to miss the point they suggested natural
laws should follow. Most prominent are symmetry principles. Related
group theory is beautiful mathematics but is not always mapping the
real or supposed physical problem it was proposed to describe. Kepler
already dreamed of the Platonic bodies (regular polyhedra) to rule
celestial mechanics of planets. After his attempt to prove this by
analyzing celestial data, finally, Kepler's laws resulted from his
investigation. Kepler's model is completely false, the interplanetary
distances it predicts are not sufficiently accurate, and Kepler was
scientist enough to accept this eventually. But it is an excellent
example of how truth and beauty do not always fit together. The
widespread string theory paradigm assumes that a simple highly
symmetric stringy structure at and beyond the Planck
scale\footnote{String theory is motivated by the requirement to
'quantize gravity'. Note that in string theory the Planck energy level
represents the ground state on top of which an infinite tower of
harmonics at $E_n=n\,\MPl\; (n=2,\cdots,
\infty)$ resides. The algebra of the spectral raising and lowering operators
can be closed to a Kac-Moody algebra (infinite dimensional analogs of
semi-simple Lie algebra) only in a particular space-time dimension
$D$. The unique supergravity string theory requires D=11, where 7 of
them are assumed to be compactified (see e.g.~\cite{LMT07}). For me it
is hard to believe that this is what shapes our real world.} could
explain what we observe down on earth, actually a rather complex real
world. That something simple looks complicated when seen from far away
is certainly not a very natural expectation. A solid macroscopic
may look to be perfectly rotational symmetric, zooming in on its
microscopic structure can uncover a lattice of atoms exhibiting all kinds
of lattice defects and domain patterns. The \textit{``the closer you
look the more there is to see''} pattern of thought looks to me less
artificial than expecting some harmonic oscillator ``heaven''. In any
case concepts like the ones we are discussing here: Naturalness,
Hierarchy and Fine-Tuning make no sense without specifying the context
in which they are addressed.

While ``solving the hierarchy problem'' seems to fail as a route to
new physics, in contrast, the concept of the minimal renormalizable
extension of Fermi's weak interaction theory turned out to be
impressively successful in constructing piece by piece the electroweak
SM. It has lead to the introduction of the massive intermediate spin 1
bosons $W^\pm$ in charged current processes, the prediction of neural
currents and the need for a $Z$ boson. The resulting $U(1)_Y\otimes
\SU(2)_L$ gauge theory, renormalizable in the massless case, requires
a scalar spin 0 boson, the Higgs boson, as a trick to generate masses
of the weak gauge bosons and the fermions, without spoiling
renormalizability. Spontaneous symmetry breaking, a mechanism known
from condensed matter physics, turned out to be the key mechanism for
a renormalizable massive gauge theory. Another example concerning
minimality versus non-minimality we have when considering GUTs, where
fermions are necessarily populating higher representations while the
fundamental ones are not occupied. Therefore the typical leptoquarks
necessarily showing up in GUTs are unnatural as an emergent
phenomenon\footnote{The unification paradigm celebrated its triumphal
success in Maxwell's electromagnetism, which unified electrical and
magnetic laws and predicted electromagnetic waves. In contrast, as we
know the electroweak theory is not a true unification, it rather
regulates the mixing of electromagnetic and weak interaction
phenomena. At the heart is $\gamma-Z$ mixing and $Z$ resonance (a kind
of ``heavy-light'') physics, which manifests itself most convincingly
in electron-positron annihilation into $Z$ bosons. All further
unification attempts so far are missing confirmation.}.

A convincing solution of the SM's hierarchy problem is known to be
provided by a supersymmetric extension of the SM. SUSY
cannot be an exact symmetry because it would predict a degenerate mass
spectrum while phenomenologically the states of the SUSY mirror world
all must be heavier than the SM particles. This leads to a very
complicated world as a broken SUSY scenario not only is doubling the
spectrum at once but also leaving too much freedom, with about 100
unknown symmetry breaking parameters. This makes such extensions not
really predictive without additional assumptions. At the end
phenomenological constraints require a SUSY version that would not be
solving the hierarchy problem really, rather it would only be shifting
the amount of fine-tuning required.

In addition, the hierarchy fine-tuning problem if being solved by a
super\-sym\-metriza\-tion of the SM creates new problems as we
know. First, a second Higgs doublet field needs to be introduced,
which as such is an interesting option. However, in order not to be in
conflict with the absence of tree-level Flavor Changing Neutral
Currents (FCNC)~\footnote{FCNCs are automatically absent in the SM by
the GIM mechanism~\cite{GIM}, as it is also highly established by
experiment.} one has to impose $R$--parity, which is not less a fine
tuning, although FCNCs can be forbidden by a simple discrete
symmetry. $R$--parity is not required by renormalizability, it is not
naturally emergent in a low energy effective theory and thus looks to
be ad hoc\footnote{Similarly, a global lepton flavor symmetry
$U(1)_e\otimes U(1)_\mu \otimes U(1)_\tau$, which would imply  exact lepton
flavor conservation is not emergent because renormalizability does not
dependent on it. That it is a surprisingly accurate approximate
symmetry is due to the smallness of the neutrino masses, which likely
results as a consequence of a see-saw mechanism at work.}. A generic SUSY extension as such
would be in contradiction with observation right away. This also
illustrates that naturalness is a doubtful concept: $R$-parity is a
symmetry which forbids FCNCs but what is natural about it?

In my opinion, the dogma surrounding the so-called hierarchy or fine-tuning problem turned out to be a complete failure. Similarly, the GUT
paradigm has not been leading to any experimentally confirmed predictions that
would support this concept.  I think the minimal renormalizable QFT
paradigm is back. Thereby ``minimal'' is crucial, many higher
renormalizable structures like a GUT extension of the SM are not
natural in that sense. Interestingly, a missing third fermion family
or an additional fourth family would spoil important properties of the
SM, such as the RG flow, and the Higgs boson then could not be a
candidate for the inflaton. Another very important special feature of
the SM is its tree level accidental custodial symmetry. The latter is
violated by many of the proposed extensions of the SM, which then
create a different fine tuning problem~\cite{Czakon:1999ha}, in all
cases that violate the tree level SM relation $\cosW\,\mz/\mw=1$. This
relation is strongly supported by experimental data, which
precisely confirm SM predictions of the radiative corrections.

One also has to keep in mind that precision tests of the SM already
revealed a test in-depth of its quantum structure. The largest
corrections come from the running fine structure constant $\alpha(s)$,
the running of the strong coupling $\alpha_s(s)$ and the large top
Yukawa $y^2_t(s)$ effects. As contribution to the $\rho=G_{\rm
NC}/G_\mu(0)$ parameter, for example, subleading corrections amount to
a 10 $\sigma$ deviation from the SM prediction when taking into
account the largest corrections only. Thus the SM is on very solid
grounds better than everything else we ever had.

On the other hand, the view that the SM is a low energy effective
theory of some cutoff system at the Planck energy scale $\MPl$ appears
to be consolidated. This also puts QFT on a firm mathematical basis. A
crucial point is that $\MPl$, providing the scale for the low energy
expansion in powers $E/\MPl$, is exceedingly large, very far from what
we can see! A dimension 6 operator at LHC energies is suppressed by
$(E_{\rm LHC}/\lpl)^2\approx 10^{-30}\,$. This seems to motivate a
change in paradigm from the view that the world looks simpler the
higher the energy to a more natural scenario which understands the
``cutoff SM'' as the ``true world'' seen from far away, with
symmetries emerging from not resolving the details of the short
distance structure. In the low energy expansion, one is ``throwing
away'' an infinite tower of shorter distance information carried by
the suppressed so-called \textit{irrelevant} operators.

The hierarchy problem requires to take the relationship between the
bare UV and the renormalized IR regime as testable physics. Here
Wilson's RG comes into play.  Kenneth Wilson 1971~\cite{Wilson:1971bg}
has been able to solve the problems surrounding the critical indices
of phase transitions in condensed matter systems, which have been
persisting for about 75 years. His work has shed new light on the role
cutoffs may play in physical laws. Wilson's renormalization
semi-group, based on integrating out irrelevant details of the short
distance structure opened the quantitative approach of constructing
low energy effective quantum field theories that derive from systems
whose short distance structure has an intrinsic cutoff, like an
atomic lattice or an atomic gas or fluid (see
e.g.~\cite{LausanneLectures1976}). The key low energy emergent
structure notably turned out to reveal renormalizable Euclidean
quantum field theory. The latter exhibits analyticity in a way which
makes it equivalent to a Minkowski quantum field theory. The latter
hence is incorporating quantum mechanics as an emergent structure. As
I will argue in the following, cutoffs in particle physics are
unavoidable in understanding the relationship between a bare and a
renormalized theory (see e.g.~\cite{Jegerlehner:1976xd}). In such
context, renormalizability is an emergent property like all structures
required in order renormalizability to be manifest. In our context
cutoffs are indispensable for understanding early cosmology in a
bottom-up way~\cite{Jegerlehner:2013cta}. This opens the possibility
of an alternative understanding of inflation, reheating, baryogenesis
and all
that~\cite{Starobinsky:1980te,Guth:1980zm,Albrecht:1982wi,Linde:1981mu,Kolb:1990vq,Weinberg:2008zzc}. As
in condensed matter physics the connection between macroscopic long
distance physics (at laboratory scales) and the microscopic underlying
cutoff system (high energy events as they were natural in the early
universe) turn out to have a physical meaning.

I remind that the SM's naturalness problems and fine-tuning problems
have been made conscious by G. 't Hooft~\cite{'tHooft:1979bh} long
time ago as a possible problem in the relationship between macroscopic
phenomena that follow from microscopic laws (a condensed matter
system inspired scenario). Soon later the ``hierarchy problem'' had
been dogmatized as a kind of fundamental principle. In fact, the
hierarchy problem of the SM had been the key motivation for a number of
types of extensions of the SM. It is therefore important to reconsider
the ``problem'' in more detail.

One of my key points concerns the different meaning a possible
hierarchy problem has in the symmetric and in the broken phase of the
SM.  In order to understand the point, we have to remember why we need
the Higgs particle in the SM. The Higgs boson is necessary to get a
renormalizable low energy effective electroweak
theory~\cite{Glashow61}. Interestingly, one scalar particle is
sufficient to solve the renormalizability problems arising from each
of many different massive fields in the SM, of which each causes the
problem independently of the others. The point is that this one
particle has to exhibit as many new forces as there are individual
massive states~\cite{Weinberg67}. All required new interactions are in
accordance with the SM symmetry structure in the symmetric phase as we
know. The taming of the high energy behavior, of course, requires the
Higgs boson to have a mass in the ballpark of the other given heavier
SM states, if it would be much heavier it would not serve its purpose
in the low energy regime. It would lead to the so-called ``delayed
unitarity'' phenomenon~\cite{Ahn:1988fx}. Note that the Higgs boson
has to cure the unphysical mass effects for the \textbf{given} gauge
boson masses $M_W$, $M_Z$ and fermion masses $M_f$ via adequate \textit{Higgs exchange
forces}, where the coupling strength is proportional to the mass of
the massive field coupled. A very heavy Higgs boson eventually would
decouple and thus miss to restore renormalizability of the massive
vector-boson gauge theory. Interestingly, in the symmetric phase the
SM gauge-boson plus chiral fermions sector is renormalizable without
the Higgs-boson and Yukawa sectors and scalars are not required at all
to cure the high energy behavior because it is renormalizable by its
own structure. Therefore, in the symmetric phase, the mass degenerate
Higgs fields in the complex Higgs doublet can be as heavy as we
like. Since unprotected by any symmetry, naturally we would expect the
Higgs particles indeed to be very heavy. In fact the ``origin'' of the Higgs
mass is very different in the broken phase, where all the masses,
including the Higgs mass itself, are generated by the Higgs
mechanism~\cite{Englert:1964et,Higgs:1964ia}. This we learn from the
relation $m^2_H=\frac13\,\lambda\,v^2$, holding in the broken phase.
In the symmetric phase, the effective Higgs mass is dynamically
generated by the Planck medium, as we will argue below. Therefore, the
usual claim that the SM requires to be extended in such a way that
quadratic divergences are absent has no foundation. Purely formal
arguments based on perturbative counterterm adjustments do not lead
any further.

The hierarchy problem in particular addresses the presence of
quadratic UV divergences related to the SM Higgs mass
term. Infinities in physical theories are the result of idealizations
and show up as singularities in a formalism or in models. UV
singularities in general plague the precise definition as well as
concrete calculations in quantum field theories (QFT)~\footnote{Taming
the infinities we encounter in the theory of elementary particles,
i.e. of quantum field theories, by completing them with a cutoff, often
called the UV--completion of a QFT is as old as QFT itself. Actually,
it took 20 years from Dirac 1928 (Dirac hole theory of relativistic
electron-photon interaction [pre-QED]) to Feynman, Schwinger and
Tomonaga in 1948 who found how to deal with the large cutoff limit and
making QED a predictive theory. For non-Abelian gauge theories
proposed by Yang and Mills in 1954~\cite{YangMills} it took another 17
years until a renormalizable formulation was found by 't Hooft in
1971~\cite{tHooft71a} (actually by circumventing a cutoff
regularization).}. A closer look usually reveals infinities to
parametrize our ignorance or mark the limitations of our understanding
or knowledge. One particular consequence of UV divergences in local
QFT is that the vacuum energy is ill-defined as it is associated with
quartically divergent quantum fluctuations.

This is another indication that tells us that local continuum QFT has
its limitation and that the need for regularization is actually the
need to look at the true system behind it. In fact the cutoff system
is more physical and does not share the problems with infinities resulting
from the idealization realized in the large cutoff limit or
lattice continuum limit. In any case, the framework of a renormalizable
QFT, which has been extremely successful in particle physics up to
highest accessible energies is not able to give answers to the
questions related to vacuum energy and hence to all questions related
to dark energy, accelerated expansion, and inflation of the universe.

Since the SM exhibits non-AF couplings like the $U(1)_Y$ coupling $g_1$
or the Higgs self-coupling $\lambda$ at scales beyond the zero of the
$\beta_\lambda$ function, also lattice
calculations~\cite{Luscher:1988gc,Lang:1985nw,Callaway:1988ya}
strongly suggest that in fact, the theory requires a finite cutoff, because
the continuum limit at infinite cutoff would be trivial.

It is thus natural to consider the SM to be what we observe as the Low
Energy Effective SM (LEESM), the renormalizable tail of the real
cutoff system sitting at the Planck scale. As a consequence all
properties required by renormalizability, gauge symmetries, chiral
symmetry, anomaly cancellation, and the related fermion family
structure, as well as the existence of an elementary scalar, the Higgs
boson, naturally emerge as a consequence of the low energy
expansion\footnote{It is interesting to note that statistical
mechanical systems with short-range exchange and long-range multipole
interactions exhibit vector bosons and graviton modes that follow
from a multipole expansion of a static
potential~\cite{Jegerlehner:1978nk}. In this sense the emergence of
gauge-bosons looks pretty natural.}. We remind that the
emergence of SM structures in a low energy expansion is a well
investigated subject (see e.g.~\cite{Bass:2017nml} and references
therein). It is often advocated as a tree-unitary requirement but is
easily reinterpretable as a low energy expansion where
non-renormalizable effects are suppressed by inverse powers in the
cutoff. These mechanisms are calculable within perturbation
theory~\cite{Veltman:1968ki,LlewellynSmith:1973ey,Bell:1973ex,Cornwall:1973tb,Jegerlehner:1978nk,Jegerlehner:1994zp,Jegerlehner:1998kt}.
As SM perturbation theory works at the $Z$ mass scale and gets better
with increasing energy these perturbative derivations of gauge
symmetry and Higgs structure attain the status of proofs.  The
infinite tower of higher order operators is suppressed to be
invisible. Only a few operators are non-irrelevant and effectively
observable, and this is what makes the world look much simpler than a
possibly chaotic Planck medium. In reality, infinities related to the
relevant operators are replaced by eventually very large but finite
numbers, and I will show that sometimes such huge effects are needed
in order to understand the real world. I will argue that cutoff
enhanced effects are responsible for triggering the Higgs mechanism
not very far below the Planck scale and the inflation of the early
universe, as outlined already in Sect.~\ref{sec:prelude}.

The history of our universe we can trace back 13.8 billion years close
to the Big Bang, when the expansion of the universe was ignited in a
``fireball'', an extremely hot and dense state when all structures, and
in the end, all atoms, nuclei, and nucleons were disintegrated into a
world of elementary particles only. So the SM provides the key
information for what has happened in the early universe, and high
energy accelerator experiments are testing processes that only took
place in nature in the early history of the universe. If the Higgs
boson is the source of dark energy that triggered inflation, its
discovery could mark a milestone in our understanding of the dynamics
of the very early universe. The origin of cold dark matter remains
a mystery, which can have many different explanations.

I think that questions concerning the early universe can be addressed
only within a LEESM ``extension'' of the SM, e.g. given by the SM
supplied with a cutoff structure in a minimal way. As we know, in a
renormalizable QFT all renormalized quantities as a function of the
renormalized parameters and fields in the limit of a large cutoff are
finite and devoid of any cutoff relicts! Here we should remember the
Bogoliubov-Parasiuk renormalization theorem that states that order by
order in perturbation theory the renormalized Green's functions and
matrix elements of the scattering matrix (S-matrix) are free of
ultraviolet divergences. The theorem specifies a concrete procedure
(the Bogoliubov-Parasiuk R-operation) for the subtraction of divergences,
establishes the correctness of this procedure, and guarantees the
uniqueness of the obtained results, modulo reparametrizations, which
are controlled by the renormalization group. In other words, in the
low energy world cutoff effects are not accessible to
experiments. Consequently, the hierarchy problem cannot be addressed
within the renormalizable, renormalized SM, which encodes all observables. In
this framework, all independent parameters are free and have to be
supplied by experiments. In this sense, within the renormalized QFT
the hierarchy problem is a pseudo-problem.

To my knowledge, the only non-perturbative definition of a
renormalizable local quantum field theory is the possibility to put in
on a lattice by discretization of space-time. This again may be taken
as an indication that the need for a cutoff actually is an indication
that the cutoff exists in the real world. In this sense,
lattice-QFT is closer to the true system than its continuum tail. Of course,
there are many ways to introduce a cutoff and actually, we cannot know
what the cutoff system looks like truly. This is not a real problem if
we are interested in the long-range patterns mainly. The only thing we
have to take care of is that the underlying system is in the
\textit{universality class} of the SM.  This in particular concerns
the observable degrees of freedom and the emergent symmetries at work,
which require the particles to be grouped predominantly in the
simplest (lowest dimensional) representation of the corresponding
symmetry groups. The simplest symmetry groups with singlets, doublets,
and triplets are the most natural ones to emerge, as realized within
the SM's $U(1)_Y\otimes \SU(2)_L \otimes
\SU(3)_c$ gauge symmetry
pattern~\cite{Veltman:1968ki,LlewellynSmith:1973ey,Bell:1973ex,Cornwall:1973tb,Jegerlehner:1994zp,Jegerlehner:1998kt}.
More on how the SM may emerge the reader may find in the Appendix.

\section{The Hierarchy Problem revisited}
\label{sec:hierarchy}
In~\cite{Jegerlehner:2013nna} already, I outlined the flaws I see in
the common reasoning concerning the hierarchy issue. As argued above, I
am addressing the hierarchy problem within the LEESM ``extension'' of the
SM.  Specifically, I have in mind an implementation of the SM on a
Planck lattice (see e.g.~\cite{Luscher:2000hn}). The only important
point is that we can perform a low energy expansion in the
corresponding cutoff. It is an accepted fact that the SM predicts a
huge gap between the renormalized and the bare Higgs boson mass. From
the LEESM point of view, this prediction is what promotes the Higgs
boson to be a promising candidate for the inflaton. The hierarchy gap
showing up is not something we have to avoid. Now, would-be infinities
are replaced by eventually very large but finite numbers, and I will
show that sometimes such huge effects provide what we need to
understand established phenomena like inflation.

One thing we should remind here: the bare suitably regularized theory
has always been the true one. Renormalization always has just been a
reparametrization. The bare theory assumed to exhibit a cutoff of
some sort, shows a cutoff dependent large-cutoff tail, sometimes
called ``preasymptote''~\cite{Jegerlehner:1976xd,Jegerlehner:2013cta},
which is equivalent to a renormalized local QFT in the universality
class of the cutoff-system. Thereby it is \textbf{not} important that the bare
cutoff system exhibits all symmetries the long-range tail will have
because most of the symmetries of the LEET are emergent. In fact, by a
reparametrization of parameters and fields of the preasymptotic theory
(renormalizable tail) the residual cutoff-dependence is completely
removable (see~\cite{Jegerlehner:1976xd} and references
therein). Because the renormalized tail has lost all information about
the cutoff, it is nonsensical to say that in the LEET we would
naturally expect the Higgs mass to be of the order of the cutoff.

However, in the LEESM ``extension'' of the SM, bare parameters turn into
physical parameters of the underlying cutoff-system being the ``true
world'' at short distances. Then the hierarchy problem is the problem
of ``tuning to criticality'', which concerns the dim $<4$
\textit{relevant operators}, in particular the mass terms.
In the symmetric phase of the SM, there is only one mass to be
renormalized, the others being forbidden by the known chiral and gauge
symmetries. For the Higgs field mass which appears in the Higgs
potential the fine-tuning to criticality has the familiar form
\bea
m_0^2(\mu_1=\mpl)&=&m^2(\mu_2=M_H)+\delta m^2(\mu_1,\mu_2)\;;\;\; \crn
\delta m^2&=&
\frac{\lpl^2}{16 \pi^2}\,C(\mu)\,,
\label{massren}
\eea
with a coefficient typically $ C=O(1)$. To keep the renormalized mass
at some small value, which can be seen at low energy, {\bf formally} $
m^2_0$ has to be adjusted to compensate the huge number $ \delta m^2$
revealed by the perturbative SM calculation such that about {\bf 35
digits} must be adjusted in order to get the observed value below the
electroweak scale. Is this a problem?

One thing is obvious: our fine-tuning relation exhibits quantities
(in the LEESM all observable in principle) at very different scales,
the renormalized ones at low energy and the bare ones when approaching the
Planck scale. As long as we have no direct access to the Planck
physics there is no proven conflict.

Actually, a closer look reveals that in the Higgs phase there is no hierarchy
problem in the SM! Why?  It is true that in the relation
(\ref{massren}) both $ m^2_0$ and $\delta m^2$ {\bf formally} may be
expected many many orders of magnitude larger than $m^2\,.$ Even
worse, in the broken phase $\delta m^2$ has a huge negative value and hence
$m^2_0$ must be tuned to be huge negative as well.
However, in the broken phase, $ m^2\propto v^2(\mu_0)$ is $
O(v^2)$ not $ O(\MPl^2)$. Since $v$ is the result of spontaneous symmetry
breaking (non-symmetric ground state) it is per se a low energy
parameter related to the emergence of long-range order. Thus in the
broken phase, the Higgs boson is expected to be natural light. That
the Higgs mass likely is $ O(\MPl)$ in the symmetric phase is what
realistic inflation scenarios are demanding.

\begin{figure}
\centering
\includegraphics[height=4cm]{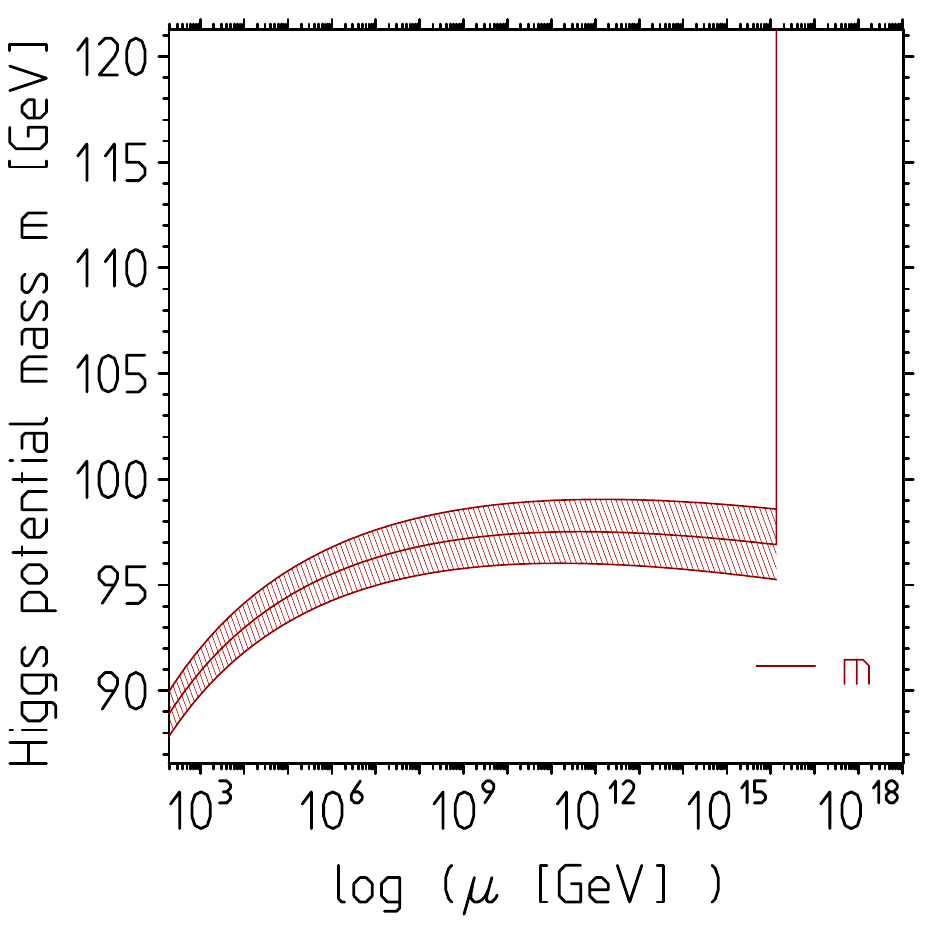}
\includegraphics[height=4cm]{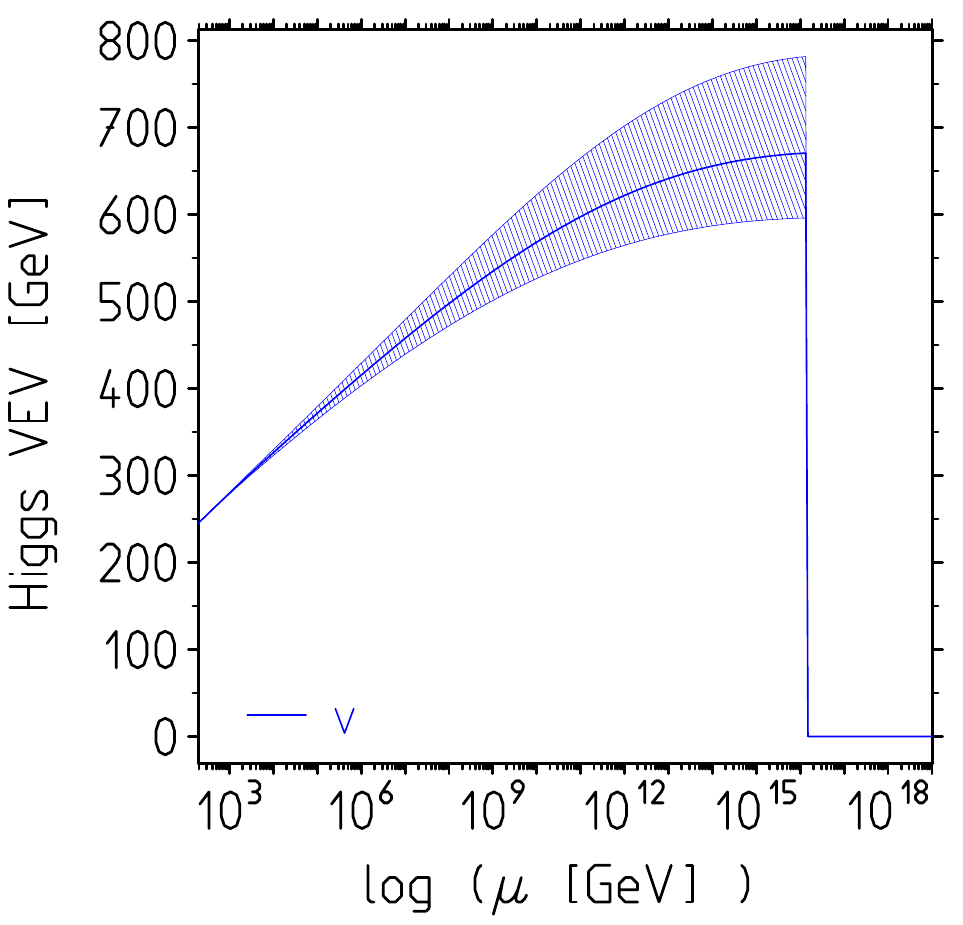}
\caption{Dimensionful SM running \MSb parameters $m$ and
$v=\sqrt{6/\lambda}\,m$. Error bands include SM parameter
uncertainties and a Higgs boson mass range $125.5\pm 1.5~\gv$ which
essentially determines the widths of the bands}
\label{fig:mandv}
\end{figure}

In the broken phase, characterized by the
non-vanishing Higgs field VEV $v(\mu) \neq 0$,
all the masses are determined by the well-known mass coupling
relations
\begin{equation}
\label{masscoupl}
\begin{array}{cclccl}
m_W^2(\mu^2)&=&\frac14\,g_2^2(\mu^2)\,v^2(\mu^2)\semis &
m_Z^2(\mu^2)&=&\frac14\,(g_2^2(\mu^2)+g_1^2(\mu^2))\,v^2(\mu^2)\semis\\[2mm]
m_f^2(\mu^2)&=&\frac12\,y^2_f(\mu^2)\,v^2(\mu^2)\semis &
{ m_H^2(\mu^2)}&{ =}&{ \frac13\,\lambda(\mu^2)\,v^2(\mu^2)}\epo
\end{array}
\end{equation}
Here we consider the parameters in the \MSb renormalization scheme,
$\mu$ is the \MSb renormalization scale, which we have to identify with
the energy scale of the physical processes or equivalently with the
corresponding temperature in the evolution of the universe.  The RG
equation for $v^2(\mu^2)$ follows from the RG equations for masses and
massless coupling constants using one of these relations. The
evolution of the \MSb versions of $m$ and $v$ are shown in Fig.~\ref{fig:mandv}. As a key
relation we
use~\cite{Jegerlehner:2001fb}
\begin{eqnarray}
\mu^2 \frac{d}{d \mu^2}\, v^2(\mu^2)
&=&3\, \mu^2 \frac{d}{d \mu^2} \left[\frac{m_H^2(\mu^2)}{\lambda(\mu^2)} \right]
\equiv
v^2(\mu^2) \left[\gamma_{m^2}  - \frac{\beta_\lambda}{\lambda} \right]\,,
\label{vev}
\end{eqnarray}
where $\gamma_{m^2} \equiv \mu^2 \frac{d}{d \mu^2} \ln m^2$ and
$\beta_\lambda \equiv \mu^2 \frac{d}{d \mu^2} \lambda \,.$ We write
the Higgs potential as $V=\frac{m^2}{2}\,H^2+\frac{\lambda}{24}H^4$,
which fixes our normalization of the Higgs self-coupling. When the
$m^2$-term changes sign and $\lambda$ stays positive, we know we have
a first order phase transition (see below).
Funny enough, the Higgs particle gets its mass from its interaction with its own
condensate! and thus gets a mass in the same way and in the same
ballpark as the heavier SM species, which couple strongest to the Higgs
field. As mentioned before the Higgs mass
cannot be much heavier than the other heavier particles if
renormalizability is to be effective at low and moderate energies. The
interrelations (\ref{masscoupl}) also show that for fixed $v$, as determined by
the Fermi constant $G_\mu=1/(\sqrt{2}\,v^2)$, the Higgs cannot get too heavy if perturbation
theory should remain applicable. Also note that the conspiracy between
those couplings relevant to stabilize the vacuum only can
work if these couplings are of comparable size.

Often an extreme point of view is taken: all particles naturally
should have masses \mbo{O(\mpl)}
i.e.  \mbo{v=O(\mpl)}. This would mean that the symmetry is not recovered at
the high (=bare) scale and the notion of spontaneous symmetry breaking
would be obsolete! Of course, this makes no sense. In a
perturbative calculation within a cutoff regulated theory formally one
finds \mbo{v=O(\mpl)} but in the broken phase $\delta m^2_{H}$ is huge
negative, which requires a non-perturbative
\textit{vacuum rearrangement} revealing the mentioned mass
coupling relations in terms of a renormalized effective $v$ also for
the Higgs particle. Since $ v\equiv 0$ above the EW phase-transition
point, it makes no sense to say that one naturally has to expect $
v(\mu=\mpl)=O(\mpl)\epo$ The Higgs-field VEV \mbo{v} is an \textit{order
parameter}, related to the spontaneous breaking of the discrete
symmetry\footnote{In the unitary gauge, we can avoid problems related
to Elitzur's theorem~\cite{Elitzur:1975im}, which claims that an order
parameter cannot be associated with SSB of a non-Abelian gauge
theory. In a physical gauge, on physical Hilbert space, Higgs ghost
fields are absent and a Mexican hat potential is a phantom as it only
exist if ghost space is taken into the display. A physical Mexican hat
potential would imply the existence of three Nambu-Goldstone bosons.}
$Z_2: H \leftrightarrow -H$, and is resulting from long range {\bf
collective behavior}. It can be as small as we like. Its value is a
function of the effective temperature (energy scale) with its maximum
at $T=0$, monotonically decreasing with increasing temperature and
vanishing at the second order phase transition point $T_c$ above which
$v(T)\equiv 0$ vanishes identically (non-analyticity).

\begin{figure}
\centering
\includegraphics[height=4cm]{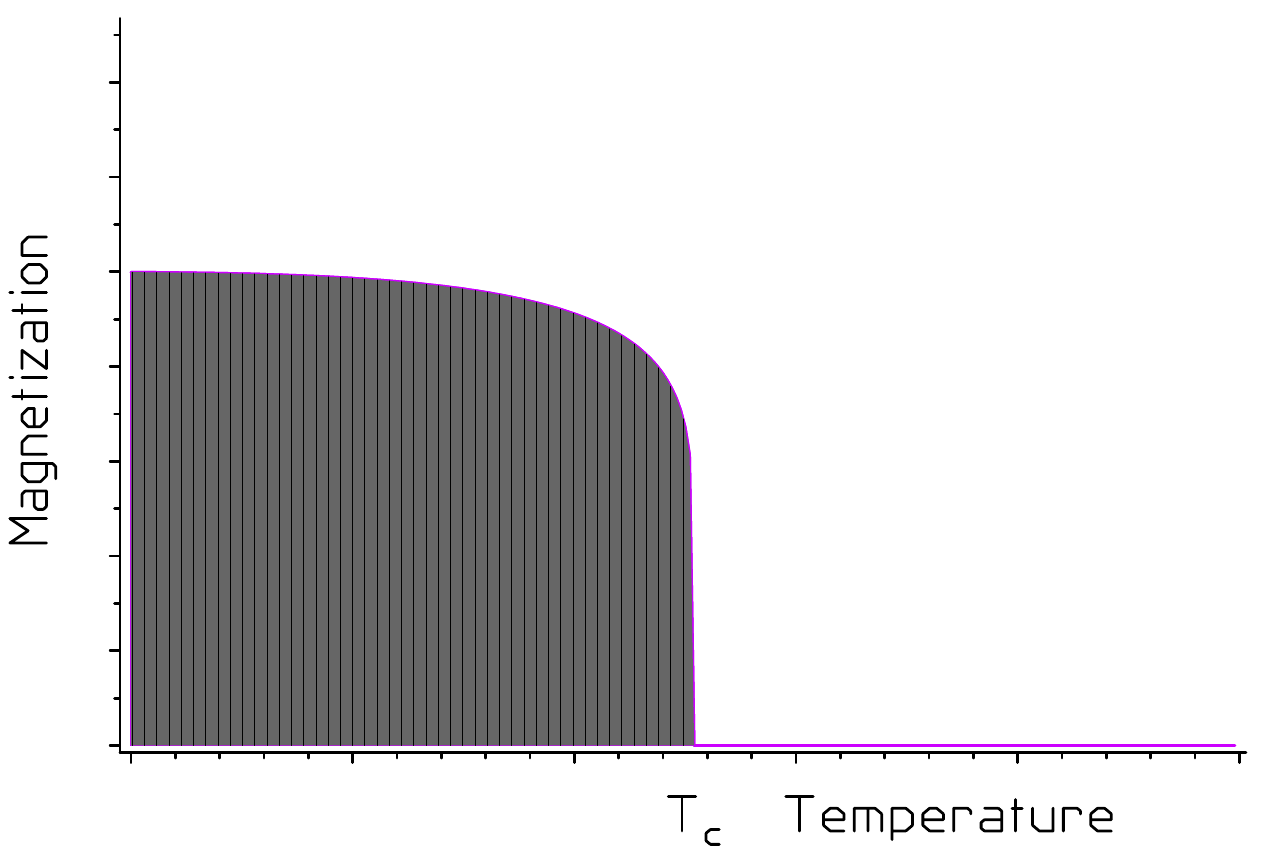}
\caption{Spontaneous magnetization \mbo{M=M(T)} as a function of
temperature \mbo{T}. $T_c$ is the critical temperature above which \mbo{M(T)\equiv 0}
for all \mbo{T>T_c}. Furthermore, \mbo{M(T) \to 0} as \mbo{T\lto T_c}
may be as small as we like depending on the distance \mbo{T-T_c} from
criticality. Note that $M(0)$ is not given by what would correspond to
the cutoff of the ferromagnetic system, even if it would be measured
in units of the cutoff}
\label{fig:ferro}
\end{figure}

A well known prototype for long range order is the magnetization in a
ferromagnetic spin system\footnote{As an example we may consider an Ising ferromagnet in
\mbo{D=2} dimensions, \mbo{J} is the nearest neighbor (n.n.) spin
coupling between the spins on a lattice
$$
H(\sigma)= -J\sum_{<ij>}\sigma_{\! i}\,\sigma_{\! j}\,\semis
P_\beta(\sigma)=\frac{\E^{-\beta H(\sigma)}}{Z_\beta}\semis
Z_\beta=\sum_\sigma \E^{-\beta H(\sigma)}\epo
$$
Here $ \beta=\frac{1}{k_B T}$ where \mbo{k_B} is the Boltzmann
constant. The Onsager solution for the critical temperature reads
$\sinh^2 \left(\frac{2J}{k_BT}\right)=1 \semis T_c=\frac{2J}{k_B\,\ln(1+\sqrt{2})}$
and the magnetization is given by
$M=\left(1-\left[\sinh 2 \beta J\right]^{-4}\right)^{\frac{1}{8}}\,,$
depending on temperature \mbo{T} and n.n. spin interaction strength \mbo{J}.
For more details see e.g.~\cite{LausanneLectures1976}} illustrated in Fig.~\ref{fig:ferro}.

The analogy shows us that \mbo{v/\mpl\ll 1} is not unnatural since
\mbo{v\neq 0} emerges only below a critical temperature, which is not
in a simple way related to \mbo{\mpl}. The EW scale is set by
\mbo{v(\mu)} and depends on $\mu$. As we learn from
Fig.~\ref{fig:mandv}, at low energy
\mbo{v(0)=1/({\small \sqrt2} G_\mu )^{1/2}\approx 246~\gv}, but
interestingly, in contrast to the magnetization of a
ferromagnet, \mbo{v(\mu)} is increasing rather than monotonically
decreasing with increasing \mbo{\mu}. This is because of the rich conspiring
dynamics of the SM encoded in the RG equation
(\ref{vev})\footnote{Fig.~\ref{fig:mandv} shows that the Higgs mass
parameter $m$ is varying little in the broken phase, while
$v=\sqrt{6/\lambda}\,m$ increases substantially because $\lambda$
decreases rapidly.}, yet \mbo{v(\mu)} is vanishing at
\mbo{\mu_0 \sim 10^{16}~\gv}:
\mbo{v(\mu) \to 0 \mathrm{ \ when \ } \mu \lto \mu_0}, as we will see later. The
second order phase transition (PT) point is a point of non-analyticity
i.e. exhibits singular behavior and physics in the ordered phase and
the disordered phase are very different.

Considering a ferromagnet one has to tune the temperature \mbo{T} to
criticality in order to find the PT point. What is tuning the
temperature to criticality in the SM? The answer is {\bf the expansion
of the universe}, which provides a scan in temperature (see
also~\cite{Hamada:2015dja}). The maximum value of \mbo{v(\mu)} is
achieved in the low energy limit at \mbo{\mu=0}. Why should the
magnitude of \mbo{v(0)} be of the order of the Planck scale, given the
fact that above the phase transition
point, in the disordered phase, the VEV is actually vanishing identically?

This shows that the Higgs boson mass renormalization equation is not a
static equation but is subject to a sophisticated dynamics driven by
the expansion of the universe.

In the symmetric phase at very high energy, we see the bare
system. There the Higgs field is a collective field exhibiting an
effective mass generated by radiative effects within the Planck system
such that $m^2_0 \approx \delta m^2$ at $\mpl$. In particle physics, a
radiatively induced mass is known from the Coleman-Weinberg
mechanism~\cite{Coleman:1973jx}, now in the symmetric phase and
applied to the Planck medium. Such a mechanism, which is natural in this
context, eliminates a possible fine-tuning problem at all
scales. There are many examples in condensed matter systems, like the
effective mass of the photon in the superconducting phase (Meissner
effect) or the effective mass of the effective field which encodes the
spin-singlet electron pairs (Cooper pairs) in the Ginzburg-Landau (GL)
model~\cite{GLtheory} of superconductivity\footnote{Originally the
Ginzburg-Landau theory of superconductivity has been proposed as a
macroscopic phenomenological effective theory describing type-I
superconductors without reference to microscopic properties. Later
Bardeen-Cooper-Schrieffer could explain superconductivity from its
microscopic structure in their BCS-theory~\cite{BCStheory}. Afterward, Gor'kov derived
the GL-theory~\cite{Gorkov} showing that in some limit
all GL parameters have a microscopic interpretation. In addition,
Abrikosov showed that GL-theory also models type-II
superconductors~\cite{Abrikosov}. The effective GL-theory thus
efficiently describes a rich variety if superconducting systems,
without the need for a detailed microscopic understanding.}. The
latter directly corresponds to the Abelian Higgs model. Emerging as an
effective field from the hot Planck system, which exhibits all types
of excitations, it is also pretty obvious that the Higgs field couples
to all these modes that we see as Yukawa and Higgs to gauge boson
couplings in the SM. That these couplings exhibit the symmetries of
the SM is again due to the fact that only the renormalizable tail can
be seen at low energies. All Planck system modes that do not
conspire, as SM degrees of freedom and their couplings do within the
SM, are not perceivable at long distances. The SM emerges as a
self-organized system.

On the one hand, we know that astronomy and astrophysics are
unthinkable without the input from laboratory physics in general and
particle physics in particular. On the other hand, it is not new that
particle physics is learning from cosmology. What is required to
explain inflation, baryogenesis, nucleosynthesis, CMB patterns, dark
matter, etc. ? If the SM has an extrapolation up to the Planck scale,
evidently one is able to confront SM predictions with physics
established to have happened in the early universe. In contrast to the
old paradigm of an empty vacuum: we know that the ground state of the
world is filled with dark energy, with a Higgs condensate and quark
and gluon condensates. All these effects have been showing up at
certain times and play a key role in the evolution of the
universe. Obviously, there are plenty of questions to be answered in
order to get a better understanding of how the universe has been
shaped after the Big Bang.

\section{Running SM parameters trigger the Higgs mechanism}
\label{sec:running}
In Sect.~\ref{ssec:run} already, we have discussed how the Higgs boson
discovery has been revealing a peculiar value for the Higgs boson
self-coupling, which largely clarified the path of extrapolating the
SM to higher energies. We remind that all dimensionless couplings
satisfy the same RG equations in the broken and in the unbroken phase
and are not affected by any power cutoff dependencies. This is as it
has to be because the Higgs mechanism (SSB) does not alter the UV
behavior. The evolution of the SM couplings in the \MSb scheme up to
the Planck scale has been investigated
in~\cite{Hambye:1996wb,Shaposhnikov:2009pv,Holthausen:2011aa,Yukawa:3,Degrassi:2012ry,Moch12,Mihaila:2012fm,Chetyrkin:2012rz,Masina:2012tz,Bednyakov:2012rb,Tang:2013bz,Buttazzo:2013uya},
and has been extended to include the Higgs-field VEV and the masses
in~\cite{Jegerlehner:2012kn,Jegerlehner:2013cta}. Except for $g_1$,
which increases very moderately, all other couplings decrease and stay
positive up to the Planck scale. This strengthens the reliability of
perturbative arguments and reveals a stable Higgs potential up to the
Planck scale~\cite{Jegerlehner:2012kn,Jegerlehner:2013cta}. While most
analyses~\cite{Yukawa:3,Degrassi:2012ry,Moch12,Masina:2012tz,Buttazzo:2013uya,Tang:2013bz}
find that for the given Higgs mass value range vacuum stability is
nearby only (meta-stability)~\footnote{Most groups are adopting
essentially the same input parameters presented
in~\cite{Yukawa:3,Degrassi:2012ry,Buttazzo:2013uya} and a radiatively
corrected effective potential and find the vacuum to lose stability at
about a surprisingly low scale of about \mbo{\mu\sim 10^{9}~\gv}
[input not independent]. Keep in mind: the Higgs boson mass
miraculously turns out to have a value very close to what was expected
from vacuum stability. It looks like a tricky conspiracy with other
couplings to reach this ``purpose''. Assuming vacuum stability, the
narrow stability window actually makes the Higgs mass to be a
predictable quantity if we consider the other SM parameters as
given. Also imposing Planck-scale boundary conditions may be argued to
fix the Higgs boson
mass~\cite{Shaposhnikov:2009pv,Holthausen:2011aa}. If the Higgs boson
misses to stabilize the vacuum, why does it just miss it almost not?},
and the SM actually fails to persist up to the Planck scale, our
evaluation of the matching conditions yields initial
\MSb parameters at the $Z$ boson mass scale which evolve preserving
the positivity of $\lambda$. Thereby the critical parameter is the
top-quark Yukawa coupling, for which we find a slightly lower value,
which is based on the analysis~\cite{Jegerlehner:2012kn}. My
\MSb input at $M_Z$ is~\cite{Jegerlehner:2013cta} $g_3=1.2200$,
$g_2=0.6530$, $g_1=0.3497$, $y_t=0.9347$ and $\lambda=0.8070$. At
\mbo{\mpl} I get $g_3=0.4886$, $g_2=0.5068$, $g_1=0.4589$,
$y_t=0.3510$ and $\lambda=0.1405$ (see Table~\ref{tab:MSbinp}).  In
view of the fact that the precise meaning of the experimentally
extracted value of the top-quark mass is not free of ambiguities,
usually, it is identified with the on-shell mass $M_t$ (see
e.g.~\cite{Jegerlehner:2012kn,Moch12,Hoang:2014oea} and references
therein), it may be premature to claim that instability of the SM
Higgs potential is a proven fact already~\cite{Bednyakov:2015sca}.  As
I have elaborated in Sect.~\ref{ssec:match}, the implementation of the
matching conditions is not free of ambiguities, while the evolution of
the couplings over many orders of magnitude is rather sensitive to the
precise values of the initial couplings. Accordingly, all numbers
presented in this article depend on the specific input parameters
adopted, as specified
in~\cite{Jegerlehner:2012kn,Jegerlehner:2013cta}. In case the Higgs
self-coupling has a zero $\lambda(\mu^2)=0$, at some critical scale
$\mu_c$ below $\MPl$, we learn from Eq.~(\ref{vev}), or more directly
from $v(\mu^2)=\sqrt{6m^2(\mu^2)/\lambda(\mu^2)}\stackrel{\lambda \to
+0}{\to} \infty$ that the SM loses its ``being well-defined'' above this singular
non-analytic point\footnote{As we have argued earlier we
consider the bare Higgs potential to be the true potential, except
that the bare parameters have to be calculated bottom-up from the
known values at low energy. A low energy reparametrization also
affects the form of the potential by radiative corrections as we know
from Coleman-Weinberg~\cite{Coleman:1973jx}. The correspondingly
modified effective potential plays a crucial role when the potential
gets unstable and actually can turn instability into
meta-stability~\cite{Bezrukov:2014ipa,Degrassi:2012ry}. This will be
discussed in Sect.~\ref{ssec:effpot} below. The Planck medium, from
which the SM derives as a long distance tail, certainly exhibits a
stable ground state. This we infer from our mere existence.}.
\begin{table}
\centering
\caption{Comparison of \MSb parameters at various scales:
Running couplings for $M_H=126~\gv$ and $\mu_0\simeq
1.4\power{16}~\gv$.  Note that \mbo{\lambda=0} is an essential
singularity and the theory cannot be extended beyond a possible zero
of \mbo{\lambda}. Remind that
\mbo{v=\sqrt{6m^2/\lambda}} i.e. \mbo{v(\lambda) \to
\infty} as \mbo{\lambda \to 0}.
Besides the Higgs boson mass \mbo{m_H=\sqrt{2}\,m} all masses \mbo{m_i
\propto g_i\,v \to \infty} would yield a different cosmology
}
\label{tab:MSbinp}
\begin{tabular}{ccccc||cc}
\hline \hline
~& \multicolumn{4}{c}{my findings [Jeg]} & \multicolumn{2}{c}{Degrassi et
al. 2013 [Deg]}\\
\hline
coupling $\backslash$ scale  & $ M_Z$ & $M_t$ & $\mu_0$ & $\mpl$ & $M_t$ &
$\mpl$ \\
\hline
$ g_3$ &   $ 1.2200$ & $1.1644$ & $ 0.5271$ & $0.4886$ & 1.1644 &\hphantom{-} 0.4873 \\
$ g_2$ &   $ 0.6530$ & $0.6496$ & $ 0.5249$ & $0.5068$ & 0.6483 &\hphantom{-} 0.5057 \\
$ g_1$ &   $ 0.3497$ & $0.3509$ & $ 0.4333$ & $0.4589$ & 0.3587 &\hphantom{-} 0.4777 \\
$ y_t$ &   $ 0.9347$ & $0.9002$ & $ 0.3872$ & $0.3510$ & 0.9399 &\hphantom{-} 0.3823 \\
$ \sqrt{\lambda}$&$ 0.8983$ & $0.8586$& $ 0.3732$ & $0.3749$ & 0.8733  & $\I\:\,$0.1131 \\
$ \lambda       $&$ 0.8070$ & $0.7373$& $ 0.1393$ & $0.1405$ & 0.7626  &
- 0.0128         \\
\hline
\end{tabular}
\end{table}

For our input parameters, Table~\ref{tab:MSbinp} shows that the
relevant running \MSb parameters at the Planck scale are of comparable
size in the range 0.51 for $g_2$ being the largest here and 0.35 for
$y_t$ being the smallest, with $\sqrt{\lambda}$ at 0.375 slightly
larger in our normalization. It tells us that approximations like the
gaugeless limit ($g_1=g_2=0$) or assuming $\lambda\approx 0$ relative
to other couplings are not viable approximations near $\mpl$.

For what follows we take up what Shaposhnikov et
al.~\cite{Bezrukov:2014ipa} say about vacuum stability in their conclusion:
\textit{``Although
the present experimental data are perfectly
consistent with the absolute stability of the Standard Model
within the experimental and theoretical uncertainties,
one should not exclude the possibility that other experiments will be
able to establish the meta-stability of the electroweak vacuum in the
future.''} But, based on a slightly modified evaluation of
\MSb parameters~\cite{Jegerlehner:2012kn} (which revealed vacuum stability), we adopt
the view:
\textit{``Although other evaluations of the matching conditions seem to favor
the meta-stability of the electroweak vacuum within the experimental
and theoretical uncertainties, one should not exclude the possibility
that other experiments and improved matching conditions will be able
to establish the absolute stability of the Standard Model in the future.''}

\begin{figure}
\centering
\includegraphics[height=4cm]{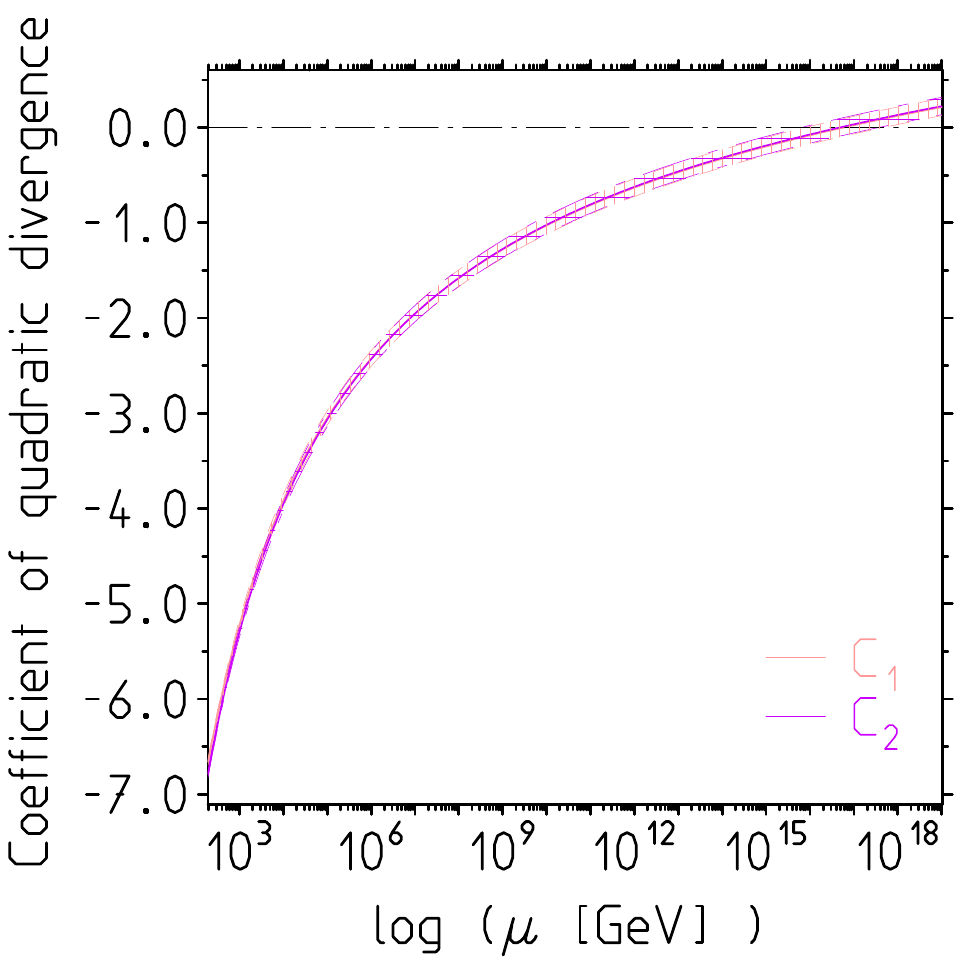}
\includegraphics[height=4cm]{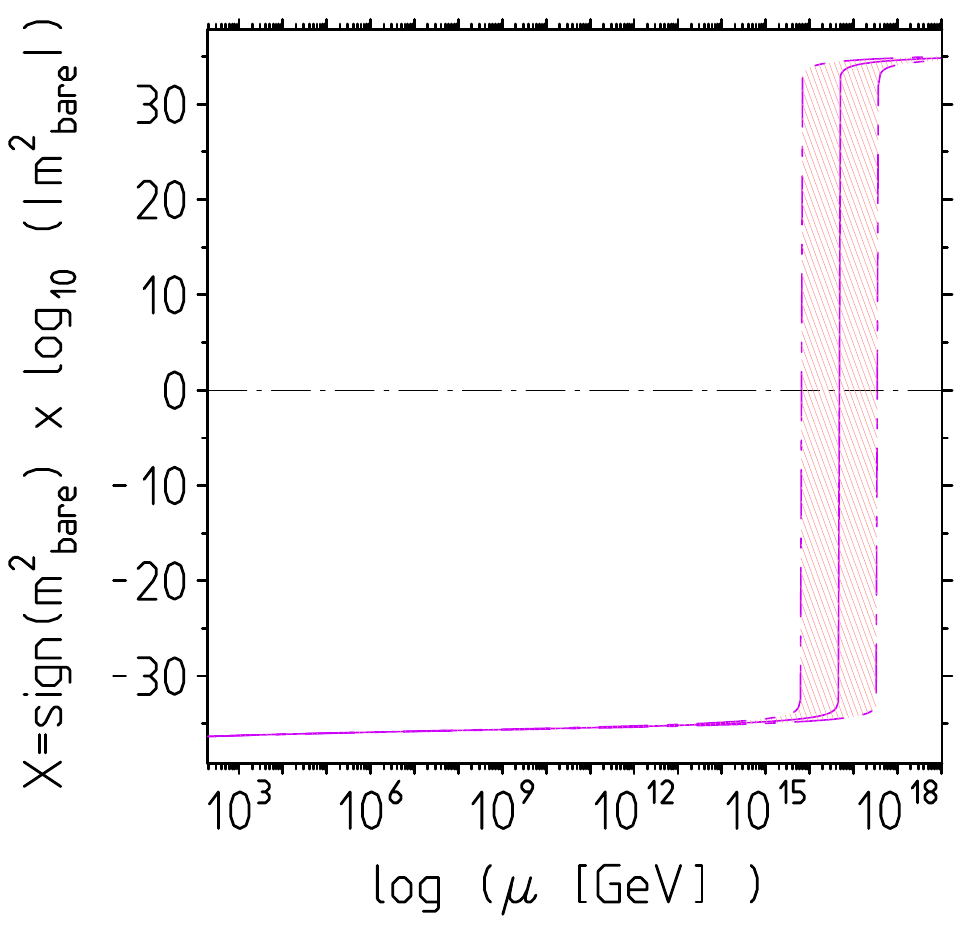}
\caption{The Higgs mechanism transition in the SM. Left: the zero in
\mbo{C_1} and \mbo{C_2} for \mbo{M_H=125.9\pm 0.4~\gv}. Right: shown is
$X=\sign(m^2_{\rm bare})\times \log_{10} (|m^2_{\rm bare}|)$, which represents
$m^2_{\rm bare}=\sign(m^2_{\rm bare})
\times 10^{X}$}
\label{fig:jump}
\end{figure}

\begin{figure}
\centering
\includegraphics[height=4cm]{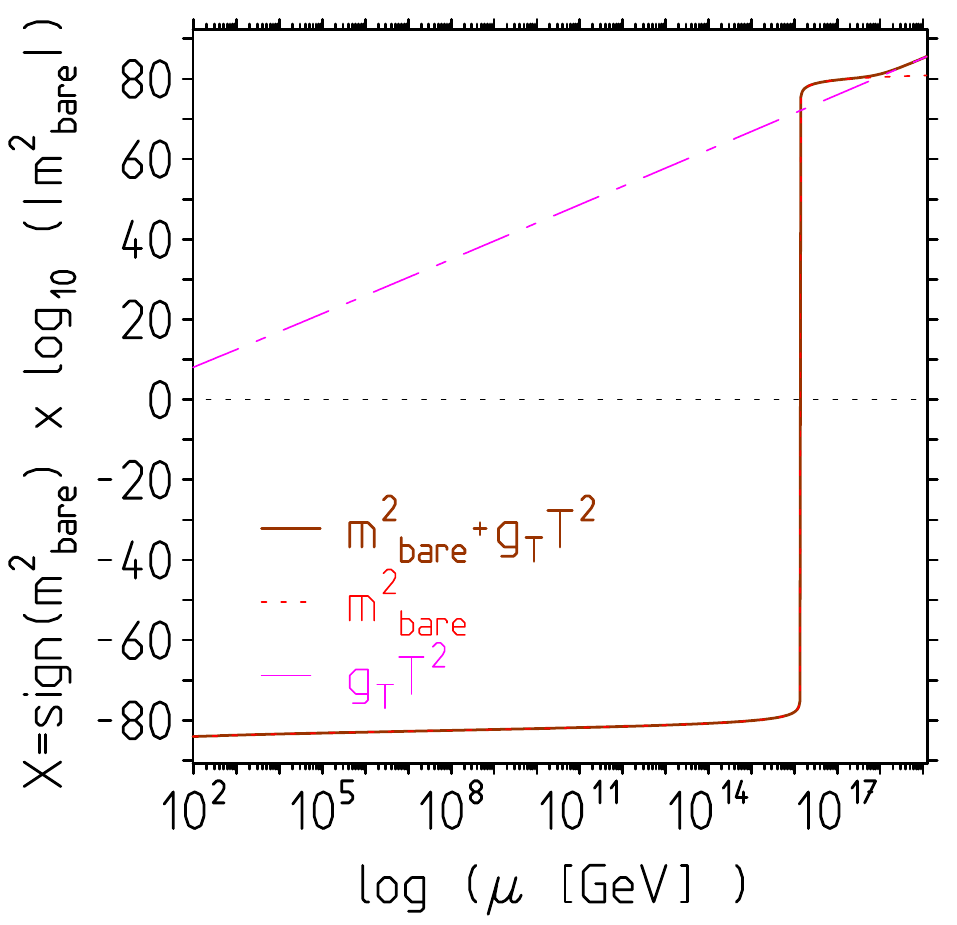}
\includegraphics[height=4cm]{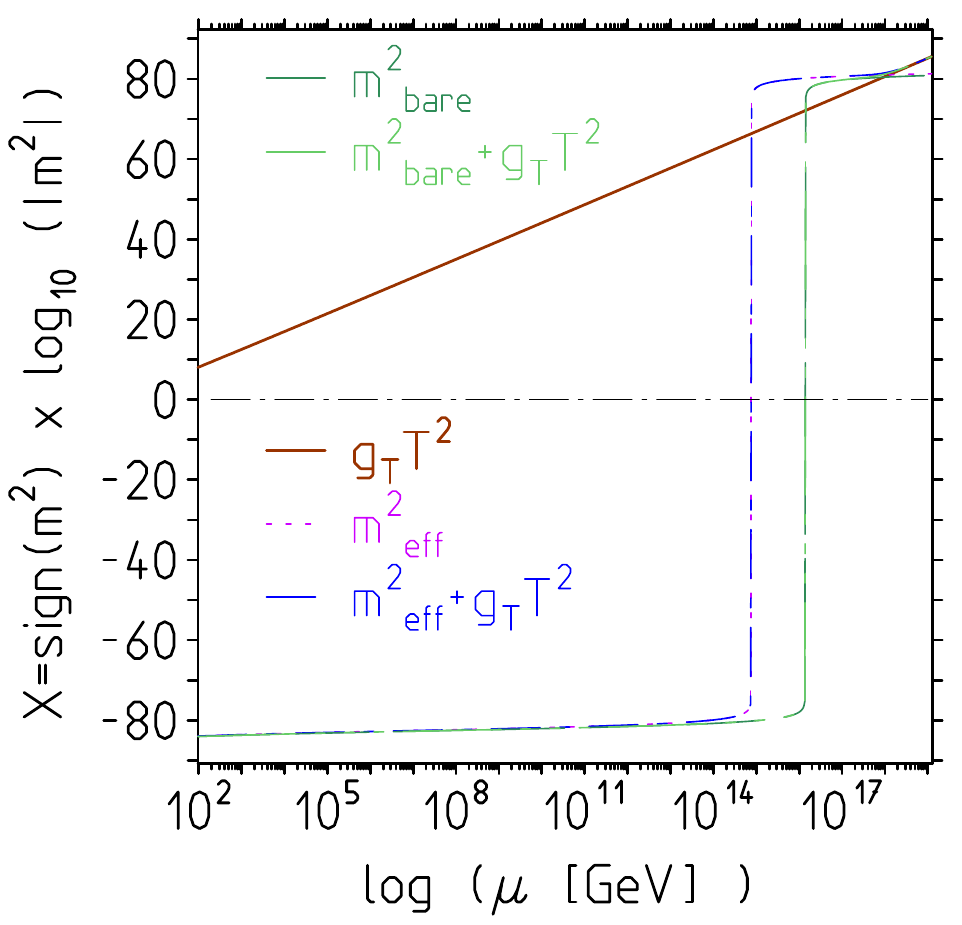}
\caption{\mbo{X} as displayed in the right panel of Fig.~\ref{fig:jump}
including leading finite temperature correction to the potential
$V(\phi,T)=\ha\,(g_T\,T^2+m_0^2)\,\phi^2+
\frac{\lambda}{24}\,\phi^4+\cdots$ with $g_T=\frac{1}{16}
\left[3g_2^2+g_1^2+4y_t^2+\frac23\,\lambda\right]$ from~\cite{Dine:1992wr} affecting the phase transition
point. Left: for the bare case [\mbo{m^2,C_1}]. Right: with adjusted
effective mass from vacuum rearrangement
[\mbo{{m'}^2,C'_1=C_1+\lambda}]. In the case
\mbo{\mu_0} sufficiently below \mbo{\mpl}, the case displayed here,
finite temperature effects affect the position of the phase transition
little, while the change of the effective mass by the vacuum
rearrangement is more efficient. The finite temperature effect with
our parameters is barely visible}
\label{fig:FT}
\end{figure}

Running couplings can affect dramatically the quadratic divergences and the
interpretation of the hierarchy problem.
Quadratic divergences have been investigated at one-loop
in~\cite{Veltman:1980mj} (see
also~\cite{Decker:1979cw,Degrassi:1992ff,Fang:1996cn}), at two loops
in~\cite{Alsarhi:1991ji,Hamada:2012bp,Jones:2013aua}.
At $n$ loops the quadratic cutoff-dependence is of the form
\be
\delta m_H^2= \frac{\Lambda^2}{16\pi^2}\,C_n(\mu)\,,
\label{quadraic1}
\ee
where the n-loop coefficient only depends on the gauge couplings $g_1$,
$g_2$, $g_3$, the Yukawa couplings $y_f$ and the Higgs self-coupling
$\lambda$. Neglecting the numerically insignificant light fermion
contributions, the one-loop coefficient function $C_1$ may be written
as
\bea
C_1=2\,\lambda+\frac32\, g_1^2+\frac92\,g_2^2-12\,y_t^2
\label{coefC1}
\eea
and is uniquely determined by dimensionless couplings. The latter are
not affected by quadratic divergences such that standard RG equations
apply. Surprisingly, as first pointed out by Hamada, Kawai and Oda
in~\cite{Hamada:2012bp}, taking into account the running of the SM
couplings, the coefficient of the quadratic divergences of the bare
Higgs mass correction can vanish at some scale, given the specific SM
couplings that became available after the Higgs boson discovery.
In contrast to our
evaluation Hamada et al. actually find the zero to lie above the
Planck scale, having adopted input \MSb parameters from~\cite{Degrassi:2012ry}.
In our analysis, relying on matching conditions for the
top-quark mass analyzed in~\cite{Jegerlehner:2012kn},
we get a scenario where $\lambda(\mu^2)$ stays positive up to the
Planck scale and looking at the relation between the bare and the
renormalized Higgs mass we find $C_1$ and hence the Higgs mass
counterterm to vanish at about $\mu_0\sim 1.4 \times 10^{16}~\gv$, not
very far \textbf{below} the Planck scale. The next-order correction,
first calculated in~\cite{Alsarhi:1991ji,Jones:2013aua} and confirmed
in~\cite{Hamada:2012bp} reads
\bea
C_2&=&C_1+ \frac{\ln (2^6/3^3)}{16\pi^2}\, [
18\,y_t^4+y_t^2\,(-\frac{7}{6}\,g_1^2+\frac{9}{2}\,g_2^2
             -32\,g_3^2) \nn \\
             &&\!\!\!\!-\frac{87}{8}\,g_1^4-\frac{63}{8}\,g_2^4 -\frac{15}{4}\,g_2^2g_1^2
             +\lambda\,(-6\,y_t^2+g_1^2+3\,g_2^2)-\frac{2}{3}\,\lambda^2]\,,
\label{coefC2}
\eea
and numerically does not change the one-loop result significantly.  The
same results apply for the Higgs potential parameter $m^2$, which
corresponds to $m^2\hat{=}\frac12\,m_H^2$ in the broken phase. For
scales $\mu <
\mu_0$ we have $\delta m^2$ large negative, which is triggering
spontaneous symmetry breaking by a negative bare mass
$m_0^2=m^2+\delta m^2$, where $m$ again denotes the renormalized mass.
The phase transition is illustrated in Fig.~\ref{fig:jump}.
The jump taking place here in the vacuum energy is given
by\footnote{Note that this is a finite prediction independent of
quadratic cutoff effects. The transition point $\mu_0$ is a matching
point where bare and renormalized quantities at scale $\mu_0$ agree, i.e.
$\lambda=\lambda(\mu_0)$ and $v=v(\mu_0)$.}
\bea
{\Delta V(\phi_0)=-\frac{m^2_\mathrm{eff}v^2}{8}=-\frac{\lambda\,v^4}{24}\sim
-9.6\power{8}~\gv^4}\approx -(176.0~\gv)^4\epo
\label{deltarhojump}
\eea
As a CC contribution it is of negative sign and 50 orders of magnitude
off relative to what corresponds to the observed
\mbo{\Lambda_{\rm CMB}} (see also ~\cite{Dreitlein74}). However, the effect is small relative to the
$O(\mpl^4)$ size $V(0)=\langle V(\phi)\rangle$, which will be
discussed in Sect.~\ref{sec:CC}.  At $\mu=\mu_0$ we have $\delta
m^2=0$ and the sign of $\delta m^2$ flips, implying a phase transition
to the symmetric phase. Finite temperature
effects~\cite{Kirzhnits:1972iw,Dolan:1973qd,Weinberg:1974hy,Kirzhnits:1976ts},
which must be included in a realistic scenario, turn out not to change
the gross features of our scenario, unless $\mu_0$ would turn out to
lie much closer to $\lpl$~\cite{Jegerlehner:2013cta}. A different
effect is due to the change in the effective mass resulting from the
Wick reordering of the Lagrangian by a non-vanishing
\mbo{\braket{\Phi^+\Phi}}. This will be discussed in Sect.~\ref{sec:CC}.
It produces a larger shift of the transition point as one may learn from
Fig.~\ref{fig:FT}, where the finite temperature effects are displayed.
What do we learn from this analysis? The Higgs mechanism is
dynamically triggered as the temperature in the universe drops below
$\mu_0$.  In the low energy phase, the Higgs boson mass $M_H$
substitutes $\sqrt{2} m$ and in fact has to be calculated from the
vacuum rearrangement (see Fig.~\ref{fig:rearrange}). Now $m_H$ turns
into an emergent mass, which is determined by the mass-coupling relation
(\ref{masscoupl}) like for all other massive particles in the Higgs
phase. At the transition point, we have $\delta m^2 =0$ and no
hierarchy problem. While above $\mu_0$ the shift $\delta m^2$ is
physical and emergent from the interaction in the Planck medium, below
$\mu_0$, the shift $\delta m^2$ looses its physical meaning. This is because at
$\mu=\mu_0$ the enhanced cutoff-effects are nullified. At this point,
the access to cutoff effects gets lost and we enter the renormalizable
renormalized low energy phase. Below $\mu_0$, we still use $\delta m_H^2$ in perturbative mass
renormalization, where it is now large negative if we still insist on
using the now physics-wise inaccessible Planck scale as a UV cutoff.
I would say that argumentations based on (\ref{massren}) now turn into
formal nonsense. Not only the magnitude of the cancellation is
arbitrary, but it also has the wrong sign, for what could be related to a
physical mass. The physical mass is determined by the curvature at the
minimum of the potential. The key outcome of our calculation is the
observation that the SM at high enough sub-Planckian energies
undergoes a transition into the symmetric
phase~\cite{Jegerlehner:2013cta}, presuming a stable vacuum.
Fig.~\ref{fig:HPMup} displays the SM prediction for the effective
Higgs mass as a function of the energy scale. This profile promotes
the Higgs boson to act as an inflaton as discussed in
Sect.~\ref{sec:prelude} already.

\subsection{Vacuum stability and effective potential}
\label{ssec:effpot}
The classical Higgs potential (\ref{potential}) for $\lambda > 0$ is
bounded from below and has a trivial minimum for $m^2 > 0$ at
$\phi_0=0$, and a non-trivial minimum at $\phi_0^2 = \frac{-6m^2}{\lambda}$
for $m^2 < 0$. When the classic potential turns unstable, because
$\lambda$ is running to be negative, the analysis of the vacuum
stability has to be based on the {\it effective potential}, which is
obtained by including the quantum
corrections~\cite{Coleman:1973jx,Weinberg:1973ua}~\footnote{Being a
part of the SM Lagrangian the Higgs potential term considered so far
gets reparametrized by a change of the effective parameters and the
effective Higgs field and by appropriate counterterms only, as long as
perturbation theory does not break down. All perturbative physics is
obtained as usual by means of the renormalizable Lagrangian, written
in terms of the quantized fields, and the corresponding Feynman
rules. Also note that the Higgs contribution to the energy-momentum
tensor of Einstein gravity is represented by the symmetric
energy-momentum tensor $$\Theta^\mu_{\:\:\nu}=\frac{\partial
\cL}{\partial (\partial_\mu
\phi)}\,\partial_\nu\phi-\delta^\mu_{\:\:\nu}\cL\,, \mathrm{ \
where \ \ }
\cL(\phi)=\frac12\,\gmunuc\,\pamu \phi\, \panu \phi-V(\phi)\,,
$$ in terms of the Higgs part of the bare SM
Lagrangian.}.  The effective potential is gauge- and scale-dependent
and not an observable. In the Landau gauge and the
\MSb~scheme it can be written
as~\cite{Ford:1991hw,Ford:1992pn,Ford:1992mv,Kastening:1991gv} (also
see~\cite{Casas:1994qy})~\footnote{As shown in~\cite{Coleman:1973jx},
the potential satisfies the RG equation
$$\left(\mu\,\frac{\partial}{\partial \mu}+\sum_i \beta_i
\frac{\partial}{\partial \lambda_i}
+\gamma\,\phi\,\frac{\partial}{\partial \phi}\right)\,V=0 $$ where
$\lambda_i=m^2,\lambda,g'=g_1,g=g_2,g_s=g_3,y_t$ with corresponding
beta-functions $\beta_i$ and $\gamma$ the anomalous dimension of the
Higgs field. The RG as usual is solved along characteristic curves
where $t$ parametrizes the position on the curve. The solution reads
$$V(\mu,\lambda_i,\phi)=Z^4_\phi(t)\,V(\mu(t),\lambda_i(t),\phi)\,,$$
with $Z_\phi(0)=1,\,\lambda_i=\lambda_i(0)$ and $\phi=\phi(0)\epo$}
\begin{eqnarray}
 V_{\rm eff}(\phi(t)) =
           \frac{1}{2} m^2(t)\,\phi^2(t)
          + \frac{1}{24} \lambda(t)\, \phi^4(t)
          + V_1 + V_2 + V_3 + V_{\rm rem}
\;,
\end{eqnarray}
with
\begin{eqnarray}
V_1&=&\kappa\,\left\{ \frac32\,
m^4_W(t)\left[\ln {\frac{m^2_W(t)}{\mu^2(t)}}-{\frac{5}{6}}\right]+
\frac34\,m_Z^4(t)\left[\ln {\frac{m^2_Z(t)}{\mu^2(t)}}-{\frac{5}{6}}\right]
-3m_t^4(t)\left[\ln {\frac{m^2_t(t)}{\mu^2(t)}} -{\frac{3}{2}}
\right]
\right. \nonumber \\ && \hspace*{2mm}\left.
+\frac14\,m^4_H(t)\left[\ln {\frac{m^2_H(t)}{\mu^2(t)}}-{\frac{3}{2}}\right]
+\frac34\,m^4_G(t)\left[\ln {\frac{m^2_G(t)}{\mu^2(t)}}-{\frac{3}{2}}\right]
\right\}\,,
\end{eqnarray}
where $\kappa=1/(4\pi)^2$ and $m_i$ are the masses of different particles in the
background of the classical Higgs source field $\phi_c$ of the
generating functional for the irreducible Higgs vertex functions,
which upon renormalization is given by
$\phi(t)=Z_\phi(t)\,\phi_c$.
Thus we have
\bea
\label{mass}
m_W^2(t)&=&{\frac{1}{4}}g_2^2(t)\phi^2(t)\comas
m_Z^2(t)={\frac{1}{4}}[g_2^2(t)+g_1^2(t)]\phi^2(t)\comas
m_t^2(t)={\frac{1}{2}}y_t^2(t)\phi^2(t)\comas
\nonumber \\
m_H^2(t)&=&m^2+\frac12\,\lambda \phi^2(t)\comas
m_G^2(t)=m^2+\frac16\,\lambda \phi^2(t)\epo
\eea
The effective potential as derived in the symmetric phase include the
would-be Higgs ghosts $G$ contribution as physical degrees of freedom,
in the broken phase Higgs ghosts are massless in the Landau gauge
(would-be Nambu-Goldstone bosons). In the symmetric phase they
contribute as three additional Higgs particles. As we know the Higgs
boson mass in the broken phase ($m^2<0$) is $M_H^2=-2m^2=\frac13
\lambda v^2$, where $v$ refers to the EW vacuum. Two-loop corrections $V_2$ have been calculated
in~\cite{Ford:1992mv,Ford:1992pn} and may be found in more condensed form
in~\cite{Degrassi:2012ry}. $V_3$ includes the leading three-loop
corrections computed in~\cite{Martin:2013gka}.  The remainder $V_{\rm rem}$
represents the higher-order contributions, which include also the
higher dimension operators starting at four
loops~\cite{Ford:1992mv,Nakano:1993jq,Burgess:2001tj}:
\begin{eqnarray}
V_{\rm rem} \sim \lambda \phi^4 \sum_{L>4} \left( \frac{\lambda^2}{M_{Pl}^2} \phi^2 \right)^{L-3} \;,
\label{rem}
\end{eqnarray}
where $L$ is number of loops.

The wavefunction renormalization of
the Higgs field takes the form
\bea
\phi(t) & = & Z_\phi(t)\, \phi_c =\exp \left\{\int_0^t \gamma(\tau) \D \tau \right\}
\phi(0) \comas \phi(0)=\phi_c\,,
\eea
where $\gamma(t)=\D \ln Z_\phi/\D t$ is the anomalous
dimension of the Higgs field:
\bea
\gamma&=&\kappa\,\left[\frac94\,g_2^2+\frac{3}{4}\,g_1^2-3\,y_t^2\right]
+\kappa^2 \left[y_t^2\, \left(\frac{27}{4}\,y_t^2-20\,g_3^2-\frac{45}{8}\,g_2^2-\frac{85}{24}\,g_1^2\right)
\right. \nonumber \\ &&  \left.
+\frac{271}{32}\,g_2^4
-\frac{9}{16}\,g_2^2g_1^2-\frac{431}{96}\,g_1^4-\frac16\,\lambda^2\right]
+ \cdots
\eea
Finally, the scale $\mu(t)$ is related to the
running parameter $t$ by
\be
\label{mu}
\mu(t)=\mu \E^t,\;\mathrm{ \ i.e. \ }\; t=\ln \mu(t)/\mu\,,
\ee
where $\mu$ is a fixed scale, that we will take equal to the physical
top-quark mass, $M_t$ as a reference point. Observable physical predictions
up to perturbative truncation errors do not depend on the choice of
the renormalization scale. This can be used in order to keep radiative
corrections moderate by choosing $\mu(t)=\phi(t)$ which avoids large
logarithms at any given $t$, since then $\ln\,{m^2_W(t)}/{\mu^2(t)}=
\ln\,g^2(t)/4$ etc. (see~\cite{Ford:1992mv}). One also may
choose $\mu(t)=\phi_c$ in which case $\ln\,{m^2_W(t)}/{\mu^2(t)}=
\ln\,g^2(t)/4+2\Gamma$ etc. The correction
$\Gamma=\int_{M_t}^\phi\,\gamma(\mu)\,\D \ln(\mu)$ stems from the
field renormalization factor $Z_\phi$.

\begin{figure}
\centering
\includegraphics[height=4cm]{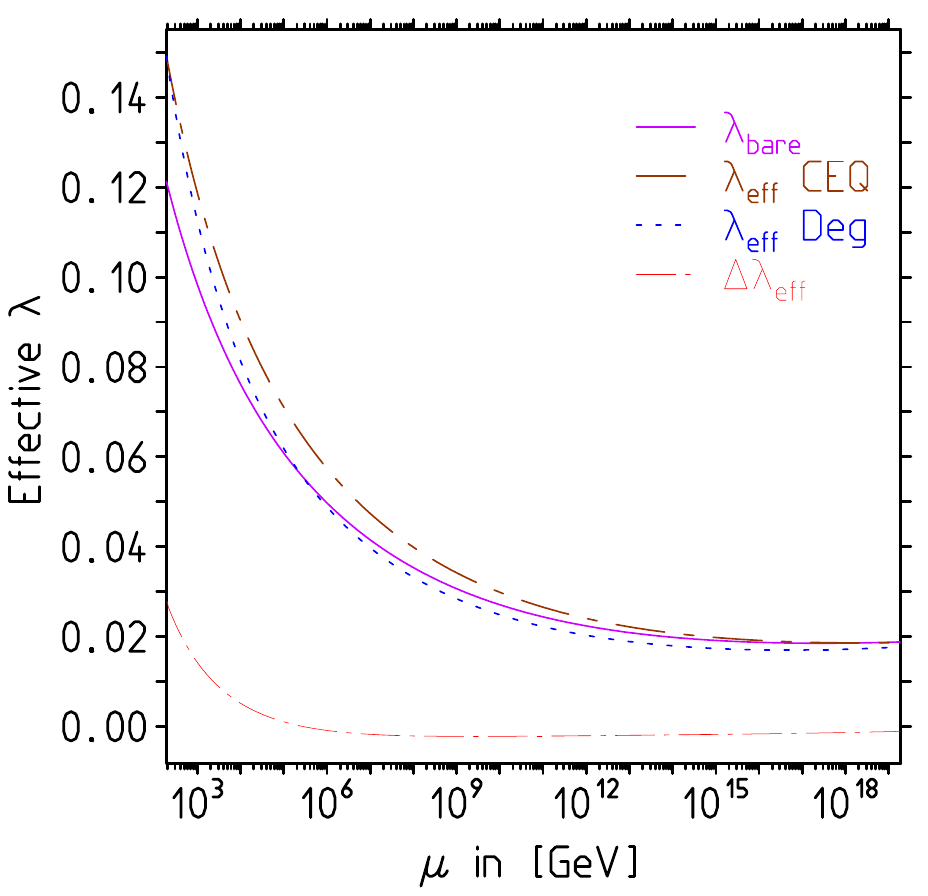}
\includegraphics[height=4cm]{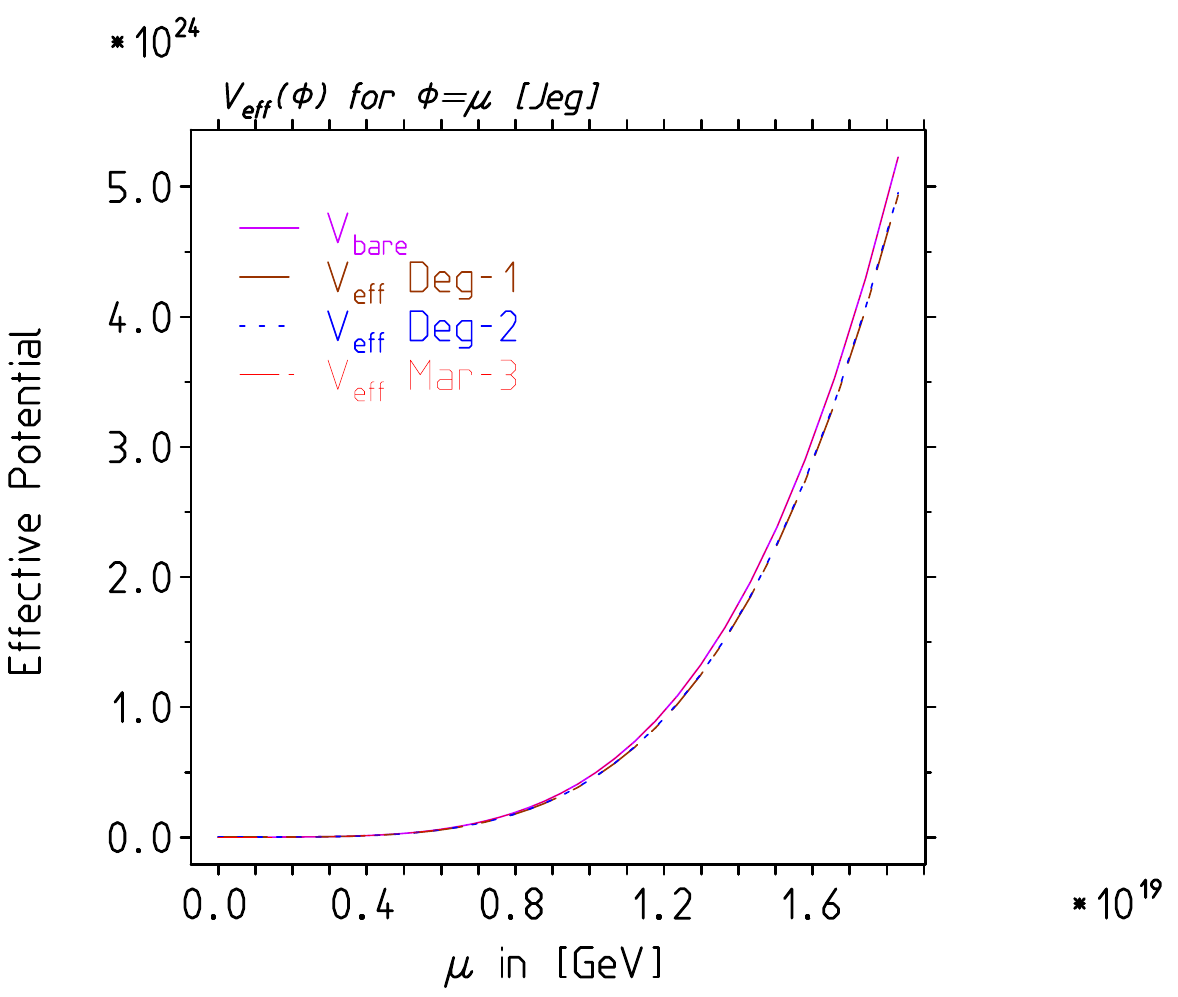}
\caption{The bare versus the effective Higgs coupling and the effective
potential for the parameter set [Jeg] of Table~\ref{tab:MSbinp}. Left:
the effective Higgs self-coupling $\lambda_{\rm eff}$ governing the SM
effective potential $V_{\rm eff}\sim \frac{\lambda_{\rm
eff}}{24}\,\phi^4$ for large fields. ``CEQ'' is one-loop improved
from~\cite{Casas:1994qy}, ``Deg'' is two-loop improved
from~\cite{Degrassi:2012ry}. The correction $\Delta \lambda_{\rm eff}$
represents the corrections included in ``Deg'' relative to
$\lambda_{\rm bare}$. Right: the bare potential compared with
different approximations of the effective potential: one-loop improved
``Deg-1'', two-loop improved ``Deg-2'' and three-loop
improved ``Mar-3''~\cite{Martin:2013gka}.}
\label{fig:VeffJeg}
\end{figure}

As elaborated in~\cite{Casas:1994qy} for high Higgs fields the
effective potential may be cast into the simple form where it is
dominated by the quartic term
\bea
V_{\rm eff}\approx  \frac{\lambda_{\rm eff}(\phi)}{24}\, \E^{4\Gamma(\phi)}\,\phi_c^4
\eea
and $\lambda(\phi)$ depends on $\phi$ the same as the running coupling
$\lambda(\mu)$ depends on the running scale $\mu=\phi_c$ with modified
coupling~\cite{Buttazzo:2013uya}
\bea
\lambda_{\rm eff}\approx\lambda+&\kappa\,&\left[
             \frac{9}{4}\,g_2^4\,(\ln\frac{g_2^2}{4}-\frac56+2\Gamma)
             +\frac{9}{8}\,(g_2^2+g_1^2)^2
             \,(\ln\frac{g_2^2+g_1^2}{4}-\frac56+2\Gamma) \right. \crn && \left.
        \hspace*{-3mm}      -18\,y_t^4\,(\ln\frac{y_t^2}{2}-\frac32+2\Gamma)
          +\frac32\,\lambda^2\,(\ln \frac{|\lambda|}{2}-\frac32+2\Gamma)
          +\frac12\,\lambda^2\,(\ln(\frac{|\lambda|}{6}-\frac32+2\Gamma)\right]  \crn
          +&\kappa^2&\,6\,y_t^4\,\left[8\,g_3^2\,(3\,r_t^2-8\,r_t+9)
          -\frac32\,y_t^2\,(3\,r_t^2-16\,r_t+23+\frac{\pi^2}{3})\right] + \cdots \,,
\eea
up to less relevant corrections. The crucial point is that for
parameters as [Deg] in Table~\ref{tab:MSbinp} the correction term is
positive all up to the Planck scale. At the EW scale the leading
positive $\lambda$-term dominates $\lambda_{\rm eff}$ up to scales
where $\lambda$ approaches a zero and there changes the sign, if such a
zero exists, which depend on the precise input values at the EW
scale. In the vicinity above the zero of $\lambda$ actually
$\lambda_{\rm eff}$ remains positive and such stabilizes the Higgs
vacuum to somewhat higher scales but also turns negative to a
metastable state before reaching the Planck scale (see Fig.~3
in~\cite{Degrassi:2012ry}).  In contrast for the parameter set [Jeg]
the correction $\Delta \lambda_{\rm eff}$ also starts positive but at
higher scales takes negative values. These are small, however, and are not
affecting the positivity of $\lambda_{\rm eff}$ itself as seen in the
left panel of Fig.~\ref{fig:VeffJeg}. In the stable vacuum scenario
radiative corrections of the effective potential are moderate and do
not affect the main pattern as long as $\lambda$ remains positive.
The cutoff power enhanced effects are always much larger than the
standard radiative corrections to the effective potential, provided
the formers are taken into account.  This we will have to remind also
for the discussion to follow in Sect.~\ref{sec:CC}.
\begin{figure}[h]
\centerline{\includegraphics[width=0.4\textwidth]{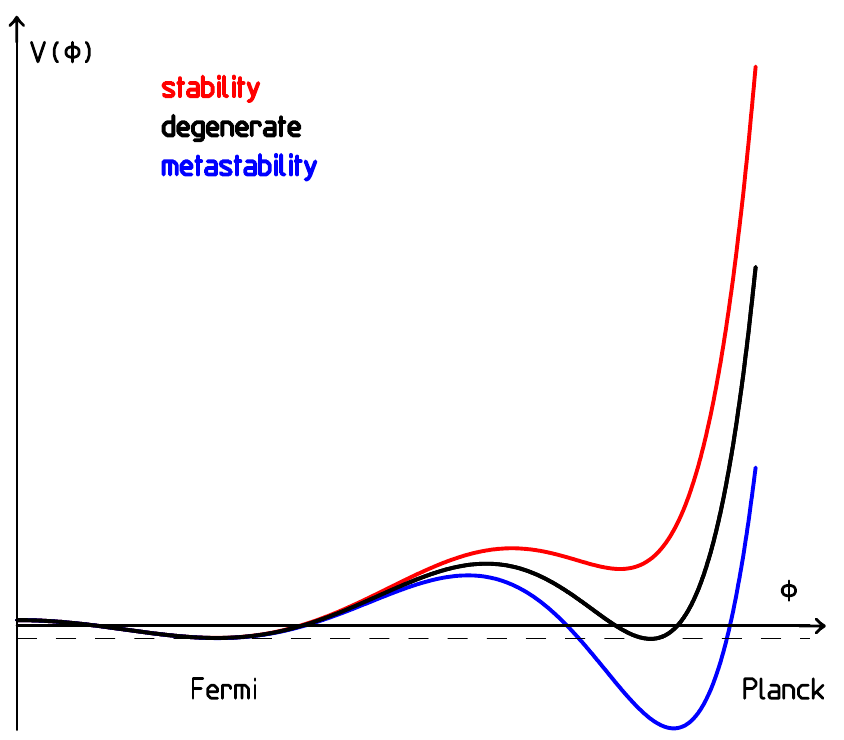}}
\caption{
The form of the effective potential for the Higgs field $\phi$ which
corresponds to the stable, critical and metastable electroweak
vacuum. The pattern displayed, admitting for two minima at non-zero
field values requires the effective potential to exhibit even powers
of $\phi$ up to $\phi^8$. $v$ is the location of the EW minimum and
$\phi_{\rm min}\gg v$ is the value of a new minimum.}
\label{stab}
\end{figure}
\begin{figure}
\centering
\includegraphics[height=4cm]{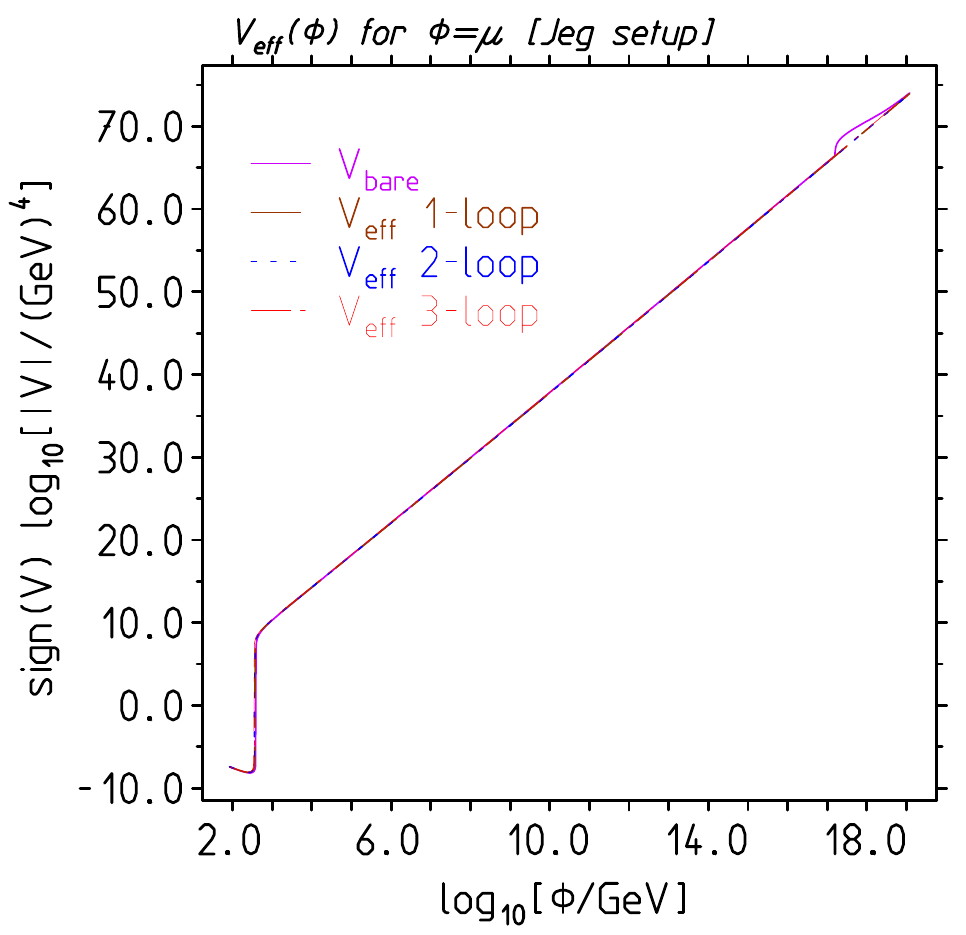}
\includegraphics[height=4cm]{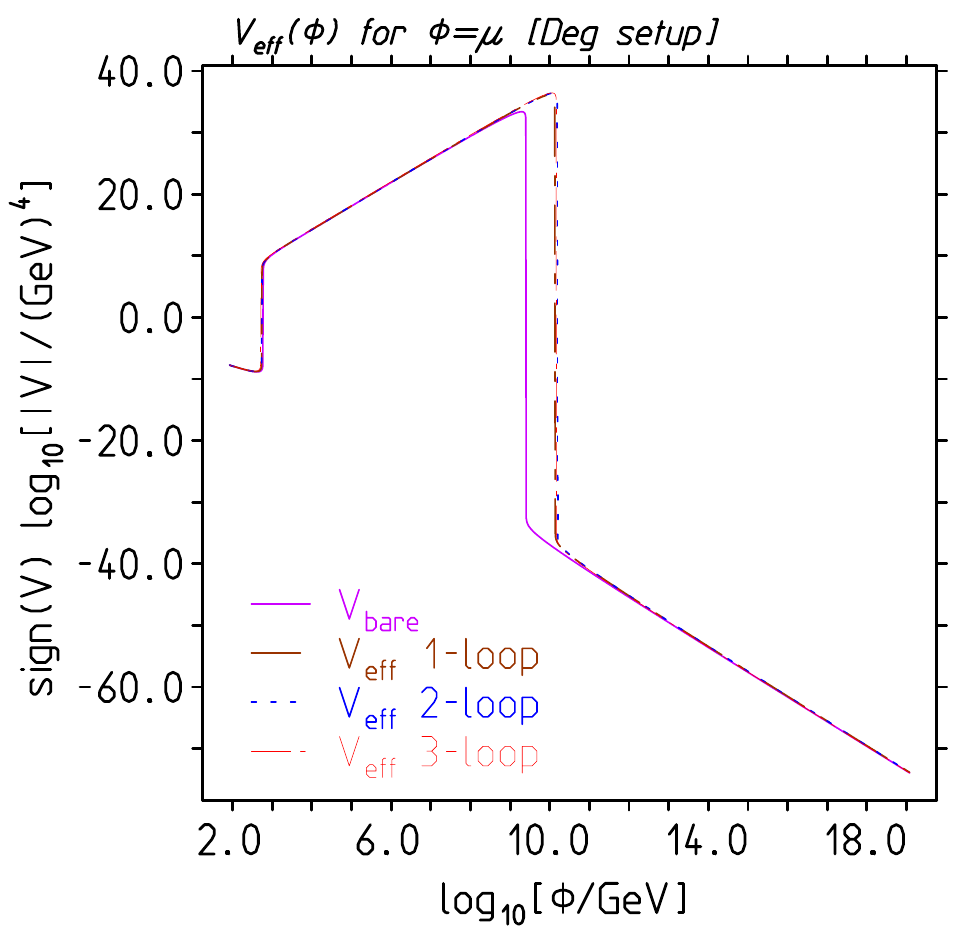}
\caption{The effective potential including 1-,2-, and
leading 3-loop~\cite{Degrassi:2012ry,Martin:2013gka} corrections, with
$\mu=\phi$ as a scale. Left: for parameter set [Jeg] (stable
vacuum). Right: for parameter set [Deg] (metastable case); the EW
vacuum is tunneling into the bottomless potential. The tunneling time
by far exceeds the age of the universe and hence looks very stable for
us.}
\label{fig:VeffLogcomp}
\end{figure}
The quantum corrections modify the shape of the potential such that a
second minimum at some higher (Planck) scale may be induced (see
Fig.~\ref{stab} and Fig.~7 in~\cite{Degrassi:2012ry}). As first
discussed in~\cite{Ford:1992mv}, a second minimum is also obtained
when a transient instability emerges above our Higgs transition point
$\mu_0$ when the bare mass term gets positive and actually gets huge
because of the quadratic cutoff-enhancement. For especially fine-tuned
parameters this may happen also if radiative corrections are not yet
included. In any case, the existence of a second minimum depends
significantly on the higher order corrections. Depending on the values
of the Higgs boson and top-quark masses the lifetime of the EW vacuum
can be larger or smaller than the age of the Universe.  The first case
corresponds to the metastable scenario.

For stationary points $\phi_0$ much larger than the electroweak scale
one has $\lambda_{\rm eff} \ll 1$ and the curvature of the potential is given
by~\cite{Espinosa:1995se}
\bea
\left.\frac{\partial^2 V_{\rm eff}}{\partial \phi^2(t)}\right|_{\phi=\phi_0}=
\frac12\,\left(\beta_\lambda-4\gamma \lambda\right)\,\phi_0^2\approx \frac12\,\beta_\lambda\,\phi_0^2\epo
\eea
Therefore, in order that the potential exhibits a second minimum, the
function $\beta_\lambda$ must have passed a zero because we know that
$\beta_\lambda$ is negative at EW scales. For the parameter set [Jeg]
a zero is found at about $\mu_\lambda\simeq 10^{17}~\gv$. What happens
for the two parameter sets [Jeg] and [Deg] is shown in
Fig.~\ref{fig:VeffLogcomp}, which also illustrates the significance of
the radiative corrections. In the stability case the effective
potential does not alter the main picture, while in the metastable
case a second minimum is also missing and the potential turns
unbounded from below way below the Planck regime. Since the tunneling rate to
the Planck regime is exceedingly low, the EW vacuum still looks to be
stable. As follows from the SM RG, because $\beta_\lambda$ contains parts
which are not proportional to $\lambda$, the Higgs self-coupling $\lambda$
is the only SM dimensionless coupling that can change the sign with increasing
energy scale.

In our case, where $\lambda (\mu) >0$ up to $\mpl$, in the early phase
of the expanding universe, the effective potential is approximated by
\bea
V(\phi)\approx V(0)+ \frac{m^2_{\rm
eff}}{2}\,\phi^2+\frac{\lambda_{\rm eff}}{24} \,\phi^4\,,
\eea
and the correction turns out not to be significant for what concerns
the scenario as such. Actually, the upshot of the two-loop analysis
in~\cite{Ford:1992mv} has been that \textit{``the requirement that the
electroweak vacuum remains stable turns out to be essentially
identical to the requirement that $\lambda$ remains positive''}.

One has to keep in mind that a metastable EW ground-state in a
globally unstable potential, as found in commonly accepted
analyses~\cite{Degrassi:2012ry,Buttazzo:2013uya,Bezrukov:2014ipa},
very likely does not model what truly happens at the Planck scale. It
could signal the need for an extension of the SM including new physics
or that the analysis underestimates uncertainties.

\section{The cosmological constant  -- dark energy provided by the
Higgs scalar}
\label{sec:CC}
It is crucial that in the early universe both terms in
the Higgs potential are positive in order to condition slow-roll
inflation during long enough time.  In fact the quadratically and
quartically cutoff enhanced terms in the Higgs potential enforce the
condition $\frac12\,\dot{\phi}^2 \ll V(\phi)$ and given the Higgs
boson pressure $p_\phi=\frac12\,\dot{\phi}^2- V(\phi)$ and the Higgs
energy density $\rho_\phi=\frac12\,\dot{\phi}^2
+ V(\phi)$, we arrive at the equation of state $w=p/\rho\approx -1$
characteristic for dark energy and the equivalent CC (see
e.g.~\cite{Straumann:1999ia,Volovik:2005zu,Sola:2013gha,Bass:2014lja}
and references therein).  A first remarkably precise measurement of
the dark energy equation of state $w=-1.01\pm 0.04$ has been obtained
by the Planck mission~\cite{Ade:2013zuv,Adam:2015rua} recently (for an
actual review see~\cite{RPP18}). A more
detailed study~\cite{Jegerlehner:2014mua} shows that the enhanced
Higgs boson effective mass alone actually does not provide a sufficient
amount of inflation, which is required to inflate the causal CMB cone to
include the full CMB sky\footnote{This is the Horizon problem: the
finite age \mbo{t} of the universe together with the finite speed of
light \mbo{c} allows us to see to distances \mbo{D_{\rm hor}=c\,t} at
most. The CMB sky is much larger [\mbo{d_{\tCMB}\simeq 4\cdot
10^{7}\ly}] than the causally connected patch [\mbo{D_\mathrm{CMB}\simeq 4
\cdot 10^5\ly}] at the time of last scattering \mbo{t_{\rm CMB}\simeq 380
000~{\rm yrs}} when the CMB decoupled from matter. As we know, no \mbo{D_\mathrm{CMB}} spot shadow
is distinguishable at the full CMB sky.}.

One important quantity we have not taken into account so far is the vacuum
energy $V(0)=\langle V(\phi)\rangle$. A key point is that in the LEESM scenario the vacuum energy is a
calculable quantity. In the symmetric phase $\SU(2)$ symmetry implies
that while \mbo{\braket{\Phi(x)}\equiv 0} the composite field
\mbo{\Phi^+\Phi(x)} is a singlet such that the invariant vacuum energy is
given just by simple Higgs field loops\\
\centerline{\includegraphics[height=9mm]{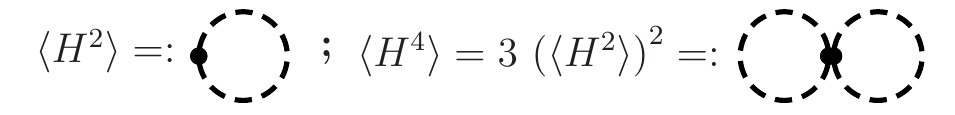}}
where
\bea
\bra{0}\Phi^+\Phi\ket{0}=\frac12\bra{0}H^2\ket{0}\equiv
\frac12\,\Xi\semis\Xi=\frac{\lpl^2}{16\pi^2}\,.
\label{vacenergy}
\eea
This provides a CC given by
\bea
V(0)=\braket{V(\phi)}=\frac{m^2}{2}\,\Xi+\frac{\lambda}{8}\,\Xi^2\epo
\eea
A Wick ordering type of rearrangement of the Lagrangian also leads to a
shift oft the effective mass
\bea
{m'}^{2}=m^2+\frac{\lambda}{2}\,\Xi\epo
\eea
For our values of the \MSb input parameters the zero in the Higgs mass
counter term and hence the phase transition point gets shifted downwards as follows
\bea
{\mu_0 \approx 1.4 \power{16}~\gv}
\to {\mu'_0\approx 7.7 \power {14}~\gv\,.}
\eea
The shift is shown in the right panel of Fig.~\ref{fig:FT}. We notice
that the SM predicts a huge CC at \mbo{\mpl}:
\bea
\rho_\phi\simeq V(\phi)
\sim 2.77\,\mpl^4\sim 6.13\power{76}~\gv^4
\eea
exhibiting a very weak
scale dependence (running couplings) and we are confronted with the
question how to get rid of this huge quasi-constant? Remember that
$\rho_\phi$ has no direct dependence on $a(t)$. An intriguing
structure again solves the puzzle.  The effective CC counterterm has a
zero, which again is a point where renormalized and bare quantities
are in agreement:
\bea
\rho_{\Lambda 0}=\rho_{\Lambda} +\delta \rho_\Lambda \semis \delta \rho_\Lambda=\frac{\lpl^4}{(16\pi^2)^2}\,X(\mu)
\label{rholambdaCT}
\eea
with $X(\mu)\simeq \frac18\,(2C(\mu)+\lambda(\mu))$ which has a zero
close to the zero of $C(\mu)$ when $2\,C(\mu)=-\lambda(\mu)$. Note
that $C(\mu)=-\lambda(\mu)$ is the shifted Higgs transition point.

Again we find a matching point $\rho_{\Lambda 0}=\rho_{\Lambda}$
between the low-energy and the high-energy world. At this point, the memory of the
quartic Planck scale enhancement gets lost, as it should be since we
know that the low energy phase does not provide access to cutoff-effects.

Crucial point is that
\bea
{X(\mu)=2C+\lambda= 5\,\lambda+3\,g_1^2+9\,g_2^2-24\,y_t^2 }
\label{Xofmu}
\eea
acquires positive bosonic contribution and negative fermionic ones,
with different scale-dependence\footnote{Unbroken SUSY would require a
perfect cancellation to happen at all
scales. Broken SUSY would largely diminish the quadratic and
quartic enhancements which are key effects in our scenery.
}. \mbo{X} can change a lot (pass a
zero), while individual couplings are weakly scale-dependent with
\mbo{y_t(M_Z)/y_t(\mpl) \sim 2.7} the biggest and
\mbo{g_1(M_Z)/g_1(\mpl)
\sim 0.76} the smallest change. Obviously, the energy dependence of any
of the individual couplings would by far not be able to sufficiently
diminish the originally huge cosmological constant. Only the existence
of a zero in the coefficient function $X(\mu)$ is able to provide the
dramatic reduction of the effective CC, by nullifying the huge
cutoff-sensitive prefactor.

\begin{figure}
\centering
\includegraphics[height=3cm]{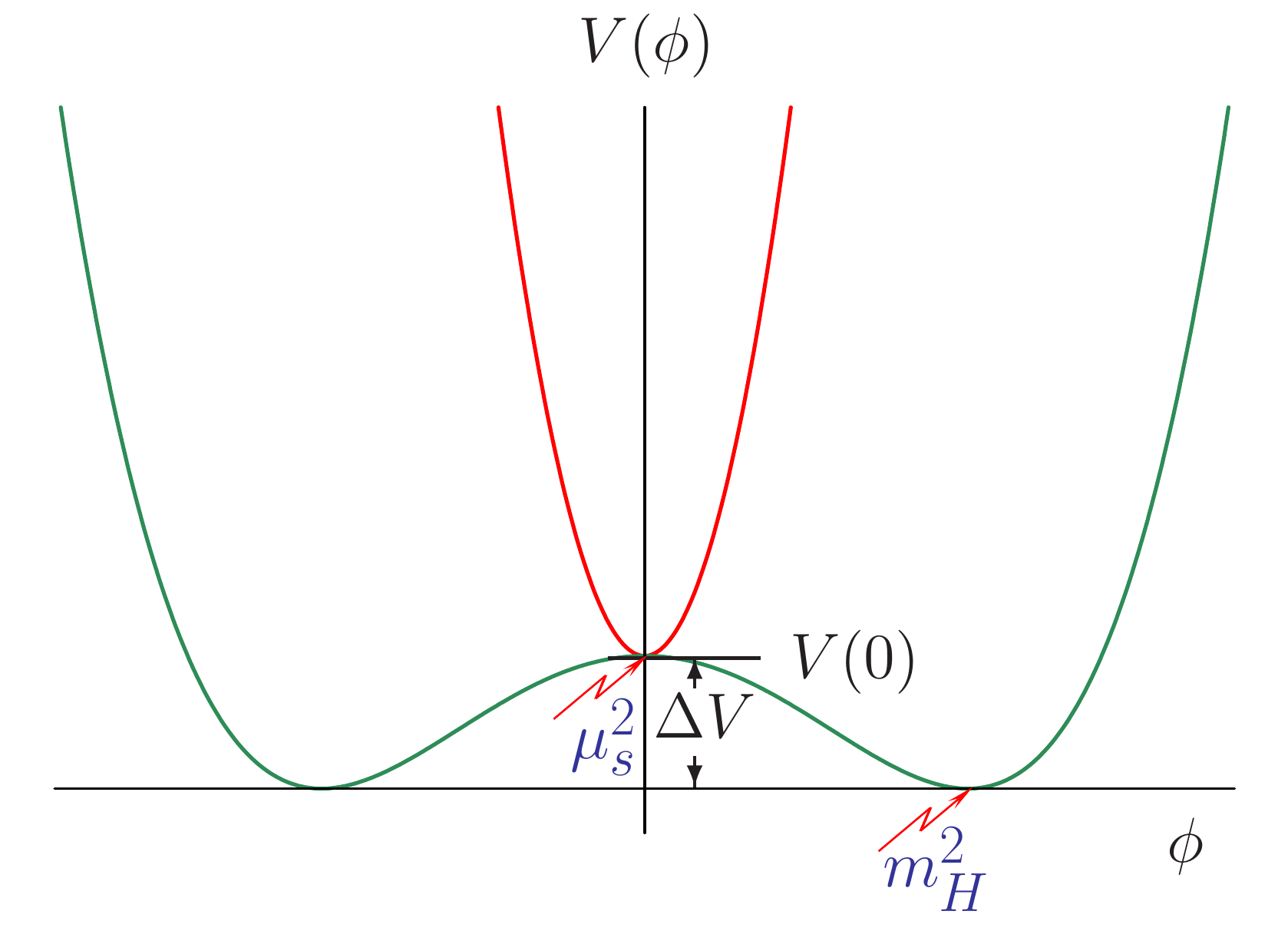}
\caption{Vacuum rearrangement by the Higgs phase transitions. The
large offset $V(0)$ essentially gets nullified at the transition point.}
\label{fig:rearrange}
\end{figure}

At the Higgs transition point, as soon as \mbo{{m'}^2 < 0} for \mbo{\mu <
\mu'_0}, the vacuum rearrangement of the Higgs potential takes place. As a
result at the minimum $\phi_v$ of the potential, we should get
$V(0)+V(\phi_v)\sim \left(0.00171~\mbox{eV}\right)^4$ about the observed
value of today's CC (see Fig.~\ref{fig:rearrange}).
How can this be? Indeed, at the zero of
\mbo{X(\mu)} we have
\mbo{\rho_{\Lambda 0}=\rho_{\Lambda\,}} and one
may expect that like the Higgs boson mass another free SM parameter is
to be fixed by experiment here\footnote{The appearance of an
non-vanishing $v$ provides a large negative contribution, which however
by far does not compensate the large positive offset
\mbo{\braket{V(\phi)}} we have from the symmetric phase. A more
accurate analysis would have to take into account subleading effects
from the chiral phase transition of QCD as-well.}.  One might expect
$\rho_{\Lambda}$ to be naturally small, since the $\lpl^4$ term is
nullified at the matching point. Note that the huge cutoff prefactors
act as amplifiers of small changes in the effective SM couplings. But
how small we should expect the low energy effective CC to be? In fact,
in the LEESM scenario, neither the Higgs boson mass nor the CC is really a
free parameters in the low-energy world. Given the other relevant SM
parameter, the Higgs self-coupling has to be constrained to a window
where the Higgs potential remains stable up to the Planck
scale. Similarly, the originally large CC, which is required to
provide a sufficient amount of inflation, has to get tuned down such
that inflation ends up at the critical density of a flat universe. The
late CC as part of the critical density then only can be a fraction of
the latter.

\subsection{A self-organized cosmological constant?}
Implications of inflation we already outlined in
Sect.~\ref{sec:prelude} after Eq.~(\ref{rhocrit}). Our analysis of the
LEESM showed that the CC is very much time dependent especially
through the running of the SM parameters and phase transitions taking
place in the evolution of the universe (see also the Quintessence
scenario advocated in~\cite{quintessence}). The typical problem is that in
general one gets a CC that is way too big, and this looks to create a
tremendous fine-tuning problem. For the SM this concerns the
contribution to the vacuum density via the Higgs-field VEV in the broken
phase, as well as the contributions from spontaneous breakdown of
chiral symmetry, which both are much too big and even of wrong
sign. Interestingly, our Higgs inflation scenario predicts a large
positive DE, which actually implies that $\rho_{\rm tot} \gg \rho_{\rm
crit}$ before inflation sets in. This means that at Planck time $k=+1$ and $\Omega_k$ in
Eq.~(\ref{omagaone}) evaluated at Planck time is large negative if
$a(\tpl)$ is of Planck size.  It is important to keep in mind that in
Big Bang cosmology $\rho_{\rm tot}$ at the beginning is always
dominated by the radiation density since $\rho_{\rm rad}\propto
a(t)^{-4}$ grows fastest when $a(t)$ gets smaller as we are going back in
time. Because they also scale with inverse powers in $a(t)$, also the
matter-density and the curvature terms first overshoot the CC supplied
by Higgs boson system. This is possible because $\rho_\Lambda
\sim (1.29\, \mpl)^4$ is of comparatively moderate size, although extremely big
relative to the critical density. However, if
inflation is at work, the final vacuum density is fixed, whatever the
initial density has been. Given that $\Omega_{\rm
tot}=\Omega_\Lambda+\Omega_{\rm DM}+\Omega_{\rm BM}+\Omega_{\rm
rad}=1$ with $1>\Omega_{\rm
DM}>\Omega_{\rm
BM}>\Omega_{\rm rad}>0$  we know that $\Omega_\Lambda$ being positive must
be of order $\Omega_{\rm tot}$, actually a fraction of it. As a
non-vanishing \mbo{\rho_{\Lambda 0}} at Planck time is needed, it is not unlikely that
the other components contributing to the total energy density do not
saturate the bound. Actually, we know that normal matter including the
tiny radiation density represents about 5\% of the critical density
only. This means that the fine-tuning is dynamically
enforced by inflation and the value of today's dark energy density
\bea
{\rho_{0\Lambda}=\mu^4_{0\Lambda}}\semis {\mu_{0\Lambda}=0.00171~\mathrm{eV}}
\eea
looks all but exotic. While $\Omega_{\rm rad}$ and very likely
$\Omega_{\rm BM}$ are essentially LEESM predictions if we include the
$B+L$ violating dimension 6 four-fermion operators, $\Omega_{\rm DM}$
is the only missing piece which remains an open problem and definitely
requires additional beyond the SM physics. This also concerns
contributions from quark- and possible gluon-condensates, which we do
not explicitly consider here.

\begin{figure}
\centering
\includegraphics[height=3cm]{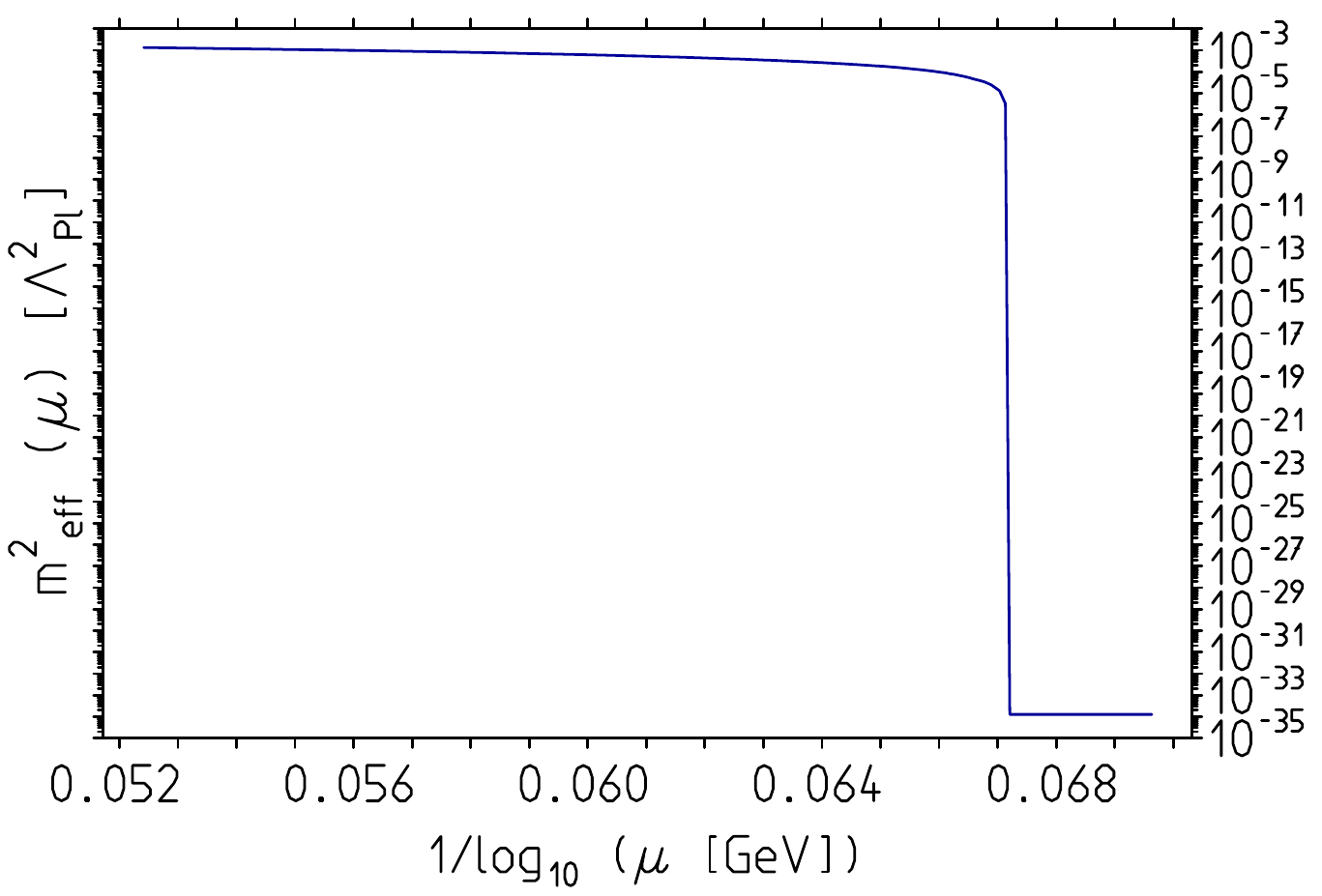}
\includegraphics[height=3cm]{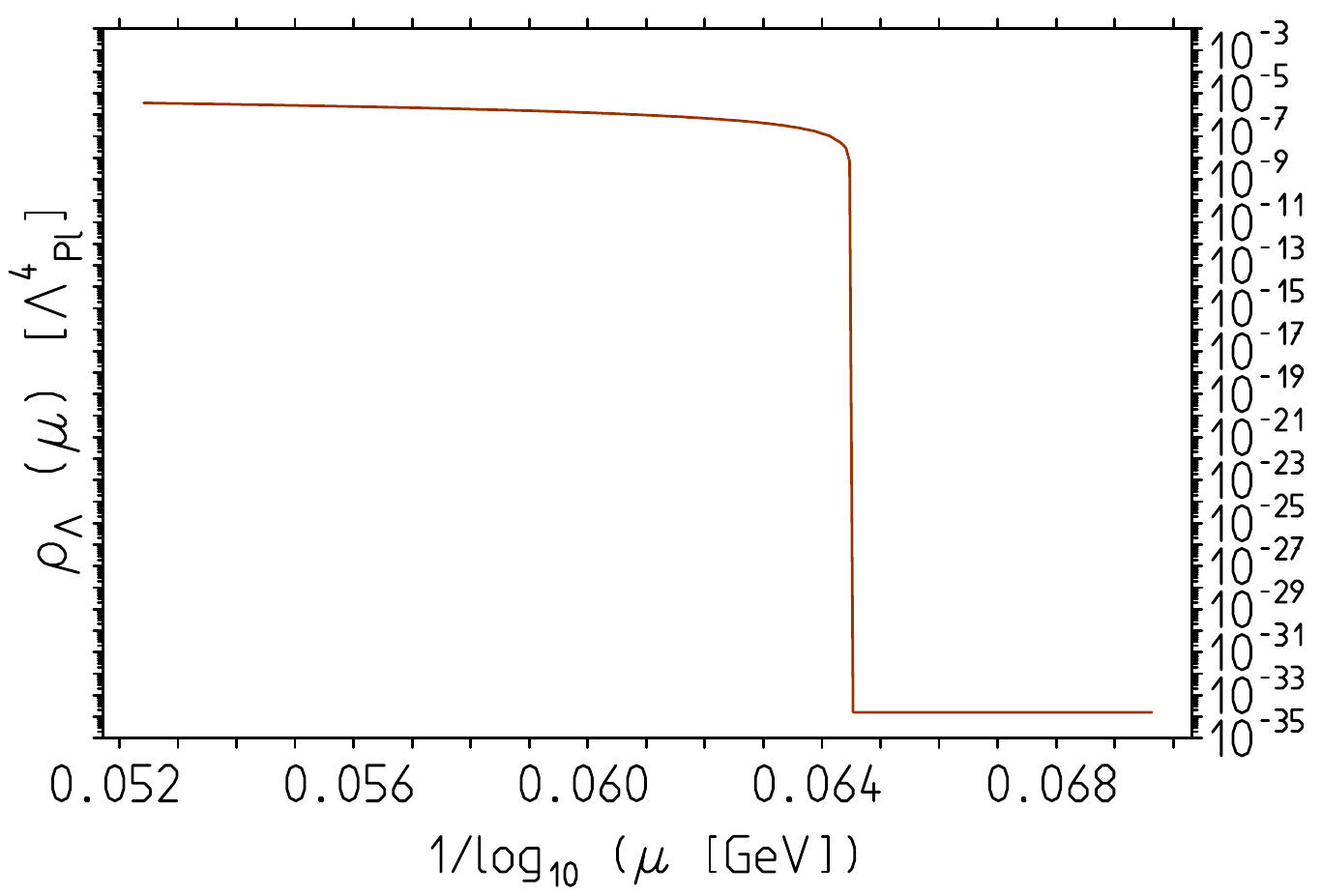}
\caption{The Higgs potential
effective $m^2_{\rm eff}$ [left] and the dark energy density
$\rho_\Lambda$ [right], in units of
\mbo{\lpl}, as functions of ``time'', represented here by $1/\log_{10}\mu$
, where $\mu$ represents the energy at that time. Below
the matching point $\mu_{\rm CC}$, where $\rho_\Lambda \simeq 1.6
\power{-47}$ in Planck mass units, we show a scaled up $\rho_\Lambda
\power{13}$ value of the present dark energy density $\mu_{0\Lambda}^4$
with $\mu_{0\Lambda}\simeq 0.00171~\mathrm{eV}$.
Note: $\rho_\Lambda(t)$ includes besides the large positive
\mbo{V(0)} also negative contributions from vacuum condensates, like
\mbo{\Delta \rho_{\rm EW}} from the Higgs mechanism and \mbo{\Delta
\rho_{\rm QCD}} from the chiral phase transition}
\label{fig:CCandm2x}
\end{figure}

Provided SM parameters indeed support a stable Higgs potential up to
$\mpl$, inflation and the CC itself are SM ingredients leading to a
highly self-consistent conspiracy which shapes the
universe. Fig.~\ref{fig:CCandm2x} shows the development of the
quadratically and the quartically enhanced terms in the symmetric
phase of the SM, and its matching to the low energy world.

\section{Inflation and reheating}
\label{sec:inflation}
\begin{figure}
\centering
\includegraphics[height=4cm]{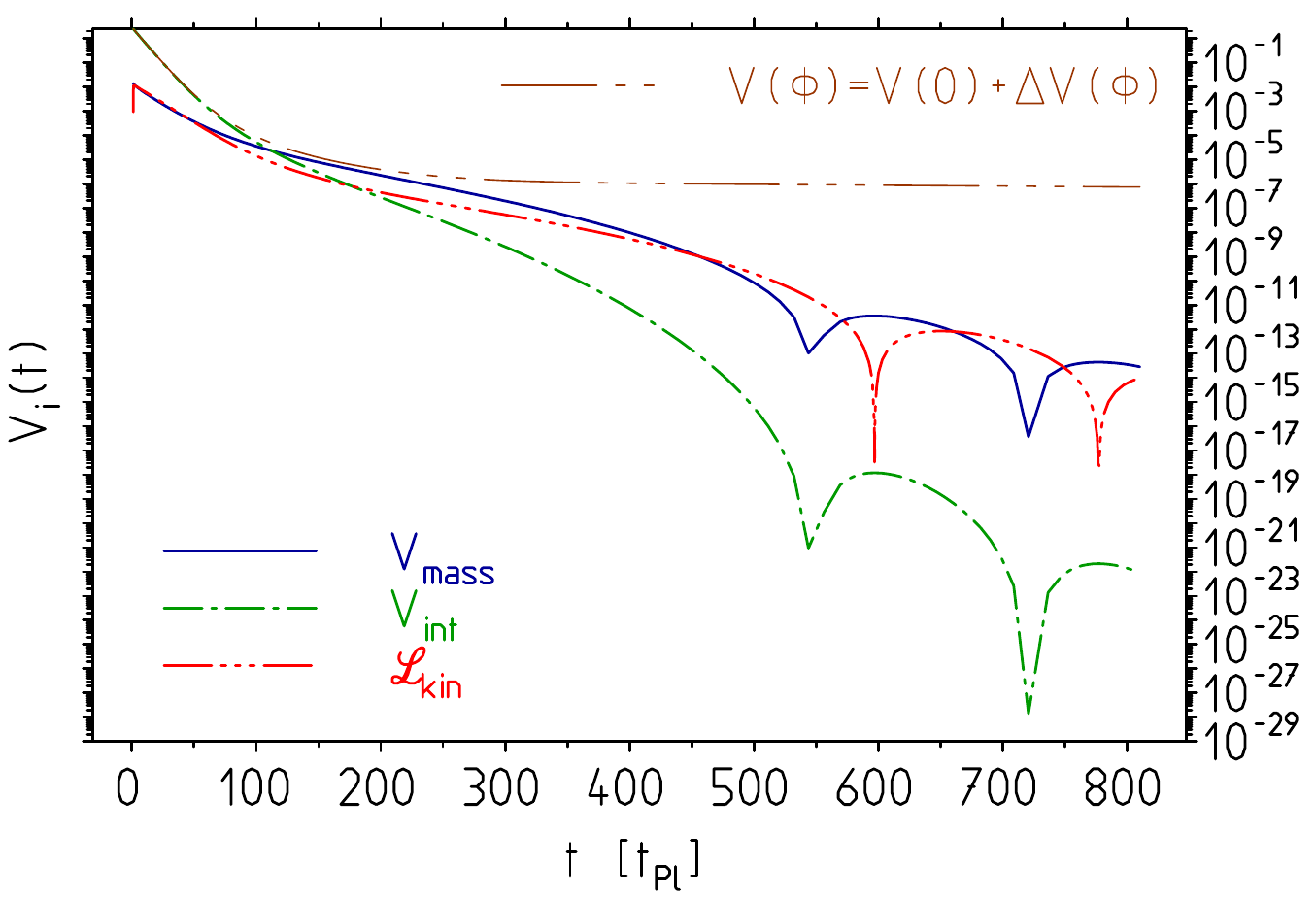}
\includegraphics[height=4cm]{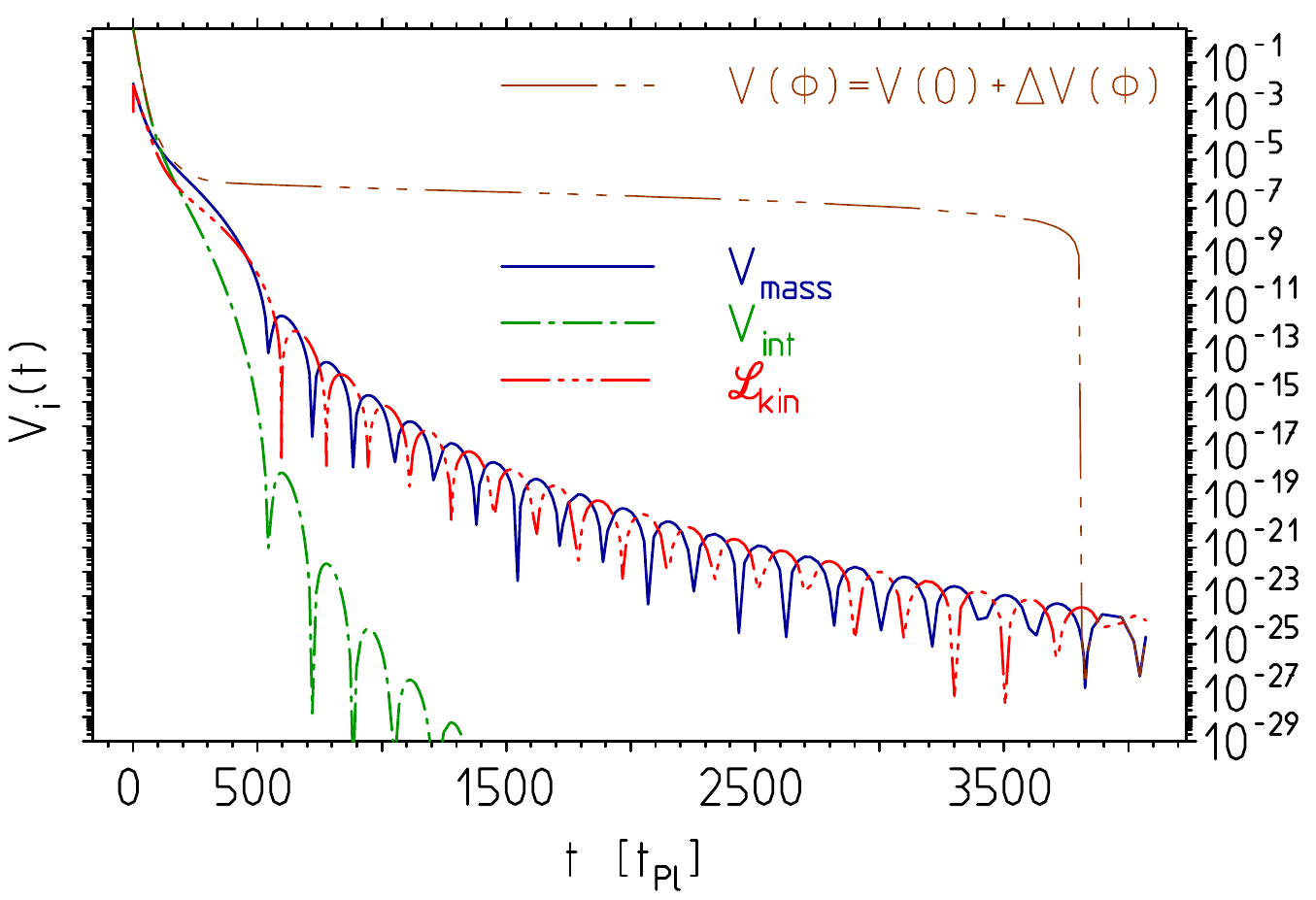}
\caption{The evolution of the universe before the Higgs phase
transition. Left: during inflation times -- the mass-, interaction-
and kinetic-term of the bare Lagrangian in units of $\mpl^4$ as a
function of time. Right: evolution until symmetry breakdown and
vanishing of the CC. After inflation the scene is characterized by a
free damped harmonic oscillator behavior}
\label{fig:Lagran}
\end{figure}

In contrast to standard scenarios of modeling the evolution of the
early universe, SM cosmology is characterized by the fact that almost
everything is known, within uncertainties of the parameters and
perturbative approximations. In LEESM cosmology the form of the
potential is given by the bare Higgs potential
$V(\phi)=\frac{m^2}{2}\,\phi^2+\frac{\lambda}{24}\,\phi^4$ as part of the SM
Lagrangian, the parameters are known, calculable in terms of the low
energy parameters, the only unknown is the magnitude of the Higgs
field. The latter must be large -- trans-Planckian -- in order to get
the required number of \textit{e-folds} $N$ given by
\bea
N\equiv \ln
\frac{a(t_\mathrm{e})}{a(t_\mathrm{i})}=\int_{t_i}^{t_e} H(t)
\D t \simeq
-\frac{8\pi}{M_\mathrm{Pl}^2}\,\int_{\phi_i}^{\phi_e}\frac{V}{V'}\,\D \phi\epo
\eea
The second form is obtained using the field equation (\ref{fieldeq}).
Note that $N$ is determined entirely by the scalar potential. Needed
is \mbo{N \gsim 60} in order to cover the CMB causal cone.  By
definition, $\exp N$ is the expansion factor $a(t_e)/a(t_i)=\exp
H\,(t_e-t_i)$, where $a(t)$ is the Friedmann-Robertson-Walker radius
of the universe at cosmic time $t$, $t_i$ denotes the begin of
inflation and $t_e$ the end of inflation and $H$ the Hubble
constant. For our set of \MSb input parameters we require
$\phi_0=\phi(\mu=\mpl) \approx 4.5\,\mpl$. Shortly after start the
slow-roll condition $V(\phi) \gg \frac12\,\dot{\phi}^2$ is well
satisfied, by the fact that in the symmetric phase the mass term as
well as $V(0)=\braket{V(\phi)}$ are huge and start to dominate
quickly. Because of the large initial field strength $\phi_0$,
however, the interaction term is actually dominating for a short time
after the initial Planck time $\tpl$. The field equation
\mbo{\ddot{\phi}+3H\dot{\phi}=-V'(\phi)} then predicts a dramatic
decay of the field, \mbo{\phi(t)=\phi_0\,\E^{E_0\,(t-t_0)}} with
\mbo{E_0=\sqrt{2\lambda}/(3\sqrt{3}\ell)\approx
4.3\power{17}~\gv\,,\,\,V_{\mathrm{int}}\gg V_{\mathrm{mass}}} and
shortly after \mbo{E_0=m^2/(3\ell\sqrt{V(0)})\approx
6.6\power{17}~\gv\,,\,\,V_{\mathrm{mass}}} \mbo{\gg V_{\mathrm{int}}}
[\mbo{\ell^2=8\pi G_N/3}], such that in almost no time, still under
slow-roll conditions, the mass term dominates and for what follows the
field equation predicts an exponential decay followed by harmonic
oscillation setting in. The universe thus undergoes an epoch of
Gaussian inflation as confirmed by observation~\cite{Ade:2013ydc}. The
time evolution is displayed in Fig.~\ref{fig:Lagran} and it is very
interesting to see which term dominates during which time
slice. Obviously, the fast decay of the Higgs
field stops inflation (see Fig.~\ref{fig:field}), in spite of the fact that a CC
$V(0)$ persists to be substantial at first.

A highly non-trivial challenge is the calculation of the spectral
indices
\bea
\eps \equiv
\frac{\mpl^2}{8\pi}\,\frac12\,\left(\frac{V'}{V}\right)^2\semis
\eta \equiv \frac{\mpl^2}{8\pi}\,\frac{V^{''}}{V}\,,
\label{indicesdef}
\eea
which sensitively depend on the form of the potential and which have
been extracted from the observed CMB radiation fluctuation
spectrum. From the theory point of view the indices are constrained by
the slow-roll criteria $\eps \ll 1$, ensuring $p \simeq -\rho$, and
\mbo{\eps,\eta
\ll 1}, which ensures that the slow-roll condition hold for long enough
time, while maintaining
\mbo{\ddot{\phi}\ll3\,H\dot{\phi}\,} before oscillations start.
These conditions also should be satisfied in order to ensure the
required amount of inflation. In our case, we are confronted with a SM
prediction modulo the unknown Higgs field $\phi_0=\phi(\mu=\mpl)$. The
calculation presented in~\cite{Jegerlehner:2014mua} shows that a
prediction of $\eta$ is delicate, but present uncertainties in
predicting the bare Higgs potential at post-inflation times certainly
do not allow us the draw definite conclusions. Given that there are
many predictions that look to work surprisingly well, I would be
surprised if the Higgs boson inflation would not work in predicting
also the spectral exponents in agreement with observation at
the end.

\subsection{Reheating by Higgs boson decays}
In the symmetric phase, all four Higgs fields are physical and very
heavy and rather unstable. The Yukawa couplings at inflation times are
pretty well known and the Higgs bosons decay predominantly (largest
Yukawa couplings) into as yet massless top-antitop pairs and lighter
fermion-antifermion pairs \mbo{H,\phi^0 \to
t\bar{t},\,b\bar{b},\,\cdots\;,H^+\to t\bar{b}\, \cdots\;,H^-
\to \bar{t}b\, \cdots} and are thereby reheating the young
universe, which just had been cooled down dramatically by inflation.
Preheating is suppressed in SM inflation as in the symmetric phase
bosonic decay channels like $H\to WW$ and $H\to ZZ$ are absent at tree
level. The CP-violating decays
\mbo{H^+\to t\bar{d}} [rate \mbo{\propto y_ty_d\,V_{td}}]\,\mbo{H^-\to
b\bar{u}} [rate \mbo{\propto y_by_u\,V_{ub}}] likely are important for
baryogenesis. Closely following the Higgs transition, where $m^2$ in
the Higgs potential changes sign, the electroweak phase transition
takes place. After it, the now heavy top-quarks decay into normal
matter as driven by CKM~\cite{CKM} couplings and phase space. At these
scales the $B+L$ violating dimension 6
operators~\cite{Weinberg:1979sa,Buchmuller:1985jz,Grzadkowski:2010es}
can still play a key role for baryogenesis and together with decays
like \mbo{t\to d e^+\nu} provide CP violating reactions during a phase
out of thermal equilibrium\footnote{We note that, in contrast to
claims that the SM cannot explain baryogenesis, the latter looks to be
well possible within the LEESM scenario, provided the EW phase-transition happens not too far below the Planck scale, at a scale
$\mu_0$ where the mentioned dimension 6 four-fermion operators can be
sufficiently effective i.e. \mbo{(\mu_0/\lpl)^2\sim 1.4\power{-6}} is
not too small. The observed baryon asymmetry is $\eta_B \sim
10^{-10}$. Remains the question of whether CP violation as given by the SM
is big enough and sufficiently efficient in the new context.}. For details
see~\cite{Jegerlehner:2013cta,Jegerlehner:2013nna,Jegerlehner:2014mua}.

A very different model of Higgs inflation, which has barely something
in common with our LEESM scenario, is the Minkowski-Zee-Shaposhnikov
et
al.~\cite{Minkowski:1977aj,Zee:1978wi,Bezrukov:2007ep,Barbon:2009ya,Bezrukov:2010jz,Bezrukov:2014bra,Bezrukov:2014ipa}
so-called \textit{non-minimal SM inflation} scenario. It is based on
the following points: i) Einstein gravity has to be extended by adding
\mbo{G_\munu \to G_\munu +\xi\, (\Phi^+\Phi)\,R\,g_\munu\,} to
Einstein's equation. On the source side
the model is assuming the renormalized low energy
\mbo{T_\munu} supplied by the renormalized SM (no relevant operator
enhancement). The new term is
a direct coupling of the gauge invariant Higgs field singlet operator
\mbo{\Phi^+\Phi} to the scalar Ricci curvature $R$.
This extra term violates the equivalence principle, yet so far without
observable consequences. ii) Choose
\mbo{\xi} large enough in order to get a sufficient amount of
inflation, which requires a rather large value \mbo{\xi \sim
10^4}. The entire inflation pattern then essentially depends on
\mbo{\xi} only (inflation ``by hand''). In case $\xi=O(1)$ the added
non-minimal coupling term is tiny and does not affect our LEESM or
standard inflation scenarios. iii) Assume quadratic and quartic SM
divergences are absent (argued by their absence in dimensional
regularization (DR) and \MSb renormalization, which is a misleading
purely formal argument in my opinion). iv) Assume the SM to be in the
broken phase at the Planck scale, which looks unnatural since SSB is a low
energy phenomenon, which assumes the symmetry to be restored at the
short distance scale!

Note: 1) It is well possible maybe even likely that such non-minimal
gravity couplings of the Higgs field exist and could play a role when
curvature is very high. However, the coupling $\xi$ would rather be
$O(1)$ than fine-tuned to be about $\xi \sim 10^{4}$. 2) Dimensional regularization and
\MSb renormalization both are possible in perturbation theory only. There
is no corresponding non-perturbative formulation (simulation on a
lattice) or measuring prescription (experimental procedure). The \MSb
scheme is based on a finite part prescription (singularities nullified by hand),
which can only be used to calculate quantities that do not exhibit
any singularities at the end. As elaborated earlier in
Sect.~\ref{sec:hierarchy}, the hierarchy problem cannot be addressed
within the dimensionally regularized SM or adopting the \MSb scheme in
a renormalized environment. In other words, dimensional
renormalization by no means explains the absence of power enhanced terms
in a LEET scenario. These terms are there and have to be accounted for.

\section{Remark on trans-Planckian Higgs fields}
\begin{figure}
\centering
\includegraphics[height=3.6cm]{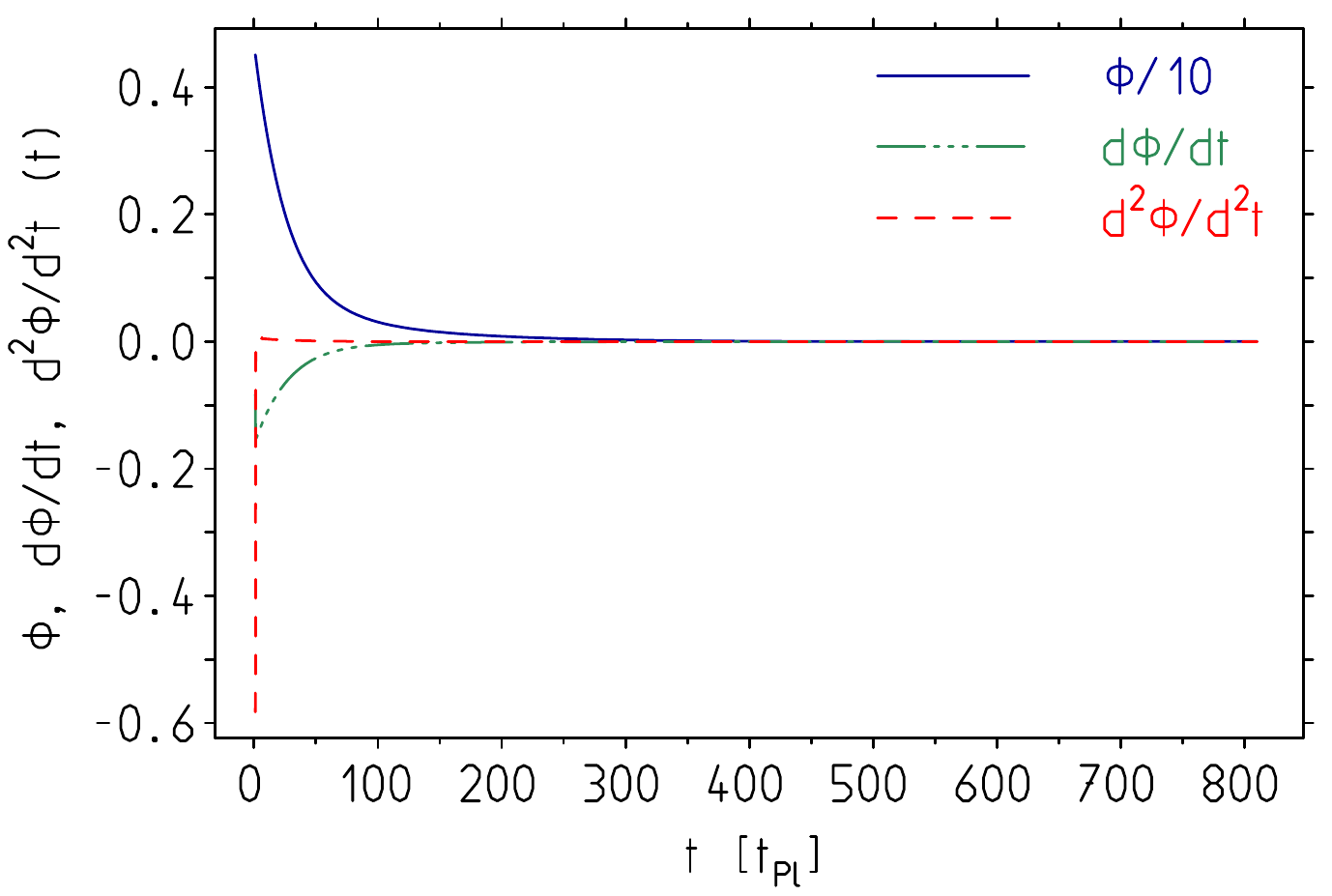}
\includegraphics[height=3.6cm]{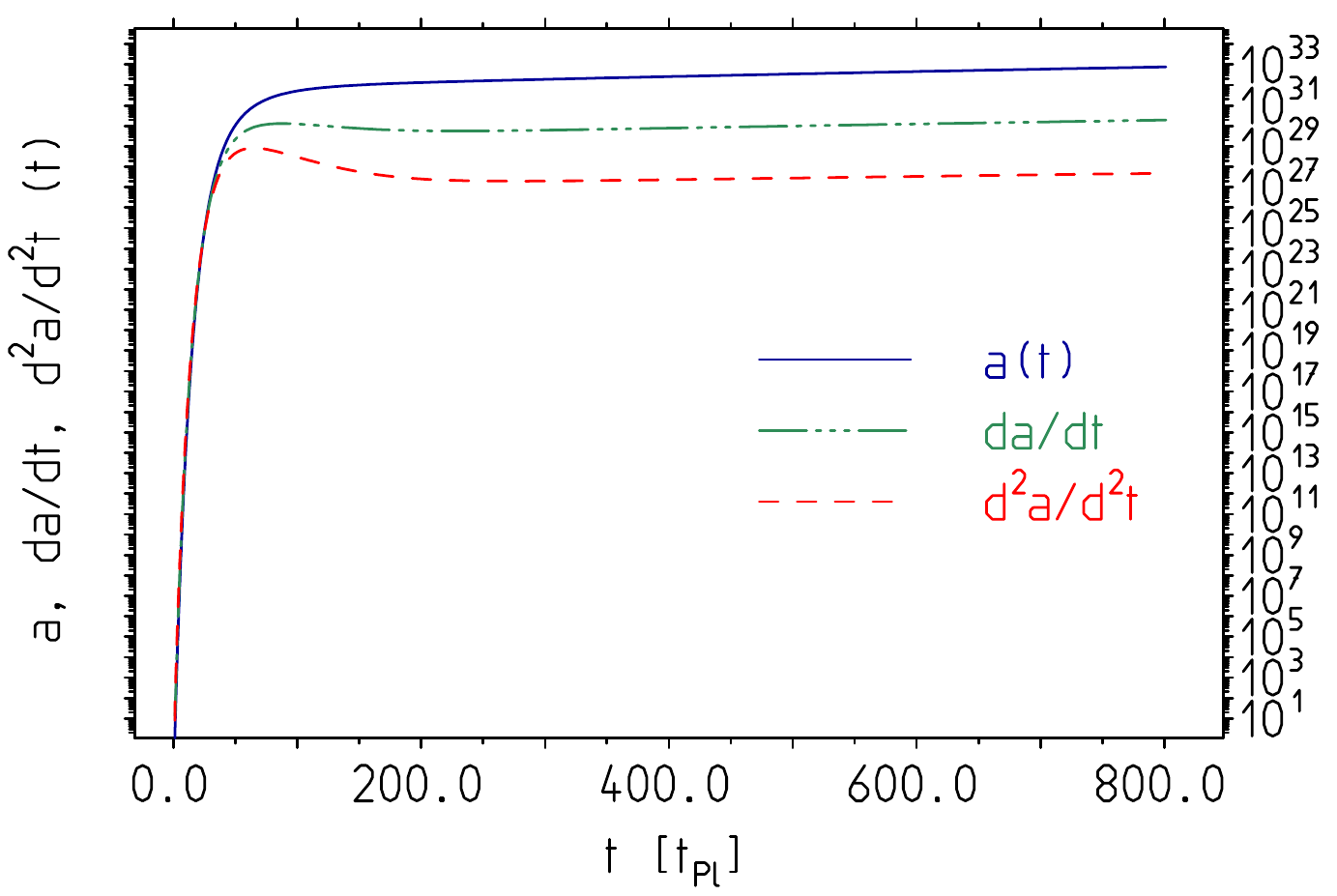}
\caption{The trans-Planckian Higgs field at $\tpl$ decays very fast
and inflation gets stopped soon. Left: the decaying Higgs field. Right:
the inflating $a(t)$.}
\label{fig:field}
\end{figure}

If the SM Higgs boson is the inflaton, sufficient inflation requires
trans-Planckian magnitude Higgs fields at the Planck scale. At the
cutoff scale, the low-energy expansion obviously gets obsolete and
likely we cannot seriously argue with field monomials and the operator
hierarchy appearing in the low energy expansion. What is important is
that the field is decaying very fast (see Fig.~\ref{fig:field}).
Formally, given a truncated series of operators in the potential, the
highest power is dominating when approaching the trans-Planckian
regime. One then expects that for some time the $\phi^4$ term of the
potential is dominating, the decay of the field is then exponential,
for higher dimensional operators it is faster than exponential, such
that the field very rapidly reaches the Planck- and sub-Planck
regime. This means that the mass term is dominating after a very short
period and before the kinetic term becomes relevant and slow-roll
inflation ends. So fears that higher order operators in low energy
effective scenarios with trans-Planckian fields would mess up things
are unfounded\footnote{As mentioned earlier, the
constructive understanding of LEETs is Wilson's renormalization
semi-group, based on integrating out short distance fluctuations. This
produces all kinds, mostly of irrelevant higher order interactions. A
typical example is an Ising model, which by itself seen as the basic
microscopic system has simple nearest neighbor interactions only and
by the low-energy expansion develops a tower of higher order
operators, which at the short distance scale are simply absent
altogether. Such operators don't do any harm at the intrinsic short
distance scale. As a minimal fairly realistic Planck system
representative in the universality class of the SM, we may consider
the lattice SM, a SM generalization of lattice QCD, which also is
expected to behave decently at the short distance
scale.}. Obviously, without the precise knowledge of the Planck
physics, very close to the Planck scale we never will be able to make
a precise prediction of what is happening. This, however, seems not to
be a serious obstacle to quantitatively describe inflation and its
properties as far as they can be accessed by observation. The LEESM
scenario in principle predicts not only the form of the effective
potential not far below the Planck scale but also its parameters and
the only quantity not fixed by low energy physics is the magnitude of
the field at the Planck scale. We also have shown that taking into
account the running of the parameters is mandatory for understanding
inflation and reheating and all that.

Trans-Planckian fields are not unnatural in a low energy effective
scenario as the Planck medium exhibits a high temperature and
temperature fluctuations imply amply of higher excitations forming a
chaotic state.
While the Planck medium will never be accessible to direct
experimental tests, a phenomenological approach to constraining its
effective properties is obviously possible, especially by CMB
data~\cite{Ade:2013uln}.

In the extremely hot Planckian medium, the Hubble constant in the
radiation dominated state with effective number $
g_*(T)=g_B(T)+\frac78\,g_f(T)=102.75$ of relativistic degrees of
freedom is given by $H=\ell \sqrt{\rho}\simeq
1.66\,\left(k_B T\right)^2\sqrt{102.75}$ $\mpl^{-1}$, at Planck time
$ H_i \simeq 16.83\,\mpl$ such that a Higgs field of size
$\phi_i \simeq 4.51\, \mpl$, is not unexpectedly large and could
as well also be larger.

Often it is argued that trans-Planckian fields are unnatural in
particular in a LEET scenario~\cite{Lyth:1996im}. I cannot see any argument against
strong fields and LEET arguments (ordering operators with respect to a
polynomial expansion and their dimension) completely lose their sense
when \mbo{E/\lpl ~\gapprox~ 1}.

As mentioned already, provided the Higgs field decays fast enough,
towards the end of inflation, we expect the mass term to be dominant
such that a Gaussian fluctuation spectrum prevails. The quasi-constant
CC $V(0)$ at these times mainly enters the Hubble
constant $H$ and does not affect the fluctuation spectrum. As the
originally large $V(0)$ get nullified at $\mu_{\rm CC}$ also the
Hubble constant suffers a jump down to a value as known from standard
$\Lambda$CDM cosmology.

\section{Remarks for the skeptic}

How do our results depend on the true UV completion? In other words,
how realistic are the numbers I have presented?

In order to answer these questions, we have to stress once more the
extreme size of the cutoff, $\MPl>>>>...$ from what we can see! This
let us look what we can explore to be ruled by fundamental principles
like the Wightman axioms (the ``Ten Commandments'' of QFT) or
extensions of them as they are imposed in deriving the renormalizable
SM. In the LEESM approach, many things are much more clear-cut than in
condensed matter systems, where cutoffs are directly accessible to
experiments. In other words, in condensed matter systems the gap between the microscopic structure
and its macroscopic manifestation is by far nowhere nearly as big as in our LEESM case. Also lattice QCD
simulations differ a lot, as cutoffs are always close-by, such that
lattice artifacts affect results throughout before the extrapolation to
the continuum has been performed.

We also have to stress that taking actual numbers too serious is
premature as long as even the realization of vacuum stability is in
question. Detailed results evidently depend sensitively on accurate
input values and on the perturbative approximations used for the
renormalization group coefficients as well as for the matching
relations needed to get the \MSb input parameters in terms of the
physical (on-shell) ones. After all, we are attempting to extrapolate
over 16 orders of magnitude in the energy scale. Such an attempt may
look to be megalomaniacal, but it is a bottom-up approach that appears
to lead to a reasonable and very possible scenario that is able to
explain and predict inflation and reheating. And it is a very modest
step in relation to attempts to construct a TOE for example.

The next question is how close to \mbo{\mpl} can we trust our
extrapolation?  It is very important to note that above the EW scale
[\mbo{v\sim 250~\gv}] perturbation theory seems to works the better
the closer we are near the Planck-cutoff, vacuum stability
presupposed. As long as we are talking about the perturbative regime
we can expand perturbative results in powers of
\mbo{E/\lpl} up to logarithms. Then we have full control over the
cutoff-dependence to order \mbo{\cO((E/\lpl)^2)}, corresponding to dim
\mbo{\geq 6} operator corrections. Effects \mbo{\cO((E/\lpl))}, related
to dim 5 operators, only show up in special circumstances e.g. in
scenarios related to generating neutrino masses and mixings and the
see-saw mechanism.

The true problem comes about when we approach the Planck scale, where
the expansion in \mbo{E/\lpl} completely breaks down. Especially, it
does not make sense to talk about a tower of operators of increasing
dimensions. This does not mean that everything gets out of control. If
the ``ether'' would be something that can be modeled by a lattice SM,
implemented similar to lattice QCD, one could still make useful
predictions, which eventually could be tested in cosmological
phenomena. In condensed matter physics it is well known that an
effective Heisenberg Hamiltonian allows one to catch essential
properties of the system, although the real structure cannot be
expected to be reproduced in the details. One also should keep in mind
that models like the mentioned Ginzburg-Landau effective theory of
superconductivity perfectly models the phase transition between type
I, type II superconductivity and normal state, without reference to
the true microscopic structure. In any case, it is always possible to
find out to what extent the description fits reality.

It is well known that long-range physics manifests itself as field theory
naturally from underlying classical statistical systems exhibiting
short-range exchange interactions (e.g. nearest-neighbor interactions
on a lattice system)~\cite{Wilson:1971bg} (for lectures on the topic
see e.g.~\cite{LausanneLectures1976}). However, the Planck system not
only shows typical short-range interactions. We know that it also
features a long-range gravitational potential, which develops
multipole excitations showing up as spin 1, spin 2 and higher modes at long
distances~\cite{Jegerlehner:1978nk}.

In our context what is important is that the quadratic and quartic
enhancements are persisting, as well as the running (screening or
anti-screening effects) of couplings and their competition and
conspiracy. Together these elements manifest themselves in the
existence of the zeros of the enhanced terms, provided these zeros are
not to close to the cutoff. A look at Fig.~\ref{fig:jump} shows that
such effects can be dramatic fairly well below the cutoff. Again, the
perturbativeness, together with the fact that leading corrections to
these results are by dim 6 operators, let us expect that results are
reliable at the
\mbo{10^{-4}} level up to \mbo{10^{17}~\gv}, which is in the
middle of the symmetric phase already. Once the phase transition has
happened, the running is anyway weak and even when cutoff-effect are
starting to play a role they cannot spoil the relevant qualitative
features concerning triggering inflation, reheating and related
phenomena.

Lattice SM simulations in the appropriate parameter range of vacuum
stability, keeping top-quark Yukawa and Higgs self-coupling to
behave asymptotically free, which requires to include simultaneously
besides the Higgs system also the top-quark Yukawa sector and QCD, could
help to investigate such problems quantitatively. Experience from
lattice QCD simulations may not directly be illustrative since usually
the cutoff is rather close and a crucial difference is also
the true non-perturbative nature of low energy QCD.

In any case, not to include the effects related to the relevant
operators (dim \mbo{< 4}) simply must give wrong results. Even
substantial uncertainties, which certainly show up closer to the
cutoff in power-like behaved quantities, seem to be an acceptable
shortcoming in comparison to not taking into account the
cutoff-enhancements at all (as usually done).

\section{Summary and conclusions}
\label{sec:summary}
A cutoff regularized SM with the Planck mass as a cutoff is considered
to exhibit the relevant features of the physical Planck world in the
sense that it resides in the same universality class with respect to
its long-range behavior. The SM we observe at low energy is then the
emergent renormalizable effective theory of the Planck medium. All
conditions that usually have to be imposed, as principles to ensure
renormalizability, are emergent as a result of the low energy
expansion. The relation between long distance physics and short
distance physics can be controlled in principle via Wilson's
renormalization semi-group. Taking into account renormalization
effects and the ``running'' of the parameters are mandatory in order to
understand what has been happening in the evolution of the
universe. The key outcome of the LEESM setting is that for
appropriately tuned parameters the quadratic as well as the quartic
``singularities'' are nullified at a specific energy, where bosonic
and fermionic effects cancel\footnote{Note that the locations of the
nearby zeros of $\delta m_H^2$ Eq.~(\ref{quadraic1}) and $\delta
\rho_\Lambda$ Eq.~(\ref{rholambdaCT}) are independent of the value of
the cutoff $\Lambda$, but they depend very sensitively on the input
parameters specified in Tab.~\ref{tab:MSbinp}.}. In the SM the
corresponding zeros happen because the bosonic and the fermionic
contributions are running differently, according to their respective
renormalization groups, and at some point turn out to be of equal
magnitude.

In this scenario the Higgs field/particle has two different functions
in our world: 1) it has to render the effective low energy electroweak
theory (massive vector-boson and fermion sector) renormalizable. In
place of fermion mass terms, we have fermion Yukawa couplings to start
with, while gauge boson mass terms enter via the kinetic Higgs term
through the covariant derivative that has to include the gauge
fields. In the broken low energy phase, the Higgs field forms a vacuum
condensate, which provides masses to all massive fields including the
Higgs boson itself. The key point is the many new Higgs field
exchange-forces necessary to render the low energy amplitudes
renormalizable. 2) in the symmetric phase there exist four very heavy
Higgs bosons ($H,\phi^0,H^\pm$) that generate a huge positive dark
energy, as required to triggers inflation. After inflation has ended
and we are out of equilibrium the Higgs bosons are decaying
predominantly into the fermions pairs with largest Yukawa couplings
(predominantly at this stage still massless top-antitop pairs), which
provides the reheating of the inflated universe. The universe cooling
further down then pushes the universe into the Higgs phase, where the
particles acquire their masses. The predominating heavy quarks decay
into the lighter ones, which later form the baryons and normal
matter. This scenario is possible because of the quadratically
enhanced Higgs boson mass which together with the quartically enhanced dark energy,
shows up in the symmetric phase of the SM before the transition
into the Higgs phase. The existence of such relevant operator effects,
in my opinion, is supported by observed inflation patterns. Both, the
hierarchy ``problem'' as well as the cosmological constant ``problem''
reflect important properties of the SM needed to understand the
evolution of the early universe (for different opinions
see~\cite{Aoki:2012xs,Blanke:2013uia,Tavares:2013dga,Masina:2013wja,Bian:2013xra}). Consolidation
of our bottom-up path to physics near the Planck scale will sensibly
depend on progress in high precision physics around the EW scale
$v$. Especially, Higgs boson and top-quark factories (like FCC-ee or ILC)
will play a key role in this context.

Concerning the presumed fine-tuning problem: the scales $\mpl$ and $v$
stand for different regimes and there is no reason why they
should not be vastly different; one is related to gravity (Planck medium)
the other to long-range order at low energies. It is the energy dependence
of the SM interactions that triggers spontaneous symmetry breaking. The
emergent SM symmetry apparently only three orders of magnitude below $\mpl$
gets broken by a non-symmetric ground state rearrangement. The
critical point nevertheless is the actual value of $v$ that is
non-vanishing only below a critical temperature. While in a condensed
matter system one is adjusting the temperature by hand, the key
problem seems to be that in particle physics we cannot adjust the
temperature. But this the expanding universe is doing for us. In fact,
the hot big bang universe provides a scan of the temperature spectrum
and automatically triggers the phase transition at some point, as the
calculations show. Temporary out of equilibrium phases do not disturb
the gross behavior of this scan, but they will have to be investigated
in any case.  For more details I refer to my original
articles~\cite{Jegerlehner:2013cta,Jegerlehner:2014mua} and to my
Krakow Lectures~\cite{KrakowLectures2014}.

The scenario I advocate requires a change of paradigm, to one where
the SM with its structure is emergent from a Planck cutoff-medium
following a minimal self-organized ``strategy'', i.e. conspiracies are
taking place to make structures to show up at large distances (for a
short history of development of the emergence paradigm
see e.g.~\cite{PCorningEmergence15}) . This looks like a version of an
anthropic principle at work. It lets look the SM to be more natural
than many of the BSM scenarios we have heard about during the last
about 45 years. Although the SM started to turn out to work well and
to work better than ever expected, we know it cannot explain a number
of observationally established facts. Yes, the SM misses dark matter,
singlet neutrinos, axions and likely more, but all of these have room
in an emergent scenario. This is in contrast to the top-bottom
philosophy behind the most popular BSM physics scenarios like string
theory, supersymmetry or grand unification, which assume that the
short distance world is intrinsically highly symmetric and symmetries
are broken spontaneously only because renormalizability is always
assumed to be a basic law of nature. We know that the SM as seen at
low energies is a spontaneously broken gauge theory, which gets more
symmetric as we go to higher energies because mass operators in a
high-energy expansion turn into irrelevant operators. This may have
lead to a wrong perception concerning what we have to expect on the
path to higher energies. This view overlooks the fact that a tower of
possible symmetry breaking irrelevant operators of the low energy
expansion turn into relevant contributions at high energies. Thus
symmetries as seen at low energy usually do not persist at higher
energies. The dream that an eternal highly symmetric ``theory of
everything'' should sit at and above the Planck scale may not be very
realistic. The opposite very probably is true, the world looks more
complex the closer we look, and symmetries emerge from not resolving
the detailed structure behind.  And the final truth remains something
we can get closer only but will never be reached.

Between my Higgs inflation scenario and the metastability scenery
favored in~\cite{Yukawa:3,Degrassi:2012ry,Buttazzo:2013uya,Bednyakov:2015sca},
the major difference is that for me understanding the relationship
between the physical low energy parameters and the bare parameter,
assumed to become the physical ones at short distances, is the
mandatory premise. Most other analyses are working with a
renormalized effective potential in the broken phase all the way up
towards the Planck scale and do not consider cutoff effects to be
physical. Power enhanced cutoff-effects, in my scenario, are triggering a phase transition
between an early symmetric and a later broken phase. Higgs system
conditioned inflation is possible only if in the early universe the SM
is in the symmetric phase.

I think the LEESM scenario has a good chance to find its confirmation
along with the lines described in this article. Many aspects need to
be checked and possibly modified. Admittedly, there are many open
questions, which should be investigated more thoroughly. One
conclusion seems to be unavoidable, namely that the SM Higgs sector
provides dark energy that affects early as well as late
cosmology. Obviously, discovering SUSY, a GUT, extra-dimensions, a
little Higgs extension, Technicolor or similar extensions of the SM
would spoil the scene. Notably, also one more fermion family or one
less would completely mess up everything. The sharp dependence of the
Higgs vacuum stability on the SM input parameters and on possible SM
extensions and the vastly different scenarios which can result as a
consequence of minor shifts in parameter space makes the stable vacuum
case a particularly interesting one and it could reveal the Higgs
particle as ``the master of the universe''. After all, it is commonly
accepted that dark energy is the``stuff'' shaping the universe both at
very early as well as at the late times.

\section{Appendix: How natural is the minimal SM?}
\label{sec:appendix}
Often it is considered that it would be more natural to have a
left-right symmetric world including mirror fermions. The following
consideration, which goes back to Veltman~\cite{Veltbrighton}, is
instructive as it helps to understand why parity violation is quite
natural and why QED conserves parity. It has a lot to do with the
assumption of a minimal Higgs system. I reproduce a version, which I
had presented in~\cite{Jegerlehner:1991dq} some time ago. Actually,
within the context of our LEESM scenario, we gain a much deeper
insight, because the assumptions made are now emergent properties
resulting from the low energy expansion.

In order to try to derive the SM let us make the following
assumptions: \\
1) local field theory\\
2) interactions follow from a local gauge principle\\
3) renormalizability\\
4) masses derive from the minimal Higgs system \\
5) $\nu _R$ which we know must exist does not carry hypercharge.\\
Note that all points besides the last one are emergent structures in a LEESM
as we may learn
from~\cite{Veltman:1968ki,LlewellynSmith:1973ey,Bell:1973ex,Cornwall:1973tb,Jegerlehner:1978nk,Jegerlehner:1994zp,Jegerlehner:1998kt}
(see also Sect.~\ref{sec:paradigms}).
We admit that the last assumption looks somewhat ad hoc, but nevertheless we
make it. From the above assumptions the following picture develops:
\begin{itemize}
\item
For the gauge interactions, the simplest non-trivial possibility is
that the fundamental massless matter fields group according to the simplest
possibilities, into doublets and triplets, which are the fundamental
representations of $\SU(2)$ and $\SU(3)$, besides possible singlets.
\item
Since fields are massless all fields can be chosen left-handed.
Left-handed particles and left-handed antiparticles
at this stage are uncorrelated.
\item
We must have {\em pairing} for particles that are going to be massive, since a
mass term (we ignore the possibility to have Majorana fields here) has the form
$\bar{\psi} \psi = \bar{\psi}_L \psi_R + \bar{\psi}_R \psi_L$.
Notice that for massive particles only, we know which left-handed antiparticle
belongs to which left-handed particle to form a Dirac field.
\item
For $\SU(3)_c$ triplets we {\em must} have pairing in order to avoid axial
anomalies. $\SU(3)$ is the simplest group having complex representations. This
allows putting particles in $3$ and antiparticles in the inequivalent $3^*$.
As a consequence a
rich color singlet structure ($\equiv$ hadron spectrum) results. Furthermore,
confinement requires $\SU(3)_c$ to be unbroken!
\item
$\SU(2)_L$ is anomaly free and hence there is no anomaly condition associated
with this group. To generate mass we have to break $\SU(2)_L$ by a Higgs
mechanism. The simplest and natural possibility is to choose one Higgs field
in the
fundamental representation of $\SU(2)_L$. There is no hypercharge for the
moment. The Higgs field may be written in the form
\bea
\Phi_b = \widetilde{\Phi}\: \chi_b   \; ; \;
\chi_b =                \left ( \begin{array}{c}
                                  0 \\
                                  1
                                          \end{array} \right ) \nn
\eea
in terms of a $2 \times 2$ matrix field ($\tau_i\,,\,\, i=1,2,3\,$ the
Pauli matrices)
\bea
\widetilde{\Phi} = \frac{1}{\sqrt{2}}(H_s +\I \tau_i \phi_i)\;.  \nn
\eea
The covariant derivative being given by
\bea
D_\mu  \Phi_b =
(\partial_\mu -\I \frac{g}{2} \tau_a W_{\mu a})\, \Phi_b \; , \nn
\eea
and the $\SU(2)$ invariant renormalizable Higgs system
\bea
{\cal L}_{\rm Higgs}=
\left (  D_\mu \Phi_b \right )^+ \left (  D^\mu \Phi_b \right )
- \lambda \left ( \Phi_b^+ \Phi_b \right )^2
+ \mu^2 \left ( \Phi_b^+ \Phi_b \right )\;,
\label{LHiggs}
\eea
exhibits an extra global $\SU(2)_R$-symmetry $\chi_b \ra V^+ \chi_b$.
One easily checks that the transformation
\bea
\widetilde{\Phi} \ra U(x)\, \widetilde{\Phi} V^+\,, \nn
\eea
with $U(x)\in \SU(2)_{L,{\rm local}}, V \in \SU(2)_{R,{\rm global}}$
leaves the Higgs Lagrangian invariant. This implies that the fields
($W^+, W_3, W^-$) form a weak isospin triplet with $M_Z=M_{W^{\pm}}$. \\
Now consider the fermions (still no hypercharge). Since $L_f$ and
$\Phi_b$ are $\SU(2)$ doublets there also \textit{must} exist singlet fermions
$R_f$, otherwise we would not be able to write down an invariant and
renormalizable fermion-Higgs coupling. Therefore $\SU(2)_L$ \textit{must} be
parity violating of V-A-type! The Yukawa term has the general form
\bea
{\cal L} _{\rm Yukawa} = - \bar{L}_f \widetilde{\Phi}
\left ( \begin{array}{c} g_{11}~~g_{12} \\
			 g_{21}~~g_{22} \end{array} \right ) R_f + h.c.\,, \nn
\eea
with 4 complex couplings $g_{ij}$ and $R_f$ a ``doublet'' having two
right-handed singlets as entries. Although we have not used hypercharge to
restrict these couplings the existence of a global $\SU(2)_R$-symmetry of the
Higgs system allows us to transform the Yukawa couplings
\bea
\widetilde{\Phi}\, (\cdot)\, R_f \ra \widetilde{\Phi} V^+ (\cdot) W R_f \nn
\eea
to standard form, $V^+ (\cdot) W$= real diagonal. Since $V \in
\SU(2)_R$ has 3 parameters and $W$ is an arbitrary unitary matrix with
4 parameters we end up with one free parameter such that the system
exhibits a global $U(1)$ invariance. This is not surprising since in
the unitary gauge we always can end up only with ${\cal L}_{\rm
Yukawa}$ in the simple standard form
\bea
{\cal L}_{\rm Yukawa}&=&- \sum_{f} m_f \bar{\psi}_f \psi_f \;(1 +
\frac{H}{v})\epo
\label{LYdiag}
\eea
\item
The global $U(1)$ which is a consequence of the minimal Higgs
mechanism may be {\em interpreted} as a global $U(1)_Y$. We are free
to assign $Y=1$ to $\Phi_b$, which means nothing else than that we
measure $Y$ in units of the $\Phi_b$- hypercharge. Then
\bea
\Phi_t = \widetilde{\Phi}\: \chi_t \; ; \;
\chi_t =                \left ( \begin{array}{c}
                                  1 \\
                                  0
                                          \end{array} \right ) \nn
\eea
has $Y=-1$ , and we may write $\widetilde{\Phi}=(\Phi_b,\Phi_t)$. Since we
have the global $U(1)_Y$ for free, we may {\em assume} this symmetry to be
{\em local}. The covariant derivative for $\widetilde{\Phi}$ now reads
\bea
D_\mu \widetilde{\Phi}= \partial_\mu \widetilde{\Phi}+\I \frac{g'}{2}B_\mu \widetilde{\Phi}
\,\tau_3 -\I \frac{g}{2} \tau_a W_{\mu a}\widetilde{\Phi} \nn
\eea
and we find back the usual Higgs Lagrangian
\bea
{\cal L}_{\rm Higgs}&=&\frac{1}{2}(\partial_\mu H \partial^\mu H)
+\frac{(H+v)^2}{2 v^2}(M_Z^2 Z_\mu Z^\mu +2 M_W^2 W_\mu^+ W^{- \mu})
\crn && -\frac{\lambda}{4} H^4 -\lambda v H^3 -\frac{1}{2} m_H^2 H^2\epo
\label{HLdiag}
\eea
The 3 real fields $\phi_a
\; a=1,2,3 $ could and have been gauged away and only 3 out of 4 gauge fields can
acquire a mass.  Hence there must exist one massless field, the
photon! Evidently we obtain the relations $g'=g \tan \Theta_W$ and
$\rho=\mw / (\mz \cosW)=1$ ! instead of $M_Z=M_{W^{\pm}}$ when
$g'=0$.\\ Now, what can we say about the hypercharge of the
fermions?\\ A left-handed doublet transforms like
\bea
L \ra e^{\I \frac{g'}{2}Y_L} L\,,  \nn
\eea
where $Y_L$ is arbitrary. By inspection of ${\cal L}_{\rm Yukawa}$ we
find for the hypercharges of the singlets: $\psi_{1R}$ must have
$Y_{1R}=Y_L+1$ and $\psi_{2R}$ must have $Y_{2R}=Y_L-1$. One
consequence is that $U(1)_Y$ {\em must} violate parity. The
astonishing thing is that the fermion current which couples to the
photon preserves parity. By inspection we find
\bea
D_\mu L_f &=& (\partial_\mu -\I\,\frac{g'}{2}Y_L B-\mu
  -\I \frac{g}{2} \tau_3 W_{\mu 3}-\cdots )\, L_f \,, \crn
D_\mu R_f &=& (\partial_\mu -\I\,\frac{g'}{2}Y_L B-\mu
  -\I \frac{g}{2} \tau_3 B_{\mu}-\cdots )\, R_f\,, \nn
\eea
and the couplings of $L_f$ and $R_f$ to $A_\mu$ read
\bea
L_f \;: && -\I\,(g \siW \ \frac{\tau_3}{2} + g' \coW \frac{Y_L}{2})\, A_\mu  \crn
R_f \;: && -\I\,(g' \coW \ \frac{\tau_3}{2} + g' \coW \frac{Y_L}{2})\, A_\mu \;. \nn
\eea
Because we have $g' \coW = g \siW = e$  we find
the Gell-Mann-Nishijima relation
\bea
Q=T_3+\frac{Y}{2}         \nn
\eea
as a consequence of a minimal Higgs structure! What we find is that,
whatever the hypercharge of $L_f$ is, $L_f$ and $R_f$ must couple
identically to photons.  Thus QED must be parity conserving!
Furthermore, the charges of the upper (1) and lower (2) components of
the doublets satisfy
\bea
Q_{Li}=Q_{Ri}\; ,\; Q_1-Q_2=1 \;\mathrm{and} \; Q_1+Q_2=Y_L \;.\nn
\eea
So far we have no charge quantization. Here we need the last assumption.
\item
If $\nu_R$ does not couple to the $U(1)$ gauge field, we have to set $Y_{\nu R}=0$ and consequently we must
have $Y_{\nu L}=-1=Y_{\ell L}=0$ and $Q_\nu=0$, $Q_\ell=-1$. For the
$U(1)_Y$ anomaly cancellation we need lepton-quark duality and
the charges of the quarks must have their
known values if they appear in three colors. One thus must have the usual
charge quantization.
\end{itemize}
We finally summarize the consequences of the assumptions stated above:
\begin{itemize}
\item
breaking $\SU(2)_L$ by a minimal Higgs automatically leads to a global
$U(1)_Y$, which can be gauged
\item
parity violation of $\SU(2)_L$
\item
$\rho =\mw /(\mz \cosW)=1$
\item
the existence of the photon
\item
parity conservation of QED
\item
the validity of the Gell-Mann-Nishijima relation
\item
family structure
\item
charge quantization
\end{itemize}
We do not know why right-handed neutrinos are sterile i.e. do not
couple to gauge bosons. In the SM of electroweak interactions,
neutrinos originally were assumed to be massless i.e. that
right-handed neutrinos did not exist. This is definitely ruled out by
the observation of neutrino oscillations\footnote{We note that
right-handed neutrinos have been expected to exist from the point of
view of the SM gauge symmetry. They have been excluded by imposing
additional global flavor conservation, because the latter had not been
observed within the achieved experimental accuracy. In fact flavor
violation is so tiny that a direct observation still is missing. The
corresponding smallness of neutrino masses very likely is a
consequence of a see-saw mechanism, which could be triggered within the
SM when the singlets would be of Majorana type. Majorana mass terms
are not protected by any symmetry and therefore would be expected to
be related to the Planck scale in a similar way as the Higgs field
mass in the symmetric phase.}.

I think this reasoning is able to help understanding how various
excitations in the chaotic Planck medium develop a pattern like the SM
as a low energy effective structure. Renormalizability as a
consequence of the low energy expansion and the very large gap between
the EW and the Planck scales plus a certain minimality (not too little
but not too much e.g. only up to symmetry triplets) determines the SM
structure without much freedom. After all, minimality is not a new
concept in physic as we know from the principle of least action.
Three fermion families are required in order CP violation emerges in a
natural way, and to make baryogenesis eventually possible within the
LEESM scenario as addressed in Sect.~\ref{sec:inflation}. We have been
emphasizing the high self-consistency of the SM where all essential
structures look to be emergent properties in the low energy effective
viewport of a cutoff-system residing at the Planck scale. ``What is
not capable of surviving at long distances does not exist there''
(Darwin revisited).

\bigskip

\noindent
Acknowledgments:\\
I thank the organizers of the
\textit{Naturalness, Hierarchy and Fine Tuning Workshop}, at the RWTH Aachen,
for the kind invitation and the support.

\end{document}